\documentclass[11pt,a4paper]{JHEP3}
\usepackage{amsmath,amssymb}
\usepackage{euscript}
\usepackage{slashed}
\usepackage{amsfonts}
\usepackage{verbatim}
\def\be{\begin{eqnarray}}
 \def\ee{\end{eqnarray}}
\def\0{\nonumber}
\def\del{\partial}

\global\long\def\pochhammer#1#2{\left(#1\right)_{#2}}
\usepackage[latin9]{inputenc}
\usepackage{amsmath}
\usepackage{amsthm}
\usepackage{amssymb}
\usepackage{esint}

\preprint{SISSA/46/2016/FISI\\ZTF-EP-16-03\\{\tt hep-th/1609.02088 } }

\title{One-loop effective actions and higher spins}

\author{ L.~Bonora$^{a}$, M.~Cvitan$^{b}$, P.~Dominis
Prester$^{c}$, S.Giaccari$^{b}$, B.~Lima de Souza$^{a}$, T.~$\bf \check{S}$temberga$^{b}$
\\\textit{${}^{a}$ International School for Advanced Studies (SISSA),\\Via
Bonomea 265, 34136 Trieste, Italy, and INFN, Sezione di
Trieste\\}%
\textit{${}^{b}$ Theoretical Physics Division of Particles and Fields, Faculty
of Science, University of Zagreb, 
Bijeni\v{c}ka 32, HR-10000 Zagreb, Croatia\\}%
\textit{${}^{c}$ Department of Physics, University of Rijeka,\\
Radmile Matej\v{c}i\'{c} 2, 51000 Rijeka, Croatia\\}%
E-mail: \email{bonora@sissa.it}, \email{mcvitan@phy.hr}, 
\email{pprester@phy.uniri.hr}, \email{blima@sissa.it},\email{sgiaccari@phy.hr},\email{tstember@phy.hr}}

\abstract{The idea we advocate in this paper is that the one-loop effective action of a
free (massive) field theory coupled to external sources (via conserved currents)
contains complete information about the classical dynamics of such
sources. We show many explicit examples of this fact for (scalar and fermion) free field
theories in
various dimensions $d=3,4,5,6$ coupled to (bosonic, completely symmetric) sources with a  number
of spins. In some cases we also provide compact formulas for any dimension. This paper 
is devoted to two-point correlators, so the one-loop effective action we
construct contains only the quadratic terms and the relevant equations of motion
for the sources we obtain are the linearized ones. }


\begin{document}
%\maketitle

\section{Introduction. Higher spins are everywhere}
\label{sec:intro}
The idea we wish to support in this paper is that the one-loop effective action of a
free (massive) field theory coupled to external sources (via conserved currents)
contains complete information about the possible classical dynamics of the
sources. We
exhibit several examples of this fact for (scalar and fermion) free field
theories in
various dimensions $d=3,4,5,6$ coupled to (bosonic) sources with a large number
of spins. In some cases we also provide compact formulas for any dimension. In this paper we
concentrate on two-point correlators, so the one-loop effective action we
construct contains only the quadratic part. Consequently the equations of motion
for the sources we obtain are the linearized ones. We postpone to a future
work the analysis of one-point and three-point correlators. But our thesis is that the
dynamics generated by the one-loop effective action (OLEA) contains all the
information we need to reconstruct complete interacting equations of motion. 

This paper is a follow-up of \cite{BCLPS}, which contains a few (mostly parity
odd) 3d examples of what has just been said. The present paper is more general and
systematic, not limited to 3d, and devoted especially to the parity even sector.

As we have just mentioned, the crucial issue here is the calculation of the
two-point functions of free massive field theories coupled to external sources.
We do it via Feynman diagrams. 
This is in principle a simple calculation, and to carry it out we resort to a
method introduced by Davydychev and collaborators, \cite{BoosDavy}.  However, as
we will see,  although we can derive from it very general formulas they are expressed
in terms of  hypergeometric functions and derivatives thereof and not easily
`readable'. For this reason it is
often very useful to expand such results near their IR and UV fixed points.
These expansions in powers of the mass $m$ in odd dimensions, and $m$ and $\log
m$ in even dimensions, allows us to single out the
dynamics of the sources and will be referred to as {\it tomography}. In other words
what we do is to describe RG trajectories of two-point current correlators that 
pass through those of free massive theories, but
we focus in particular on their IR and UV expansions, where the physical content is more
easily recognizable. 

Such IR and UV expansions are necessary also for another reason: one has to check 
that the IR and UV limits of the one-loop effective action are well defined.
We find in fact divergent and non-conserved term in the limit $m\to\infty$. These terms 
are local and can be subtracted. We subtract as well the IR finite terms, which are also local.
This is to make the OLEA well defined and scheme independent.
The results obtained in this way, in particular in the even parity sector, transferred to the OLEA, allow
us to find the linearized Fronsdal eom's (see \cite{Fronsdal,Curtright,Singh}) for all the source fields we have considered,
in the nonlocal form introduced by Francia and Sagnotti, \cite{FS}. In 3d we
consider also the odd parity sector, and confirm the connection with
Pope and Townsend's generalizations of Chern-Simons theory, already pointed out in \cite{BCLPS}. 

In this paper, for the purpose of comparison, we also analyze massless free theories beside the corresponding massive ones. The difference 
between the two is that the latter allow us to control not only the UV but also the IR, while in the former 
only the UV is regularized. This explains the difference in the results. In general in the massless case we 
do not get all the information we can extract from the massive theory and  
many results are scheme dependent. Briefly stated, at least for the purpose of this paper, to make 
sure we get a complete information we must use massive models.

The subject of this paper is inspired by the idea of exploring theories with infinite many fields,
\cite{Maldacena}, in particular string theories and Vasiliev-type higher spin theories, 
\cite{Vasiliev}. As shown in the body of the paper higher spin fields appear naturally in the 
one-loop effective action of the simplest free theories in any dimension and it is possible to make contact
with the literature on classical higher spin theories, \cite{solvay2004,sorokin}. Other sources of 
inspiration have been \cite{Sakharov,Maldacena-Zhiboedov,closset,campoleoni,giombi,GMPTWY}. The idea of exploring 
the one-loop effective action is far from new: the list of works which may have some overlap with 
our paper includes \cite{Jackiw, Appelquist,Vuorio,Babu, Delbourgo,Dunne, Gama,bekaert}, but is likely to be incomplete. 
>From a technical point of view this paper continues the line of research started with \cite{BL, BGL,BGLBled13,BDL} 
with more powerful techniques (a new Mathematica code). 

The paper is organized as follows. In the next section we introduce the massive scalar and fermion
 model and define the relevant OLEA's. Section 3 is meant to explain the motivation for this research by means 
of simple concrete examples. We also introduce the issue of higher spin Fronsdal eom and their various forms. 
In section 4 we introduce a new representation of higher spin eom's in momentum space and their general form, 
which is independent on the dimension of space-time. Section 5 contains a short summary of Davydychev's 
method to compute one-loop Feynman diagrams. In section \ref{3dEA} to \ref{sec:6d} we analyze the one-loop scalar and fermion 
model two-point functions and their IR and UV expansion (tomography) in 3, 5, 4 and 6 dimensions, respectively. 
In section  \ref{sec:spins}
we produce the formulas for two-point correlators of spin 1, 2, 3 currents in any dimensions. Section \ref{Conclusion} is devoted to 
the conclusion. Appendix A contains the demonstration of a result used in section 4 and Appendix B
the analysis of the massless scalar and fermion models.

\section{Free field theory models}

In this paper we limit ourselves to two type of models, the free scalar and free
fermion, although it is not hard to extend the analysis to other models. 
By the first we mean the complex scalar theory defined by the Lagrangian
\be 
L= \partial_\mu \phi^\dagger \partial^\mu \phi-m^2 \phi^\dagger
\phi\label{scalarlag}
\ee
in any dimension. On shell the current 
\be
 J_\mu = {i}\left(\phi^\dagger \partial_\mu \phi -\partial_\mu \phi^\dagger \phi\right)
\label{scalarcurrent}
\ee
is conserved. We can couple it to a gauge field via the action term $\int d^dx\,
A^\mu(x) J_\mu(x)$. 
The scalar-scalar-gluon vertex with momenta
$p,\,p',\,k$, respectively, ($p$ incoming and $p',\,k$ outgoing), and the propagator
are, respectively,
\be
{i}(p+p')_\mu \delta(p-p'-k)\quad,\quad \frac i{p^2-m^2}\label{ssgvertex}
\ee
But, of course we can define infinite many completely symmetric (on shell)
conserved currents, of which (\ref{scalarcurrent}) is only the simplest example:
\be
J_{\mu_1\ldots\mu_s} ={i^s} \phi^\dagger \stackrel
{\leftrightarrow}{\partial_{\mu_1}} \ldots 
\stackrel {\leftrightarrow}{\partial_{\mu_s}}\phi\label{scalarcurrents}
\ee
They couple minimally to external spin $s$ fields, $a^{\mu_1\ldots \mu_s}$. The
on-shell current conservation
implies (to the lowest order) invariance under the gauge transformations
\be
\delta a_{\mu_1\ldots\mu_s}= \partial_{(\mu_1} \Lambda_{\mu_2\ldots
\mu_s)}\label{gaugesym}
\ee  
where round brackets stand for symmetrization. 

In the case $s=2$ the conserved current
is the energy-momentum tensor and the external source is the metric fluctuation
$h_{\mu\nu}$, where $g_{\mu\nu}=\eta_{\mu\nu}+h_{\mu\nu}$. In this case the
action is the integral of (\ref{scalarlag}) multiplied by $\sqrt g$.
The vertex for an incoming scalar with momentum $p$ and outgoing scalar with
momentum $p'$ and an outgoing spin-$s$ field with momentum $k$
is
\be
V_{sst}: \quad {i} (p+p')_{\mu_1} \ldots (p+p')_{\mu_s} \delta^{(d)}(p-p'-k)
\label{Vsst}
\ee   

The free fermion model is represented by a Dirac fermion coupled to a
gauge field. The action is 
\be
S[A]=\int
d^{3}x\,\left[i\bar{\psi} \gamma^{\mu}D_{\mu}\psi-m\bar{\psi}
\psi\right],\quad D_{\mu}=
\partial_{\mu}+ A_\mu \label{actionA}
\ee
where $A_\mu= A_\mu^a(x) T^a$ and $T^a$ are the generators of a gauge algebra in
a given representation determined by $\psi$. We will use the antihermitean
convention, so
$[T^a,T^b]=f^{abc} T^c$, and the normalization ${\rm tr}(T^a T^b)={ -}\delta^{ab}$. 

The current
\be
J^a_\mu(x) = i\bar \psi \gamma_\mu T^a \psi\label{Jmu}
\ee
is (classically) covariantly conserved on shell as a consequence of the gauge
invariance of (\ref{actionA})
\be
(DJ)^a = (\del^\mu \delta^{ac} + f^{abc} A^{b\mu})
J_\mu^c=0\label{currentconserv}
\ee

The next example involves the coupling to gravity
\begin{equation}
S[h]=\int
d^{3}x\,e\left[i\bar{\psi}E_{a}^{\mu}\gamma^{a}\nabla_{\mu}\psi-m\bar{\psi}
\psi\right],\quad\nabla_{\mu}=
\partial_{\mu}+\frac{1}{2}\omega_{\mu
bc}\Sigma^{bc},\quad\Sigma^{bc}=\frac{1}{4}\left[\gamma^{b},\gamma^{c}\right]
.\label{actiong}
\end{equation}
The corresponding energy momentum tensor
\begin{equation}
T_{\mu\nu}^{(g)}=\frac{i}{4}\bar{\psi}\left(\gamma_{\mu}\!\stackrel{\leftrightarrow}{
\nabla}_{\nu}+\gamma_{\nu}\!\stackrel{\leftrightarrow}{\nabla}_{\mu}
\right)\psi.\label{EMtensorg}
\end{equation}
is covariantly conserved on shell as a consequence of the diffeomorphism
invariance of the action,
\be
\nabla^\mu T_{\mu\nu}(x)=0.\label{Tconserv}
\ee
If we expand the metric around the flat spacetime, $g_{\mu\nu}(x) = \eta_{\mu\nu} + h_{\mu\nu}(x)$, then, contrary to spin-1 case, interaction is not linear in the gauge field, which is $h_{\mu\nu}$. However, for the purposes of this paper, only linear term matters, and it is given by coupling the flat space energy-momentum tensor
\begin{equation}
T_{\mu\nu}=\frac{i}{4}\bar{\psi}\left(\gamma_{\mu}\!\stackrel{\leftrightarrow}{
\partial}_{\nu}+\gamma_{\nu}\!\stackrel{\leftrightarrow}{\partial}_{\mu}
\right)\psi.\label{EMtensor}
\end{equation}
to the metric fluctuation $h_{\mu\nu}$. 

Again we can couple the fermions to more general fields. Consider the free
action
\be
S_0=\int
d^{3}x\,\left[i\bar{\psi} \gamma^{\mu}\del_{\mu}\psi-m\bar{\psi}
\psi\right], \label{actionfree}
\ee
and the spin three conserved current 
\begin{eqnarray}
J_{\mu_1\mu_2\mu_3} &=&  {-
\frac 12 \bar \psi \gamma_{(\mu_1} \partial_{\mu_2}
\partial_{\mu_3)}\psi
-
\frac 12  \partial_{(\mu_1}
\partial_{\mu_2}\bar \psi \gamma_{\mu_3)} \psi
+}\frac 5{3} \partial_{(\mu_1} \bar \psi \gamma_{\mu_2}
\partial_{\mu_3)}\psi 
\0\\
&&
{-} \frac 13 \eta_{(\mu_1\mu_2} \partial^\sigma \bar \psi
\gamma_{\mu_3)}
\partial_\sigma \psi {+}\frac {m^2}3 \eta_{(\mu_1\mu_2} \bar \psi \gamma_{\mu_3)} \psi \label{J3}
\end{eqnarray}
Using the equation of motion one can prove that 
\begin{eqnarray}
&&\partial^\mu J_{\mu\nu\lambda} =0 \label{conservJ3}\\
&& J_{\mu}{}^\mu{}_\lambda = {- }\frac 49 m \left( -i \partial_\lambda \bar \psi
\psi +i \bar \psi \partial_\lambda \psi +2 \bar\psi \gamma_\lambda \psi
\right)\label{trJ3}
\end{eqnarray}
Therefore, the spin three current (\ref{J3}) is conserved on shell and its
tracelessness is softly broken by the mass term. Similarly to the gauge field
and the metric, we can couple
the fermion $\psi$ to a new external source $b_{\mu\nu\lambda}$ by adding to
(\ref{actionfree}) the term
\begin{equation}
\int d^3x J_{\mu\nu\lambda} b^{\mu\nu\lambda} \label{3rdordercoupling}
\end{equation}
Due to the (on shell) current conservation this coupling is invariant (to lowest
order) under the
infinitesimal gauge transformations
\begin{equation} \label{Bgaugetransf}
\delta b_{\mu\nu\lambda} = \partial_{(\mu} \Lambda_{\nu\lambda)}
\end{equation}
In the limit $m\to 0$ we have
also invariance under the local transformations
\begin{equation}
\delta  b_{\mu\nu\lambda}=\Lambda_{(\mu} \eta_{\nu\lambda)} \label{Bweyltransf}
\end{equation}
which are usually referred to as (generalized) Weyl transformations and which 
induce the tracelessness of $J_{\mu\nu\lambda}$ in any couple of indices. 

We can construct on-shell conserved currents for any spin $s$, but their form is
more complicated than in the scalar case. The explicit expressions can be found
in \cite{BCLPS}.

We notice that to lowest order in the external sources the relevant action,
in all cases above, takes the form
of the free action + a linear interaction term such as (\ref{3rdordercoupling}).
We make the identification
$a_\mu=A_\mu, a_{\mu\nu} { \sim}h_{\mu\nu}, a_{\mu\nu\lambda} { \sim} b_{\mu\nu\lambda}$, with
the obvious exception of the non-Abelian field in (\ref{actionA}). The latter
will be the only case in which we consider non-Abelian external sources.\footnote{Also note that the nonlinearity present in spin-2 case, which is forced by the consistency requirements, is a signal that we should expect the same for higher-spin fields. However, this is not relevant in our two-point calculations.}

\subsection{Generating functions and effective actions}
\label{ssec:genfunct}

In both scalar and fermion cases, the generating function for the external
source $a_{\mu_1 \ldots \mu_s}$ is
\begin{multline} \label{Was}
W[a,s] =W[0] +\sum_{n=1}^\infty \frac {i^{n-1}}{n!} \int \prod_{i=1}^n d^3x_i
a^{\mu_{11}\ldots \mu_{1s}}(x_1)  \ldots a^{\mu_{n1}\ldots \mu_{ns}}(x_n) \\ 
\times \langle 0|{\cal T} J^{(s)}_{\mu_{11}\ldots\mu_{1s}}(x_1)\ldots
J^{(s)}_{\mu_{n1}\ldots\mu_{ns}}(x_n)|0\rangle . 
\end{multline}
In particular $a_\mu=A_\mu, a_{\mu\nu}={ \frac14} h_{\mu\nu}$ and
$J^{(2)}_{\mu\nu}={ 2}T_{\mu\nu}$ with
$a_{\mu\nu\lambda}= b_{\mu\nu\lambda}$.
The full one-loop 1-pt correlator for $J_{\mu_1\ldots\mu_s}$ is
\begin{multline} \label{Japmux}
\langle\!\langle J^{(s)}_{\mu_1\ldots \mu_s}(x)\rangle\!\rangle = \frac{\delta
W[a,s]}{\delta a^{\mu_1\ldots \mu_s}(x)}
= \sum_{n=0}^\infty \frac {i^{n}}{n!} \int \prod_{i=1}^n d^3x_i 
a^{\mu_{11}\ldots \mu_{1s}}(x_1)  \ldots a^{\mu_{n1}\ldots \mu_{ns}}(x_n) \\
\times \langle 0|{\cal T} 
J^{(s)}_{\mu_{1}\ldots\mu_{s}}(x)J^{(s)}_{\mu_{11}\ldots\mu_{1s}}(x_1)\ldots
J^{(s)}_{\mu_{n1}\ldots\mu_{ns}}(x_n)|0\rangle . 
\end{multline}
The full one-loop conservation law for the energy-momentum tensor is
\be \label{emconserv}
\nabla^\mu\langle\!\langle T_{\mu\nu}(x)\rangle\!\rangle=0.
\ee
A similar covariant conservation should be written also for the other currents,
but for $s>2$ we will content ourselves with the lowest nontrivial 
order in which the conservation law reduces to
\be 
\del^{\mu_1} \langle\!\langle J^{(s)}_{\mu_1\ldots
\mu_s}(x)\rangle\!\rangle=0\label{Jpconserv}
\ee

\vskip 0.3cm

{\bf Warning}. One must be careful when applying the previous formulas for
generating functions.
If the expression $\langle 0|{\cal T}
J^{(s)}_{\mu_{11}\ldots\mu_{1s}}(x_1)\cdots
J^{(s)}_{\mu_{n1}\ldots\mu_{ns}}(x_n)|0\rangle$ in (\ref{Was}) is meant to
denote
the $n$-th point-function calculated by using Feynman diagrams, a factor $i^n$
is already
included in the diagram themselves and so it should be dropped in (\ref{Was}).
When the
current is the energy-momentum tensor an additional precaution is necessary: the
factor
$\frac {i^{n-1}}{n!}$ must be replaced by $\frac {i^{n-1}}{2^n n!}$.    
The factor 
$\frac 1{2^n}$  is motivated by the fact that when we expand the action 
$$S[\eta+h]= S[\eta] + \int d^dx \frac {\delta S}{\delta
g^{\mu\nu}}\Big{\vert}_{g=\eta} h^{\mu\nu}+\cdots,$$ 
the factor $ \frac {\delta S}{\delta g^{\mu\nu}}\Big{\vert}_{g=\eta}= \frac 12
T_{\mu\nu}$. 
Another consequence of this fact will be that the presence of vertices with one
graviton in 
Feynman diagrams will correspond to insertions of the operator
$\frac{1}{2}T_{\mu\nu}$ 
in correlation functions.

\vskip 0.3cm

Our purpose in this paper is to compute the effective action for the external
source fields at the quadratic order. As a consequence the first task is to
compute the two-point functions
\be 
\langle 0|{\cal T} J^{(s)}_{\mu_{1}\ldots\mu_{s}}(x)\,
J^{(s)}_{\nu_{1}\ldots\nu_{s}}(y)|0\rangle \label{2ptJs}
\ee
or their Fourier transforms
\be
\tilde J_{\mu_1\ldots\mu_s \nu_1\ldots \nu_s}(k)= \langle 0|{\cal T}\tilde
J^{(s)}_{\mu_{1}\ldots\mu_{s}}(k)\, \tilde J^{(s)}_{\nu_{1}\ldots\nu_{s}}(-k)|0\rangle 
\label{2ptJsk}
\ee
In the sequel we compute them by using the Feynman diagram technique. For all
two-point functions the only relevant diagram is the bubble diagram with one
spin $s$ line of ingoing momentum $k$ and one with the same outgoing momentum
and one scalar or fermion circulating in the internal loop. For instance the 2pt
function for the current $J^a$ in the fermion model is
\begin{multline}\label{Jmunuab}
\tilde J_{\mu\nu}^{ab}(k)= \langle \tilde J_\mu^a(k)   \tilde J_\nu^b(-k)\rangle =
\int \frac{d^3p}{(2\pi)^3} {\rm Tr}
\left({\gamma_\nu T^b}  \frac 1{\slashed{p} -m } {\gamma_\mu T^a}
\frac 1{\slashed{p}-\slashed{k} -m }\right) 
\end{multline}
while for the e.m. tensor it is
\be
\tilde T_{\mu\nu\lambda\rho}(k)\label{Tmnlr} &=&{ \frac14} \langle \tilde T_{\mu\nu}(k) \tilde
T_{\lambda\rho}(-k)\rangle
\\ &=& - \frac 1{64} \int \!\! \frac {d^3p}{(2\pi)^3}
{\rm Tr}\left(\frac 1{\slashed{p}-m}(2p-k)_\mu \gamma_\nu \frac
1{\slashed{p}-\slashed{k}-m}(2p-k)_\lambda \gamma_\rho\right),\0
\ee
where symmetrization of indices $(\mu,\nu)$ and $(\lambda,\rho)$ and the factor $\frac14$ is introduced accordingly to the above warning.

\section{An appetizer in 3d}
\label{sec:appetizer}

In \cite{BCLPS} we calculated in particular the two-point function of the
current $J^a$ in the fermion model
as well as its IR and UV limit. In the parity violating part we found a
well-known result: when Fourier antitransformed and inserted in the generating
function of the OLEA (\ref{Was}) it gives rise to the linearized version 
of the gauge CS action in 3d (which is in fact conformal invariant).  In
\cite{BCLPS} we did the same for the two-point correlator of the e.m. tensor for
the fermion model, and proceeding the same way we found the linearized
version of the gravity CS action. Something that was also known before,
\cite{Vuorio}. Repeating the same thing
for the spin 3 current above we found instead a previously unknown result: the
UV limit in particular leads to a linearized action that corresponds to a spin 3
CS generalization postulated long ago by Pope and Townsend, \cite{pope,blencowe,deWitFreedman,Damour}.

These were the results found in the parity odd part (in \cite{BCLPS} we were
mostly interested in the latter). But the even parity parts of the two-point correlators have perhaps even more
interesting interpretations, so let us briefly analyze the parity even parts of the linearized effective actions obtained 
from 2-point current correlators in the free massive Dirac fermion quantum field theory in 3d in \cite{BCLPS}.

\subsection{Spin one and two - parity even sectors}

The UV limit of the two-point function of the $J^a$ currents are nonlocal conformal
correlators, according to expectations, see \cite{closset}. The same is true for
the e.m. tensor two-point function. But now let us focus on the IR limits. 
According to \cite{BCLPS}, for  the $J^a$ current two-point function, for large
$m$
we have 
\be \label{JmunuEvenIRUV}
\tilde J_{\mu\nu }^{ab(even)}(k)= \frac {i}{4\pi}\frac{1 }{3 |m|} \delta^{ab}
(k_\mu k_\nu - k^2 \eta_{\mu \nu} ) 	
\ee
This term is local. Fourier anti-transforming it and inserting it into
(\ref{Was}) it gives rise to the action
\be
S\sim \frac 1m \int d^3x\, \left(A^a_\mu \partial^\mu\partial^\nu A^a_\nu -
A^a_\nu \square A^{a\nu}\right)\label{SYMlocal}
\ee
which is the lowest term in the expansion of the YM action
\be
S_{YM}={ -} \frac 1{g_{YM}} \int d^3x\, {\rm Tr}
\left(F_{\mu\nu}F^{\mu\nu}\right)\label{SYM}
\ee
where $g_{YM}\sim |m|$.

Now let us go to the IR limit of the even part of the 2pt e.m. tensor
correlator.
Eq.(3.36) of \cite{BCLPS} says
\begin{multline} \label{IRlimitTT}
\langle T_{\mu\nu}(k)T_{\lambda\rho}\left(-k\right)\rangle^{IR} _{even}  = 
\frac{i |m|}{96\pi}
\left[ \frac{1}{2}\left( \left(k_\mu k_\lambda \eta_{\nu\rho} + \lambda
\leftrightarrow \rho \right) + \mu \leftrightarrow \nu \right) - \right. \\
 \left.\phantom{\frac{1}{2}} - \left( k_\mu k_\nu \eta_{\lambda\rho} + k_\lambda
k_\rho \eta_{\mu\nu}  \right)
 -\frac{ k^2}2 \left( \eta_{\mu\lambda} \eta_{\nu\rho} + \eta_{\mu\rho}
\eta_{\nu\lambda} \right) + k^2 \eta_{\mu\nu}\eta_{\lambda\rho} \right].
\end{multline}
This is a local expression multiplied by $|m|$. In fact Fourier
anti-transforming it and inserting it into (2.19) it gives rise to the action
\be
S \sim |m| \int d^3x \left( -2 \partial_\mu h^{\mu\lambda} \partial_\nu
h^\nu_\lambda- 2h\, \partial_\mu\partial_\nu h^{\mu\nu}- h^{\mu\nu} \square
h_{\mu\nu} + h\square h\right)\label{Slocal} 
\ee
which is the linearized Einstein-Hilbert action:
\be
S_{EH}=\frac 1{2\kappa}\int d^3x \,\sqrt{g}\, R \label{SEH}
\ee
where $\kappa \sim \frac 1{|m|}$. 

These results for spin-1 and -2 are known have been known for a long time, see for instance \cite{Sakharov}. Now, we ask the same question for the 2pt correlator 
of the 3-current (sec. 3.3). What action, if any, does it represent for the external source field?

\subsection{Linearized equations for spin 3 in parity even sector}
\label{ssec:fronsdal}

Before presenting our results in 3d, let us briefly review the status of the linearized equations for the massless spin 3 field described by the completely symmetric field $\varphi_{\mu\nu\lambda}$. Historically the first formulation of equations for the \emph{unconstrained} free massless spin 3 field was given by Fronsdal \cite{Fronsdal}
\be \label{fronsdal3}
{\cal F}_{\mu\nu\lambda}\equiv \square \varphi_{\mu\nu\lambda}-
\partial_{\underline \mu} \partial \!\cdot\!
\varphi_{\underline\nu\underline\lambda}+ \partial_{\underline
\mu}\partial_{\underline \nu} \varphi_{\underline\lambda}'=0
\ee
where underlined indices mean the sum over the minimum number of terms necessary to completely symmetrize the expression in $\mu,\nu$ and $\lambda$,
i.e. for instance
\be
\partial_{\underline \mu} \partial  \!\cdot\! \varphi_{\underline\nu\underline\lambda} =  
\partial_\mu \partial_\alpha \varphi^\alpha{}_{ \nu \lambda}
 + \partial_\nu \partial_\alpha \varphi^\alpha{}_{ \lambda \mu}
 + \partial_\lambda \partial_\alpha \varphi^\alpha{}_{ \mu \nu} \;, \0
\ee 
and where a prime $'$ means that the tensor is traced over a pair of indices. In some formulas we shall use shorter notation in which all indexes are suppressed.

Under the gauge variation (\ref{Bgaugetransf}), $\delta \varphi_{\mu\nu\lambda} = \partial_\mu
\Lambda_{\nu\lambda} + \mathrm{perm.}$, the Fronsdal kinetic tensor transforms as 
$\delta\mathcal{F}_{\mu\nu\lambda} = 3 \partial_\mu\partial_\nu \partial_\lambda \Lambda'$. It follows that the Fronsdal equation is invariant only on \emph{restricted} gauge transformations satisfying $\Lambda' = 0$ (this requirement holds for all higher spins). Also, the Fronsdal tensor is not divergence-free, $\partial \cdot \mathcal{F} \ne 0$, so one cannot directly couple the spin 3 field to a conserved (i.e., divergence-free) current using the Fronsdal equation. As we construct effective actions and corresponding equations for the higher spin fields by (minimally) coupling to conserved currents, it is obvious that Fronsdal's formalism is not suited for our purposes.

The formulation appropriate for our purposes was proposed in \cite{FS}, and analyzed in more detail in \cite{Francia1} (for a review, see \cite{Francia2}). It was shown that there is a one parameter class of equations for unconstrained spin 3 field, which are order 2 in derivatives, fully gauge invariant, and ready to be coupled to the external conserved current. These equations are most elegantly expressed by using gauge invariant linearized spin 3 Riemann tensor defined by
\be
R_{\mu_1\nu_1\mu_2\nu_2\mu_3\nu_3} = \partial_{\mu_1} \partial_{\mu_2} \partial_{\mu_3}\,
 \varphi_{\nu_1 \nu_2 \nu_3} \quad (\,\mbox{antisymmetrised in all } (\mu_j,\nu_j)\, )
\ee
The spin 3 equations are parametrized by real number $a$ and given by
\be
&& \mathcal{G}(a)_{\mu\nu\lambda} \equiv \mathcal{A}(a)_{\mu\nu\lambda} - \eta_{\underline{\lambda\nu}}\,\mathcal{A}(a)'_{\underline{\mu}} = 0
\label{impeqaa} \\
&& \mathcal{A}(a)_{\mu\nu\lambda} \equiv \frac{1}{\Box}\, \partial \cdot R'_{\underline{\mu\nu\lambda}}
 + a\, \frac{\partial_{\underline{\nu}} \partial_{\underline{\lambda}}}{\Box^2}\,
  \partial \cdot R^{\prime\prime}_{\underline{\mu}}  
\label{impeqab}
\ee
where spin 3 Ricci tensors are defined by
\be
R'_{\mu\nu\rho\sigma} &\equiv& \eta^{\alpha\beta}\, R_{\mu\nu\rho\alpha\sigma\beta}
 = 2 \partial_{[\mu} \mathcal{F}_{\nu]\rho\sigma}
\nonumber \\
R''_{\mu\nu} &\equiv& \eta^{\rho\sigma}\, R'_{\mu\nu\rho\sigma}
 = 2 \partial_{[\mu} \mathcal{F}'_{\nu]}
\ee
while their divergences are defined by\footnote{The Riemann tensor symmetries guarantee that the definitions for Ricci's and corresponding divergences (after symmetrization is taken into account) are essentially unique, in the sense that different choices for contracting indexes can differ only by a sign, or are vanishing \cite{Damour}.}
\be
\partial \cdot R'_{\mu\nu\lambda} = \partial_\alpha R'^\alpha{}_{\mu\nu\lambda} \qquad,\qquad
\partial \cdot R''_{\mu} = \partial_\alpha R''^\alpha{}_\mu
\ee

What is the difference between equations with different $a$? First of all, it can be shown that regardless the value of $a$, the free field equation (\ref{impeqaa})-(\ref{impeqab}) is equivalent to Fronsdal equation (\ref{fronsdal3}). They start to differ when interactions are introduced. Note that equations (for any $a$) are non-local. From the purely mathematical side, the equation for $a=0$ plays a special role because it is the least singular on-shell\footnote{In momentum space the on-shell condition is $k^2 = 0$.}, and because of this it was originally promoted in \cite{FS}. However, it was later shown in \cite{Francia1} that equations with different parameters $a$ propagate different set of excitations when coupled to a conserved external current $J_{\mu\nu\lambda}$,
\be \label{GaJ}
\mathcal{G}(a) = J \qquad,\qquad \partial \cdot J = 0
\ee
In particular, it was shown that only equation with $a = 1/2$ propagates \emph{spin 3 massless excitations and nothing else}, if one does not introduce additional constraints on $\varphi$ or $J$. For $a=1/2$ the tensor $\mathcal{A}$ can be also written as 
\be
\mathcal{A}(1/2) = \mathcal{F} - \frac{\partial^3}{\Box^2}\, \partial \cdot \mathcal{F}'
\ee
Let us emphasize that this by itself does not mean that the equation with $a = 1/2$ is the "right one" to be used for the consistent coupling to the dynamical matter.

The non-locality of equations (\ref{impeqaa})-(\ref{impeqaa}) can be 'cured' by multiplying with $\Box^r$ with $r$ large enough. It is obvious that the equation with $a=0$ is special in that $r=1$ already does the job, while for $a\ne0$ one needs $r = 2$. In this way one cures non-locality, but the price paid is that equations become higher-derivative (order 4 for $a=0$ and order 6 for $a\ne0$). This opens up an additional question when one considers coupling to the conserved current $J$: should we do this as in (\ref{GaJ}), or should we couple the current in the local way,
\be \label{lfseqa}
\Box^r \mathcal{G}(a) = J \qquad,\qquad \partial \cdot J = 0
\ee
with $r$ large enough?

The moral of the above analysis is that, due to several reasons, there is a large degeneracy in formulating equations of motion for the free massless spin 3 field, and it is not obvious that all formulations can be used as a basis for constructing consistent interacting quantized theories. It would be advantageous to know which formulation(s) are more promising, before embarking into such enterprise. We shall now argue that the induced action method may give us a hint.

In section 3.2.4 of \cite{BCLPS} it was shown that the parity even part of the spin 3 two-point current correlator for a massive Dirac fermion in 3d is given by
\be
\begin{aligned} \label{J3J3FullEven}
\widetilde{\mathcal{J}}^{(even)}_{\mu_1\mu_2\mu_3\nu_1\nu_2\nu_3}(k) & =  \tau_b
\left(\frac{k^2}{m^2}\right) |k|^5 \pi^{(k)}_{\mu_1 \mu_2}  \pi^{(k)}_{\mu_3 \nu_1}
\pi^{(k)}_{\nu_2 \nu_3} 
+ \tau'_b \left(\frac{k^2}{m^2}\right) |k|^5 \pi^{(k)}_{\mu_1 \nu_1}  \pi^{(k)}_{\mu_2
\nu_2} \pi^{(k)}_{\mu_3 \nu_3}
\end{aligned}
\ee
where $\tau_b$ and $\tau_b'$ are form factors presented in \cite{BCLPS}, and
\be
\pi^{(k)}_{\mu\nu}= \eta_{\mu\nu} - \frac {k_\mu k_\nu}{k^2}\label{proj}
\ee
are projectors which guarantee conservation. From (\ref{Japmux}) it follows that the linearized effective equation in momentum space for the background spin 3 field minimally coupled to a conserved current in free QFT with massive Dirac field in 3d, is given by
\be \label{ls3eome}
\widetilde{\mathcal{J}}_{\mu_1\mu_2\mu_3\nu_1\nu_2\nu_3}(k)\, \widetilde{\varphi}^{\nu_1\nu_2\nu_3}(k)
 = \langle\!\langle \widetilde{J}^{(3)}_{\mu_1\mu_2\mu_3}(k)\rangle\!\rangle
\qquad,\qquad k \cdot \widetilde{J}^{(3)}(k) = 0
\ee
The form factors contain branch-cuts, which means that this equation is strongly non-local. The fact that there are two independent conserved structures present in (\ref{J3J3FullEven}), and so in (\ref{ls3eome}), is directly connected with the one-parameter degeneracy introduced in (\ref{impeqab}).

In the IR region ($|k^2|/m^2 < 4$) the form factors are analytic, as expected, and the equation is weakly nonlocal (infinite sum of local terms) when expanded around $|k|/m = 0$. Using the expansions of form factors from \cite{BCLPS}, we obtain that the leading term in the IR is given by
\be \label{s32pe}
\widetilde{\mathcal{J}}^{(even)}_{\mu_1\mu_2\mu_3\nu_1\nu_2\nu_3}(k) \sim |m|\, k^4
 \left( \pi^{(k)}_{\mu_1\mu_2}  \pi^{(k)}_{\mu_3\nu_1} \pi^{(k)}_{\nu_2\nu_3}
 - \pi^{(k)}_{\mu_1\nu_1}  \pi^{(k)}_{\mu_2\nu_2} \pi^{(k)}_{\mu_3 \nu_3} \right)
\ee
Observe that this is \emph{the lowest derivative conserved local expression}, which is 
\emph{unique}. Now, plugging (\ref{s32pe}) into (\ref{ls3eome}) and Fourier antitransforming, we obtain for the linearized induced equation in the coordinate space
\be \label{s3leome}
|m|\, G_{\mu\nu\rho}(x) \sim \langle\!\langle J^{(3)}_{\mu_1\mu_2\mu_3}(x)\rangle\!\rangle
\qquad,\qquad \partial \cdot J^{(3)} = 0
\ee
where $G$ is the conserved symmetric local tensor linear in $\varphi$, which is 4th-order in derivatives. As there is a unique such tensor, we can conclude (without doing any calculations) that it must be proportional to $\Box\, \mathcal{G}(0)$, with $\mathcal{G}(0)$ defined in (\ref{impeqaa})-(\ref{impeqab}). Explicitly written,
\be
G_{\mu\nu\lambda} = \partial_\alpha F^\alpha{}_{(\mu\nu\lambda)}
\ee
where
\be \label{fdef}
F_{\alpha\mu\nu\lambda} \equiv R'_{\alpha\mu\nu\lambda} - \frac{1}{2} R''_{\alpha\mu} \eta_{\nu\lambda}
 = 2 \partial_{[\alpha} \left( \mathcal{F}_{\mu]\nu\lambda} - \frac{1}{2} \mathcal{F}'_{\mu]} \eta_{\nu\lambda} \right)
\ee

The result (\ref{s3leome})-(\ref{fdef}) is, in some sense, natural. First of all, it is the lowest derivative linear local parity invariant equation satisfying unrestricted gauge invariance and conservation condition. Also, the equation is of the same form as in spin 1 case, and we can identify the tensor $F$ as spin 3 Maxwell tensor, while $G$ appears to be spin 3 Riemann tensor (it is the lowest derivative local conserved gauge invariant parity even rank-3 tensor).\footnote{Conventions for naming objects in higher-spin metric-like formalism is notorious for its inconsistency. In the literature different objects are called Ricci tensor and Riemann tensor. We believe that our conventions are natural generalizations of spin 1 and 2 cases.}

Let us connect our result with the known constructions, reviewed above. It is obvious that our result (\ref{s3leome})-(\ref{fdef}) is the same as (\ref{lfseqa}) with $a=0$ and $r=1$, i.e., we have obtained a local version of the equation proposed in \cite{FS}. As we already mentioned, this equation does not propagate only spin 3 massless excitations, unless the conserved spin 3 current of the Dirac theory has some special properties which takes care of the redundant modes. This is the question we plan to investigate in the future. 

Let us now briefly comment the UV limit ($m/|k| \to 0$). After subtracting IR divergent terms (for a full explanation of this issue, see below) form factors in the UV limit tend to constants, which gives rise to a non-local correlator. However one of the subleading terms gives
a combination of the two conserved quantities 
\be
A \quad&:&\quad k^2 \pi_{\mu_1 \nu_1}  \pi_{\mu_2 \nu_2} \pi_{\mu_3 \nu_3}\0\\
B \quad&:&\quad k^2\pi_{\mu_1 \mu_2}  \pi_{\mu_3 \nu_1} \pi_{\nu_2 \nu_3} \0\\
\ee
which is not the same combination as the one present in IR limit (\ref{s32pe}). So, the corresponding induced linearized equation is also different. 

A priori, one could freely linearly combine terms A and B and construct one parameter candidate equations for the free spin 3 field. For example, A by itself gives the following equation
\be \label{SBeom}
\square \varphi_{\mu\nu\lambda}- \partial_{\underline \mu} \partial \!\cdot\!
\varphi_{\underline\nu\underline\lambda}+ \frac 1{\square}\partial_{\underline
\mu}\partial_{\underline \nu} \partial \!\cdot\!\partial
\!\cdot\!\varphi_{\underline\lambda}-
\frac 1{\square^2}\partial_{\mu}\partial_{\nu}\partial_\lambda \partial
\!\cdot\!\partial \!\cdot\! \partial \!\cdot\!\varphi = 0 
\ee
By combining with the traced equation, it can be shown that it is equivalent to the Fronsdal equation. The same can be shown for generic linear combination of A and B. There is though the special case, the combination $4B-3A$, which is traceless, for which the equation is
\be 
&&\square \varphi_{\mu\nu\lambda}-3 \partial_{ \mu} \partial \!\cdot\!
\varphi_{ \nu \lambda}+ \frac 34\partial_{ 
\mu}\partial_{  \nu} \varphi_{ \lambda}'-\frac 34\frac 1{\square}
\partial_\mu\partial_\nu \partial_\lambda \partial \!\cdot\! \varphi'-\frac 14\frac
1{\square^2} \partial_\mu\partial_\nu \partial_\lambda\partial \!\cdot\!\partial
\!\cdot\! \partial \!\cdot\! \varphi\label{fronsdal3ter}\\
&& + \frac 94\frac 1{\square}
\partial_\mu\partial_\nu \partial \!\cdot\!\varphi_\lambda-\frac 34
\eta_{\mu\nu}\square \varphi'_\lambda
+\frac 34 \eta_{\mu\nu}\partial_\lambda\partial \!\cdot\! \varphi' 
+  \frac 34 \eta_{\mu\nu}\partial \!\cdot\! \partial \!\cdot\! \varphi_\lambda- \frac
34  \eta_{\mu\nu}\frac
1{\square^2} \partial_\lambda\partial \!\cdot\!\partial\!\cdot\! \partial
\!\cdot\!\varphi  =0\0
\ee

In conclusion, we see that our simple analysis, based solely on the classification of possible conserved structures, recovers the Francia-Sagnotti analysis and gives an efficient method for analyzing higher spin actions. But, we emphasize that the induced action method, out of many possibilities, picks particular equations which are already coupled to particular external currents. 

\vskip 0.5cm
{\bf Comment.} The previous results are limited to 3d and to the lowest spins.
They are nevertheless enough to stir our interest and motivate a more in depth analysis. 
It is also clear enough that equations in the coordinate space are not always the best 
fit to generalizations to higher spins.
Writing down the actions and equations of motion in the explicit form used so far becomes 
rapidly unwieldy with increasing spins and dimensions. Fortunately a language
much sleeker than this
and the formalism used so far in higher spin theories is at hand. We simply must
go to momentum space
and use the projector (\ref{proj}). Before plunging into the analysis of the
results for 2pt correlators coming from Feynman
diagrams, we'd better prepare the ground with a general analysis of their
expected structure.

\section{Universal EOM and conserved structures for spin $s$.}
\label{sec:universal}

Our starting point is the 2-pt functions of symmetric conserved currents. We
expect
them to be conserved too, i.e. we expect to find 0 if we contract any index with
the external momentum
$k$. We exclude the presence of anomalies. In fact we will come across also
some non-conservations, but they can be fixed by subtracting local counterterms.
This aspect of our analysis is interesting in itself, but we will illustrate it
later on in any detail. For the time being we ignore this fact and suppose 
that all 2-pt functions we deal with are conserved.

This said, the form of the conserved structures is universal, in the sense that
is does not
depend on the dimension $d$ of spacetime. For spin $s$ they can be easily
constructed by means of the
projector (\ref{proj}) and polarization vectors $n_1, n_2$:
$n_{1\mu }, n_{2\nu }$.
 
For spin $s$ let us write down the structures:
\be
\tilde A^{(s)}_0(k\!\cdot\!n_1\!\cdot\!n_2)&=&\frac 1{(s!)^2} (n_1 \!\cdot\!
\pi^{(k)}\!\cdot
\!n_2)^s\label{A0}\\
\tilde A^{(s)}_1(k\!\cdot\!n_1\!\cdot\!n_2)&=&\frac 1{(s!)^2} (n_1 \!\cdot\!
\pi^{(k)}\!\cdot
\!n_2)^{s-2} (n_1 \!\cdot\! \pi^{(k)}\!\cdot \!n_1)(n_2 
\!\cdot\! \pi^{(k)}\!\cdot \!n_2)\label{A1}\\
\ldots && \ldots\ldots\0\\
\tilde A^{(s)}_l(k\!\cdot\!n_1\!\cdot\!n_2) &=& \frac 1{(s!)^2} (n_1 \!\cdot\!
\pi^{(k)}\!\cdot
\!n_2)^{s-2l} (n_1 \!\cdot\! \pi^{(k)}
\!\cdot \!n_1)^l(n_2 \!\cdot\! \pi^{(k)}\!\cdot \!n_2)^l\label{Al}\\
\ldots && \ldots\ldots\0
\ee
where $n\!\cdot\! \pi^{(k)}\!\cdot \!m= n^\mu \pi_{\mu\nu}m^\nu$.
There are $\lfloor s/2 \rfloor$ independent such terms.

Let us set
\be 
\tilde E^{(s)}(k\!\cdot\!n_1\!\cdot\!n_2)= \sum_{l=0}^{\lfloor s/2 \rfloor} a_l \tilde
A^{(s)}_l(k\!\cdot\!n_1\!\cdot\!n_2)\label{E}
\ee
where $a_l$ are arbitrary constants. The explicit conserved structures are
obtained by differentiating  $s$ times $E^{(s)}$ 
with respect to $n_1$ and $s$ times with respect to $n_2$. One obtains in this
way conserved tensors
$\tilde E^{(s)}_{\mu_1\ldots \mu_s,\nu_1\ldots\nu_s}(k)$. Conservation is a
consequence of 
the transversality property 
\be
k^\mu \pi_{\mu\nu}=0\label{transverse}
\ee
and $\tilde E^{(s)}$ writes
\be
\tilde E_{\mu_1\ldots \mu_s,\nu_1\ldots\nu_s}(k)= \sum_{l=0}^{\lfloor s/2 \rfloor} a_l \tilde
A_{l,\mu_1\ldots \mu_s,\nu_1\ldots\nu_s}(k)\label{E(k)}
\ee
This is the most general conserved structure for spin $s$ (for a proof, see
Appendix
\ref{sec:proof}).

By Fourier anti-transforming and inserting into (\ref{Was}), one can construct 
the effective action corresponding to (\ref{E(k)}) multiplied by $k^2$ for the
spin $s$
field $B_{\mu_1\ldots \mu_s,\nu_1\ldots\nu_s}$ as follows
\be
S_E \sim \int d^dx \, B^{\mu_1\ldots \mu_s }
\square E(\partial)_{\mu_1 \ldots \mu_s,\nu_1\ldots\nu_s} B^{\nu_1\ldots
\nu_s}\label{SE}
\ee
where $E(\partial)$ is the formal Fourier transform of $\tilde E(k)$, i.e. the
same expression 
with $k_\mu$ replaced by $-i\partial_\mu$. The eom is of course
\be
\square E(\partial)_{\mu_1 \ldots \mu_s,\nu_1\ldots\nu_s} B^{\nu_1\ldots
\nu_s}=0\label{eomE}
\ee
After canonically normalization, it depends on $\lfloor s/2 \rfloor-1$ arbitrary constants.
This is the most general linearized
eom for a completely symmetric spin $s$ field.

>From $\tilde E^{(s)}(k)$ we can obtain the most general traceless combination,
by taking the trace of (\ref{E(k)}) and imposing it to vanish.
This can be done by differentiating the implicit expressions
(\ref{A0}),...,(\ref{Al}),... with respect to
$\frac{\partial}{\partial n_{1\mu}} \frac{\partial}{\partial n_1^\mu}$. The
resulting equation is the recurrence relation
\be
a_l=-\frac{(s-2l+2)(s-2l+1)}{2l(2(s-l-1)+d-1)} a_{l-1}\label{recurr}
\ee
Setting $a_0=1$ the solution is 
\be
a_l= \frac {(-1)^l}{2^{2l}l!} \frac {s!}{(s-2l)!} \frac {\Gamma\left(s+\frac
{d-3}2-l\right)}{\Gamma\left(s+\frac {d-3}2\right)}\label{al}
\ee
Replacing this in (\ref{E(k)}) we obtain a traceless conserved structure. In
turn this gives rise to a traceless eom.

\subsection{Eom's from conserved structures}

Any conserved structure (\ref{E}) in coordinate space is in general a non-local
differential operator. To each  there corresponds a quadratic Lagrangian and a
linearized eom.
For the EOM it is enough to differentiate $s$ times with respect to $n_2^\nu$
and saturate the exposed indices with the spin $s$ tensor field 
$a^{\nu_1 \ldots \nu_s}$, multiply by $k^2$, set the result to zero and then
differentiate also $s$ times wrt  $n_1^\mu$. 
For the Lagrangian one saturates the lhs of the EOM with  $a^{\mu_1 \ldots
\mu_s}$ and divide by 2.

Therefore we can represent the eom symbolically as
\be 
 k^2\sum_{l=0}^{\lfloor s/2 \rfloor} a_l \tilde
A^{(s)}_l(k\!\cdot\!n_1\!\cdot\!n_2)=0\label{eom-s}
\ee
In the following instead of contracting the $n_2$ indices with the field $a$, we
will always leave $n_2$ free and operate only on $n_1$. The operation will be
essentially tracing two $n_1$ indices. For instance tracing $(n_1 \!\cdot\!
\pi^{(k)}\!\cdot \!n_1)$ over $n_1$ gives $d-1$. 

Let us consider the spin 3 case. In this compact notation, the most general eom
will be
\be
k^2\left(a (n_1 \!\cdot\! \pi^{(k)}\!\cdot \!n_2)^3+ b(n_1 \!\cdot\!
\pi^{(k)}\!\cdot \!n_1) (n_1 \!\cdot\! \pi^{(k)}\!\cdot \!n_2)(n_2 \!\cdot\!
\pi^{(k)}\!\cdot \!n_2)\right)=0\label{eomspin3}
\ee
Taking the trace over $n_1$ gives
\be
(6a+ (d+1)b)(n_1 \!\cdot\! \pi^{(k)}\!\cdot \!n_2)(n_2 \!\cdot\!
\pi^{(k)}\!\cdot \!n_2)=0\label{traceeomspin3}
\ee
Thus, unless $6a+ (d+1)b=0$, i.e. for generic coefficients $a$ and $b$, the
second piece of (\ref{eomspin3}) vanishes on shell and we can simply drop it.
Therefore the relevant eom for spin 3 is
\be
k^2  (n_1 \!\cdot\! \pi^{(k)}\!\cdot \!n_2)^3 =0\label{eomspin3rel}
\ee
i.e. (\ref{SBeom}).

Now we wish to prove that this is general, that is, for any spin $s$, for
generic coefficients, the eom can be reduced to the form
\be
k^2  (n_1 \!\cdot\! \pi^{(k)}\!\cdot \!n_2)^s =0\label{eomspinsrel}
\ee
The strategy consists in taking the trace of (\ref{eom-s}) wrt to $n_1$ the
maximum number of times and replacing the results in (\ref{eom-s}). For
instance,
for spin 4 we have to trace twice.

Tracing $p$ times (\ref{eom-s}) we get  
\be
&&\sum_{l=p}^{\lfloor s/2 \rfloor}[ c_{l-1}^{(p)} (s-2l+2)(s-2l+1)+ 2 c_l^{(p)} (l-p+1)
(s+d-2p-1)]\0\\
&&\quad\quad\quad\cdot(n_1 \!\cdot\! \pi^{(k)}\!\cdot \!n_1)^{l-2} 
(n_1 \!\cdot\! \pi^{(k)}\!\cdot \!n_2)^{s-2l}(n_2 \!\cdot\! \pi^{(k)}\!\cdot
\!n_2)^l=0\label{ktraceeom-s}
\ee
where $c_l^{(0)}=a_l$, $c_{l}^{(1)}=a_{l-1} (s-2l+2)(s-2l+1)+2l a_l (s+d-3)$,
etc.
The complete expression for $c_l^{(p)} $ is not easy to compute, but these
coefficients are generically non-vanishing. 
It is however possible to infer the important property that 
\be
c_l^{(p)}=0, \quad\quad l<p\label{importantproperty}
\ee

For $s=2n$ after $n$ tracings, i.e. $p=n$, we arrive at
\be
k^2(n_2 \!\cdot\! \pi^{(k)}\!\cdot \!n_2)^n=0\label{ntraceeom-s}
\ee
Now let us consider $p=n-1$. Using (\ref{ktraceeom-s}) and (\ref{ntraceeom-s})
we arrive at
\be
2k^2(d+1)c_{n-1}^{(n-1)}(n_1 \!\cdot\! \pi^{(k)}\!\cdot \!n_2)^2l 
(n_2 \!\cdot\! \pi^{(k)}\!\cdot \!n_2)^{n-1}=0\0
\ee
So, generically,
\be
k^2(n_1 \!\cdot\! \pi^{(k)}\!\cdot \!n_2)^2l 
(n_2 \!\cdot\! \pi^{(k)}\!\cdot \!n_2)^{n-1}=0\label{n-1traceeom-s}
\ee
Now we proceed by induction. Suppose after $q$ traces, i.e. $p=n-q+1$, we have
\be 
k^2(n_1 \!\cdot\! \pi^{(k)}\!\cdot \!n_2)^{2i}
(n_2 \!\cdot\! \pi^{(k)}\!\cdot \!n_2)^{n-i}=0,\quad\quad i=0,\ldots,q-1\0
\ee
Then, at level $p=n-q$, we remain with
\be 
2k^2c_{n-q}^{(n-q)}(d+2q-1) \,(n_1 \!\cdot\! \pi^{(k)}\!\cdot \!n_2)^2q
(n_2 \!\cdot\! \pi^{(k)}\!\cdot \!n_2)^{n-q}=0\label{n-ptraceeom-s}
\ee
from which the conclusion (\ref{eomspinsrel}) follows.

For $s=2n+1$, we start from $p=n$ 
\be 
2k^2c_{n}^{(n)}(d-1) \,(n_1 \!\cdot\! \pi^{(k)}\!\cdot \!n_2)
(n_2 \!\cdot\! \pi^{(k)}\!\cdot \!n_2)^{n}=0\label{soddtraceeom-s}
\ee
and we can repeat the induction procedure arriving at the same conclusion
(\ref{eomspinsrel}).

\vskip 1cm

The next task is to recover the Fronsdal equation from (\ref{eomspinsrel}).

To this end we take the trace of (\ref{eomspinsrel}), i.e. apply to it
$\frac {\partial}{\partial n_1^\mu} \eta^{\mu\nu} \frac {\partial}{\partial
n_1^\nu}$.  This is easily seen to give
\be
 {\rm tr}(n_1 \!\cdot\! \pi^{(k)}\!\cdot \!n_2)^2=2 (n_2 \!\cdot\! n_2)-2\frac
{(n_2 \!\cdot\! k)^2}{k^2}
\label{tracesquare}
\ee
and, in general,
\be
{\rm tr} (n_1 \!\cdot\! \pi^{(k)}\!\cdot \!n_2)^s=s(s-1) (n_1 \!\cdot\!
\pi^{(k)}  n_2)^{s-2} {\rm tr}  (n_1 \!\cdot\! \pi^{(k)}\!\cdot
\!n_2)^2=0\label{eomspinstrace}
\ee
 
Using this we can easily calculate all the traces of (\ref{eomspinsrel}). The
end result is
\be
{\rm tr} (n_1 \!\cdot\! \pi^{(k)}\!\cdot \!n_2)^s\sim \left((n_2 \!\cdot\!
n_2)-2\frac {(n_2 \!\cdot\! k)^2}{k^2}\right)^{\frac s2}=0\label{traceseven}
\ee
for even $s$, and 
\be
{\rm tr} (n_1 \!\cdot\! \pi^{(k)}\!\cdot \!n_2)^s\sim \left((n_2 \!\cdot\!
n_2)-2\frac {(n_2 \!\cdot\! k)^2}{k^2}\right)^{\frac {s-1}2}(n_1 \!\cdot\!
\pi^{(k)}\!\cdot \!n_2)=0\label{tracesodd}
\ee
for odd $s$. These two equations have to be understood as follows: any solution
that satisfies 
(\ref{eomspinsrel}) also satisfies either (\ref{traceseven}) or
(\ref{tracesodd}). Therefore we can replace
these two eqs. into (\ref{eomspinsrel}). The viceversa is not true in general:
i.e. if a solution satisfies 
 (\ref{traceseven}) or (\ref{tracesodd}), it may not satisfy 
(\ref{eomspinsrel}). For the time being we assume that
Eq.(\ref{traceseven}) and (\ref{tracesodd}) imply that in (\ref{eomspinsrel}) we
can make the replacement $ (n_2 \!\cdot\! k)^2= k^2  (n_2 \!\cdot\! n_2)$ (see the comment below).
 
The result of this substitution is:        
\be
&&k^2(n_1 \!\cdot\!  n_2)^s -s (n_1 \!\cdot\!  n_2)^{s-1} (n_1\!\cdot \!
k)(n_2\!\cdot \! k)+
\left(\begin{matrix} s \\ 2\end{matrix}\right)(n_1 \!\cdot\!  n_2)^{s-2}
(n_1\!\cdot \! k)^2(n_2\!\cdot \! n_2)\0\\
&&+ \sum_{l=3}^s(-1)^l \left(\begin{matrix} s \\ l\end{matrix}\right)(n_1
\!\cdot\!  n_2)^{s-l} \frac{(n_1\!\cdot \! k)^{l} (n_2\!\cdot\!
k)^{l-2}}{(k^2)^{l-2}}(n_2\!\cdot \! n_2) =0 \label{fronsdalnonlocaleomspins}
\ee
The first line gives the spin $s$ Fronsdal operator. Therefore
(\ref{fronsdalnonlocaleomspins})  identifies
the spin $s$
nonlocal Fronsdal equation. The compensator takes the form 
\be
\alpha(n_1,n_2)= \sum_{l=3}^s(-1)^l \left(\begin{matrix} s \\
l\end{matrix}\right)(n_1
\!\cdot\!  n_2)^{s-l} \frac{(n_1\!\cdot \! k)^{l-3} (n_2\!\cdot\!
k)^{l-2}}{(k^2)^{l-2}}(n_2\!\cdot \! n_2)\label{compensatorszpins}
\ee

\subsection{Conserved odd parity structures}

It is easy to obtain also all the odd parity structures. The spin 1 odd parity
conserved Lorentz structure (linear in $n_1\!\cdot\!n_2\!\cdot\!k$) can only be
\be
\tilde C^{(1)}_0(k\!\cdot\!n_1\!\cdot\!n_2)=   (n_1\epsilon n_2\!\cdot\!k),
\quad\quad  (n_1\epsilon
n_2\!\cdot\!k)= \epsilon_{\mu\nu\lambda} n_1^\mu n_2^\nu k^\lambda\label{C0}
\ee
It is easy to realize that, for higher spin, the epsilon tensor can only appear
in the form $ (n_1\epsilon n_2,k)$  in every single term, thus it can be
factored
out. What remains is an even spin structure of one order less. So the most
general odd conserved Lorentz structure will be a combination of 
\be 
\tilde C^{(s)}_0(k\!\cdot\!n_1\!\cdot\!n_2)&=&  (n_1\epsilon
n_2\!\cdot\!k)\tilde
A^{(s-1)}_0(k\!\cdot\!n_1\!\cdot\!n_2)\0\\
\tilde C^{(s)}_1(k\!\cdot\!n_1\!\cdot\!n_2)&=& (n_1\epsilon n_2\!\cdot\!k)
\tilde
A^{(s-1)}_1(k\!\cdot\!n_1\!\cdot\!n_2)\0\\
\ldots && \dots\0\\
\tilde C^{(s)}_l(k\!\cdot\!n_1\!\cdot\!n_2)&=& (n_1\epsilon n_2\!\cdot\!k) 
\tilde
A^{(s-1)}_{l}(k\!\cdot\!n_1\!\cdot\!n_2)\0\\
\ldots && \dots\label{Bl}
\ee
where $A_0^{(0)}=1$, by definition.
Let us define 
\be 
\tilde O^{(s)}(k\!\cdot\!n_1\!\cdot\!n_2)= \sum_{l=0}^{\lfloor s/2 \rfloor} c_l \tilde
C^{(s)}_l(k\!\cdot\!n_1\!\cdot\!n_2)\label{O}
\ee
from which we can derive
\be
O_{\mu_1\ldots \mu_s,\nu_1\ldots\nu_s}(\partial)= \sum_{l=0}^{\lfloor s/2 \rfloor} c_l 
C_{l,\mu_1\ldots \mu_s,\nu_1\ldots\nu_s}(\partial)\label{O(k)}
\ee
The odd parity action is supposed to be local (and higher derivative)
\be
S_O = \int d^dx \, B^{\mu_1\ldots \mu_s }
\square^{s-1} O(\partial)_{\mu_1 \ldots \mu_s,\nu_1\ldots\nu_s} B^{\nu_1\ldots
\nu_s}\label{SO}
\ee
Therefore the odd eom is
\be 
\square^{s-1} O_{\mu_1\ldots \mu_s,\nu_1\ldots\nu_s}(\partial) B^{\nu_1\ldots
\nu_s}=0\label{eomO}
\ee

The tracelessness condition (for spin $s>1$) implies a recursion relation for
the coefficients $c_l$:
\be
c_l=-\frac{(s-2l+1)(s-2l)}{2l(2(s-l-2)+d+1)}c_{l-1}  \label{recurodd}
\ee
Setting $c_0=1$ the solution is:
\be
c_l= \frac {(-1)^l}{2^{2l}l!} \frac {(s-1)!}{(s-2l-1)!} \frac
{\Gamma\left(s+\frac {d-3}2-l\right)}{\Gamma\left(s+\frac
{d-3}2\right)}\label{cl}
\ee
 
\vskip 1cm

{\bf A comment on the non-local Fronsdal equation}

In the previous derivations of eqs.(\ref{fronsdalnonlocaleomspins}) and (\ref{eomO}),
we have simplified a few steps by disregarding a number of alternatives. First,
we have stated that several passages are {\it generic}, that is they do not hold
in some very specific cases,  
leaving out in this way several (probably pathological) possibilities. Moreover, we have disregarded solutions that satisfy (\ref{traceseven}) or (\ref{tracesodd}), but not (\ref{eomspinsrel}). Therefore our conclusions concerning eqs.(\ref{fronsdalnonlocaleomspins}) and (\ref{eomO}) are generic. They do not address more subtle questions, in particular the one pointed out in \cite{Francia1,Francia2}: the non-locality of the Fronsdal equation contains a large freedom, so an important issue is to select the form of the equation that gives rise to the correct propagator for the higher spin field, and not all non-local equations which give rise to the Fronsdal equation upon gauge fixing also give the correct propagator\footnote{We thank the referee of this paper for stressing the importance of such an issue. On the other hand our intention in this section is to stress the universal features of the non-local Fronsdal equation.}. We cannot say, on the basis of our previous derivation, that our non-local Fronsdal equations have the property of generating the correct propagator, but we can verify this {\it a posteriori}, by analyzing the effective actions we obtain for the massive scalar and fermion models in various dimensions. We will return to this issue in the concluding section.

\section{The general method}
\label{sec:method}

In this section we illustrate the method to compute the 2-pt functions with
Feynman diagrams. 
On first reading one can skip this section and go directly to the results in the
next one.
The integrals we have to compute in this paper are like the ones in
(\ref{Jmunuab}) and (\ref{Tmnlr}),
that is of the general form
\begin{equation}\label{J2initial}
\tilde J_{\mu_1\ldots\mu_p}(d;\alpha,\beta;q_1,q_2,m)=\int\frac{d^d
p}{(2\pi)^d}\frac{p_{\mu_1}\ldots
p_{\mu_p}}{\left((p+q_1)^2-m^2\right)^\alpha\left( (p+q_2)^2-m^2
\right)^\beta}
\end{equation}
where, eventually, $q_1=0, q_2=-k$.
We will use the method invented by \cite{BoosDavy} to reduce the tensor integral
to a sum of scalar ones
\begin{multline}
\tilde J_{\mu_{1}\dots\mu_{p}} \left(d; \alpha,\beta,\gamma;q_1,q_2,m \right) =
\sum_{\substack{\lambda,\kappa_{1},\kappa_2\\2\lambda+\sum\kappa_{i}=p}}
\left(-\frac{1}{2}\right)^{\lambda}\left(4\pi\right)^{p-\lambda}\left\{
\left[\eta\right]^{\lambda}\left[q_{1}\right]^{\kappa_{1}} 
\left[q_{2}\right]^{\kappa_{2}} \right\}
_{\mu_{1}\dots\mu_{p}}\\
\times\pochhammer{\alpha}{\kappa_{1}}\pochhammer{\beta}{\kappa_{2}} 
\tilde I^{(2)}(d+2(p-\lambda);\alpha+\kappa_1,\beta+\kappa_2 ;q_1,q_2,m)
,\label{eq:GeneralTensorIntegralApdx}
\end{multline}
where the symbol $\left\{
\left[\eta\right]^{\lambda}\left[q_{1}\right]^{\kappa_{1}}\dots\left[q_{N}\right
]^{\kappa_{N}}\right\} _{\mu_{1}\dots\mu_{M}}$
stands for the complete symmetrization of the objects inside the curly brackets,
for example
\[
\left\{ \eta q_{1}\right\}
_{\mu_{1}\mu_{2}\mu_{3}}=\eta_{\mu_{1}\mu_{2}}q_{1\mu_{3}}+\eta_{\mu_{1}\mu_{3}}
q_{1\mu_{2}}+\eta_{\mu_{2}\mu_{3}}q_{1\mu_{1}}.
\]
The basic integral is now  the scalar one
\begin{equation}\label{J2scalar}
\tilde I^{(2)} (d;\alpha,\beta;q_1,q_2,m)=\int\frac{d^d
p}{(2\pi)^d}\frac{ 1}{\left((p+q_1)^2-m^2\right)^\alpha\left( (p+q_2)^2-m^2
\right)^\beta}
\end{equation}

For instance, the bubble integral for the $s=1$ current in the scalar model
\be
\tilde J_{\mu\nu}(k) = \int \frac{d^dp}{(2\pi)^d} 
\frac {(2p-k)_\mu(2p-k)_\nu}{(p^2-m^2)((p-k)^2-m^2)}\label{2ptscalar}
\ee
reduces to 
\be
\tilde J_{\mu\nu}(m,k) &=& -\frac{8\pi}{(2\pi)^{d+2}}\eta_{\mu\nu}\,\tilde
I^{(2)}(d+2;1,1)+
8\frac{(4\pi)^2}{(2\pi)^{d+4}}k_\mu k_\nu\,\tilde  I^{(2)}(
d+4;1,3)\label{2ptscalar1}\\
&&+ \frac{16\pi}{(2\pi)^{d+2}}k_\mu k_\nu\,\tilde  I^{(2)}(d+2;1,2)+
\frac{1}{(2\pi)^{d}}k_\mu k_\nu\,\tilde  I^{(2)}(d;1,1)\0
\ee
The integral $\tilde I^{(2)}(d;\alpha,\beta;k,m)$ can be cast into the form of a
hypergeometric series
\be
\tilde I_{IR}^{(2)}(d;\alpha,\beta;k,m)&=&
{ 2^{-d}\pi ^{-d/2}} i^{1-d}\,\left(-m^2\right)^{-\alpha -\beta +\frac{d}{2}}   
\frac{\Gamma \left(-\frac{d}{2}+\alpha +\beta \right) }{\Gamma (\alpha +\beta
)}\0\\
&&\times\,\,{}_3F_2\left(\begin{matrix}\alpha ,\beta ,-\frac{d}{2}+\alpha +\beta
\\
\frac{\alpha +\beta }{2},\frac{ \alpha +\beta +1}2\end{matrix} \Big{\vert}
\frac{k^2}{4 m^2}\right)\label{I2IR}
\ee
This representation is valid for large $m$ compared to $k$. When $m$ is small
compared to $k$
another representation is available
\be
\tilde I_{UV}^{(2)}(d;\alpha,\beta;k,m)&=&
{ 2^{-d}\pi ^{-d/2}}i^{1-d}\left(k^2\right)^{-\alpha -\beta +\frac{d}{2}}
\Bigg{\{}\frac{\left(\Gamma \left(\frac{d}{2}-\alpha \right) 
\Gamma \left(\frac{d}{2}-\beta \right) \Gamma \left(-\frac{d}{2}+\alpha +\beta
\right)\right)}{\Gamma (\alpha ) \Gamma (\beta ) 
\Gamma (d-\alpha -\beta )} \0\\
&&\quad\times\,\, _3F_2\left(\begin{matrix}-\frac{d}{2}+\alpha +\beta ,\frac{
-d+\alpha +\beta +1}2,
\frac{        
-d+\alpha +\beta +2}2\\-\frac{d}{2}+\alpha +1,-\frac{d}{2}+\beta
+1\end{matrix}\Big{\vert}
\frac{4 m^2}{k^2}\right)\label{I2UV}\\
&& + \left(-\frac{m^2}{k^2}\right)^{\frac{d}{2}-\alpha } 
\frac{\Gamma \left(\alpha -\frac{d}{2}\right) }{\Gamma (\alpha )}
\,\, _3F_2\left(\begin{matrix}\beta ,\frac{-\alpha +\beta +1}2,\frac{-\alpha
+\beta +2}2\\
\frac{d}{2}-
\alpha +1,-\alpha +\beta +1\end{matrix}\Big{\vert} \frac{4 m^2}{k^2}\right) \0\\
&&+\left(-\frac{m^2}{k^2}\right)^{\frac{d}{2}-\beta }\frac{\Gamma \left(\beta 
-\frac{d}{2}\right)}{\Gamma (\beta )} \,\,
{}_3F_2\left(\begin{matrix}\alpha ,\frac{ \alpha -\beta +1}2,
\frac{ \alpha -\beta +2}2\\-\beta +\frac{d}{2}+1,
\alpha -\beta +1\end{matrix}\Big{\vert} \frac{4 m^2}{k^2}\right) \Bigg{\}}\0
\ee

In the sequel we consider also massless models. The relevant results can be obtained
from the massive models by taking the $m\to 0$ limit. But they can also be obtained by setting $m=0$ from the very beginning. In such a case the basic integral is
\be\label{J2scalarm=0}
\tilde I^{(2)} (d;\alpha,\beta;q_1,q_2,0)&=&\int\frac{d^d
p}{(2\pi)^d}\frac{ 1}{\left((p+q_1)^2\right)^\alpha\left( (p+q_2)^2
\right)^\beta}\0\\
&=& 2^{-d}\pi ^{-d/2} i^{1-d} (k^2)^{\frac d2-\alpha-\beta}\frac{\Gamma\left(\frac d2-\alpha\right)\Gamma\left(\frac d2-\beta\right)\Gamma\left(\alpha+\beta-\frac d2\right)}{\Gamma(\alpha)\Gamma(\beta)\Gamma(d-\alpha-\beta)}\0
\ee

\subsection{Guidelines for the calculations}

We will now set out to do explicit calculations and derive results for two-point
functions in the scalar and fermion model in different dimensions. The method just
outlined is the most
convenient for our purposes, but it is nevertheless one out of many. In fact,
even within  it there are  different possibilities or schemes. We expect that
our results may depend on such schemes, but also to find a criterion to extract
the scheme independent part. In most cases this is conservation and finiteness. In particular,
by suitably choosing the scheme we will be able, for instance, to obtain both
finiteness and conservation for spin 1 current in any dimension in the fermion model. The same is not
as easy for higher spin currents.
In generic spin current correlators and, therefore, 
in the corresponding one-loop effective actions, we will find, beside non-conserved terms,
also terms that diverge in the IR limit $m\to\infty$. Fortunately these
terms are finite in number and easy to identify 
by expanding the OLEA near the IR and the UV. Not only, all the nonconserved and
all IR divergent terms are local. It is thus possible to subtract all the terms
that diverge in the IR, which include, in particular, all the nonconserved ones
and  recover both conservation and finiteness in the IR. 

In this process a particular attention has to be paid to the terms of order 0 in $m$,
in even dimensions. In some cases they are local and conserved, and appear both in the IR and the UV.
Even in this case we follow the attitude of subtracting the IR term from the corresponding UV one,
on the assumption that physical information is contained in the difference between the UV and the IR,
not in their absolute values.

Finally it should be added that the resulting IR and UV expansions are both convergent.

The calculations in the sequel are mainly carried out using a new Mathematica code \cite{Blima}.

To somewhat abbreviate the following formulas, at times we use the compact notation
\be
\,\Pi_{\mathfrak a}^{(2)}(k,n_1,n_2) &=&  \left(n_1\cdot \pi ^{k}\cdot n_2\right){}^2
+{\mathfrak a} (n_1\cdot \pi ^{k}\cdot n_1) ( n_2\cdot
\pi ^{k}\cdot n_2),\label{Pi2}\\
\Pi^{(3)}_{\mathfrak a}(k,n_1,n_2) &=&  \left(n_1\cdot \pi ^{k}\cdot n_2\right){}^3
+{\mathfrak a}(n_1\cdot \pi ^{k}\cdot n_1)( n_1\cdot \pi ^{k}\cdot n_2)( n_2\cdot
\pi ^{k}\cdot n_2),\label{Pi3}
\ee
where ${\mathfrak a}$ is some constant.

The symbol $k$ used in the above formula and in the sequel deserves an explanation: $k$ saturated with $n_1,n_2$ represents the vector $k_\mu$, while in the other cases
it represents the modulus $|k|$. Finally, contrary to (\cite{BCLPS}), 
the latter is $k\equiv |k|=\sqrt{k^2}$.

\section{3d scalar effective field action tomography}
\label{3dEA}

In this section we start the analysis of the two-point functions of spin higher
than 1 currents.

Before reporting on the general spin $s$ case we would like to analyze in detail
a few low spin cases. It is in in general possible to obtain compact expressions
of the one-loop effective actions. However
expanding it in powers of $m$ near the IR and UV limits (an operation we call
{\it tomography})
provides the most interesting information.

It is possible to use the parameter $m$ to cut to slices the two-point function
of currents of any spin.
Let us consider the case of a massive scalar model (msm) in 3d. The basic formulas are
(\ref{scalarlag},\ref{scalarcurrent},\ref{scalarcurrents},\ref{Vsst})
and (\ref{2ptscalar}) together with the analogous ones for higher spins, with
$d=3$.
 
\subsection{3d msm: spin 1 current}

This case is well known and simple, but it is excellent for pedagogical
purposes. 
The exact 2-pt correlator for $s=1$ is  
\be
(n_1 \!\cdot\! \tilde J(k) \!\cdot\! n_2)&=&\frac{i}{8 \pi  k^3} \left(-4 m^2
\coth ^{-1}\left(\frac{2 m}{k}\right)
+2 k m+k^2 \coth ^{-1}\left(\frac{2 m}{k}\right)\right)   \left(k\cdot
n_1\right) 
\left(k\cdot n_2\right) \0\\
&&+\frac{i}{8 \pi  k}\left(n_1\cdot n_2\right) \left(4 m^2 \coth
^{-1}\left(\frac{2 m}{k}\right)
+2 k m-k^2 \coth ^{-1}\left(\frac{2 m}{k}\right)\right)\label{n1Jn2}
\ee
We can expand (\ref{n1Jn2}) in power of $\frac
k{m}$ (IR) or of $\frac mk$ (UV).
In the IR case we find
\be 
{\cal O}(m):&& \frac{i m \left(n_1\!\cdot\! n_2\right)}{2 \pi}\label{IRscalar1currentOm}\\
{\cal O}(m^{-1}): && -\frac{ik^2}{24 \pi m }\left(n_1\!\cdot\!
\pi^{(k)}\cdot n_2\right)    \label{IRscalar1currentOm-1}\\
{\cal O}(m^{-3}): &&-\frac {ik^4} {480 \pi m^3} \left(n_1\!\cdot\!
\pi^{(k)}\cdot n_2\right)  \label{IRscalar1currentOm-3}\\
.... && ....\0
\ee
while the even powers of $m$ vanish. 
The first is a (non-conserved and divergent in the IR limit) local term $\sim
\eta_{\mu\nu}$, which must be subtracted away.
The other terms are all conserved and proportional to the conserved structure
\be
  n_1\!\cdot\! \pi ^{(k)}\!\cdot\! n_2,  \label{k-projector}
\ee

The UV expansion is instead
\be 
{\cal O}(m^0): && -\frac k{16 }  \left(n_1\!\cdot\!
\pi^{(k)}\cdot n_2\right) \label{UVscalar1currentOm0}\\
{\cal O}(m):&&  \frac {im}{2 \pi  k^2} \left(k\!\cdot \!n_1\right) \left(k
\!\cdot \!n_2\right) \label{UVscalar1currentOm}\\
{\cal O}(m^2):&&\frac{m^2} {4k} \left(n_1\!\cdot\!
\pi^{(k)}\cdot n_2\right)\label{UVscalar1currentOm2}\\
{\cal O}(m^3):&& \frac {2i m^3 }{3 \pi  k^2}  \left(n_1\!\cdot\!
\pi^{(k)}\cdot n_2\right)\
\label{UVscalar1currentOm3}\\
....&&....\0
\ee
In fact we have ${\cal O}(m^{2n})=0$ for $n\geq 2$. The only nonvanishing terms
with even powers of $m$ are ${\cal O}(m^0), {\cal O}(m^2)$. 
For these terms see the comment below.

Except (\ref{UVscalar1currentOm}) the other terms are conserved and proportional
to (\ref{k-projector}). The terms proportional to (\ref{k-projector})
are all non-local in the UV, and local in the IR, in particular
(\ref{IRscalar1currentOm-1}) is local and corresponds to the YM action in 3d,
see (\ref{JmunuEvenIRUV}).

The two nonconserved terms are (\ref{IRscalar1currentOm}) in the IR and
(\ref{UVscalar1currentOm}) in the UV. The first is local and the second is
nonlocal, 
but their divergence is the same and local:
\be 
-\frac i{2 \pi } \left(k\!\cdot\! n_2\right)\0
\ee
This means that we can cancel it by subtracting a local term,  $\sim m \int
d^3x\, {\rm tr}( A^2)$.
This amounts to subtracting the IR contribution (which is local) from the UV
one. Indeed we get
\be
{\cal O}_{UV}(m) -{\cal O}_{IR}(m)=- \frac{i m}{2 \pi }\, \left(n_1\!\cdot\!
\pi^{(k)}\cdot n_2\right)\label{UV-IR}
\ee
So the term of order $m$ in the UV and IR conjure up to reform again the same
conserved structure as all the other terms. 
Taking the UV and IR limits splits apart this conserved structure.
The conclusion is that, {\bf up to a local term} {\it we can view the effective
action as a sum of infinite many terms, all proportional to 
$ n_1\cdot \pi^{(k)}\cdot n_2$   with coefficients
proportional to various monomials of $m$ and $k$}. In compact form:
\be
\frac i{8 \pi  k} \left(4 m^2 \coth ^{-1}\left(\frac{2 m}{k}\right)-
2 k m-k^2 \coth ^{-1}\left(\frac{2 m}{k}\right)\right) n_1\!\cdot\! \pi
^{(k)}\cdot n_2\label{subtractedJevenEA}
\ee

\subsection{3d msm: e.m. tensor}

We have to consider 
\be 
(n_1^2\cdot \tilde T(k) \cdot n_2^2)= n_1^\mu n_1^\nu \tilde
T_{\mu\nu\lambda\rho}(k) n_2^\lambda n_2^\rho\0
\ee
Expanding in the IR we have
\be 
 {\cal O}(m^3):&& {+} \frac {2im^3} {3 \pi }\left(2  \left(n_1\!\cdot
\!n_2\right){}^2+  
\left(n_1\!\cdot \!n_1\right)  \left(n_2\!\cdot \!n_2\right)\right)
\label{IRemOm3}\\
{\cal O}(m): &&  - \frac{im}{6 \pi } \left(- \left(n_2\!\cdot \!n_2\right)
\left(k \!\cdot \!n_1\right){}^2-
4  \left(n_1\!\cdot \!n_2\right) \left(k \!\cdot \!n_2\right)  \left(k \!\cdot
\!n_1\right)\right.\0\\
&& {+}2 k ^2 \left. \left(n_1\!\cdot \!n_2\right){}^2+k ^2 \left(n_1\!\cdot
\!n_1\right)  \left(n_2\!\cdot \!n_2\right)
- \left(n_1\!\cdot \!n_1\right) \left(k \!\cdot
\!n_2\right){}^2\right)\label{IRemOm0}\\
  {\cal O}(m^{-1}):&& \frac{i k^4}{60 \pi m } \Pi^{(2)}_{\frac 12}(k,n_1,n_2)   \label{IRemOm-1}\\
  {\cal O}(m^{-3}):&& 
\frac{i k^6}{1680 \pi m^3} \Pi^{(2)}_{\frac 12}(k,n_1,n_2)  \label{IRemOm-3}\\
....&&....\0
\ee
while all the even powers vanish.
The ${\cal O}(m^3)$ and  ${\cal O}(m)$ terms are non-conserved, while the other
terms are all conserved and proportional to the same structure.

In the UV we have
\be 
{\cal O}(m^0):&& \frac{ k^3}{32}\Pi^{(2)}_{\frac 12}(k,n_1,n_2)   \label{UVemOm0}\\
{\cal O}(m): && \frac{i m}{2 \pi  k^2} \left(k\!\cdot\!n_1\right){}^2
\left(k\!\cdot\!n_2\right){}^2 \label{UVemOm}\\
{\cal O}(m^2):&&- \frac{m^2k}{4} \Pi^{(2)}_{\frac 12}(k,n_1,n_2)    \label{UVemOm2}\\
{\cal O}(m^3):&& \frac{2 i m^3}{3 \pi  k^4}\left(-3  \left(k\!\cdot
\!n_1\right){}^2  \left(k  n_2\right){}^2\right. 
+k ^2  \left(n_1\!\cdot\!n_1\right) \left(k\!\cdot \! n_2\right){}^2\0\\
&& {+}4 k ^2  \left(n_1\!\cdot \!n_2\right) \left. \left(k \!\cdot \!n_1\right) 
\left(k\!\cdot \!n_2\right)
+k ^2  \left(n_2\!\cdot \!n_2\right)  \left(k\!\cdot
\!n_1\right){}^2\right)\label{UVemOm3}\\
{\cal O}(m^4):&&\frac{m^4}{2 k}  \Pi^{(2)}_{\frac 12}(k,n_1,n_2)  \label{UVemOm4}\\
{\cal O}(m^{2p}):&&0 ,\quad\quad {\rm for}\quad\quad p\geq 3 \0
 \ee
In fact we have ${\cal O}(m^{2m})=0$ for $m\geq 3$. The only nonvanishing terms
with even powers of 
$m$ are ${\cal O}(m^0), {\cal O}(m^2),{\cal O}(m^4)$ (again, about these terms,
see
the comment below)

All the terms are conserved except ${\cal O}(m)$ and ${\cal O}(m^3)$.
But putting together the analogous non-conserved terms in the UV and IR (that is
subtracting the local IR terms from the (nonlocal) UV ones) we recover
conservation.
\be 
{\cal O}_{UV}(m) -{\cal O}_{IR}(m)&=& \frac{im k^2}{3 \pi }  \Pi^{(2)}_{\frac 12}(k,n_1,n_2) 
  \label{emOm}\\ 
{\cal O}_{UV}(m^3) -{\cal O}_{IR}(m^3)&=& {-}\frac{4 im^3}{3 \pi } \Pi^{(2)}_{\frac 12}(k,n_1,n_2) \label{emOm3}
\ee
So we find a result analogous to the 1-current. 
{\bf Up to local terms} {\it the effective action is a sum of infinite
many terms, all proportional to the same conserved structure (\ref{emOm3})
with coefficients proportional to various monomials of $m$ and $k$.} 
They form a convergent series both in the IR and in the UV. In compact
form:
\be
&&\frac{i} {48 \pi  k}\Bigg{(}48 m^4 \coth ^{-1}\left(\frac{2 m}{k}\right)+2 k m
\left(5 k^2-12 m^2\right)-
24 k^2 m^2 \coth ^{-1}\left(\frac{2 m}{k}\right)\0\\ 
&&\quad\quad+3 k^4 \coth ^{-1}\left(\frac{2 m}{k}\right)\Bigg{)}  
\Pi^{(2)}_{\frac 12}(k,n_1,n_2) 
\label{subtractedspin2EA}
\ee

It should be noticed that the massless model case gives the result:
\be 
(n_1^2 \!\cdot\! \tilde T(k) \!\cdot\! n_2^2)=\frac{k^3 }{32} \Pi^{(2)}_{\frac 12}(k,n_1,n_2)  \label{emmassless}
\ee
This is conserved {\it but not traceless}, which
is not surprising because  a scalar massless model in $d\geq 3$ is not
conformal invariant.  

Eq.(\ref{emOm}) is conserved. It does not coincide with the linearized Einstein-Hilbert action
(in particular it is nonlocal), but this is simply a nonlocal version
of the same, in the same sense as we have already seen for spin 3 and higher in section \ref{sec:appetizer}.

\subsection{3d msm: spin 3 current}

For the 3-spin current we have in the IR
\be 
{\cal O}(m^5):&&\frac{8i m^5}{5 \pi } \left(2 \left(n_1\!\cdot 
\!n_2\right){}^3+
3 \left(n_1\!\cdot  \!n_1\right) \left(n_2\!\cdot  \!n_2\right)\left(n_1\!\cdot 
\!n_2\right)\right)\label{IRJ3Om5}\\
{\cal O}(m^3):&&\frac{2 im^3} {3 \pi } \left(3
\left(n_1\!\cdot\!n_1\right)\left(n_1\!\cdot\!n_2\right)\left(k\!\cdot\!n_2\right){}^2
+3 \left(2\left(n_1\!\cdot\!n_2\right){}^2+
\left(n_1\!\cdot\!n_1\right)\left(n_2\!\cdot\!n_2\right)\right)\left(k\!\cdot\!n
_1\right)\left(k\!\cdot\!n_2\right)\right.\0\\
&&+\left.
\left(n_1\!\cdot\!n_2\right)
\left(3\left(n_2\!\cdot\!n_2\right)\left(k\!\cdot\!n_1\right){}^2-k^2
\left(2\left(n_1\!\cdot\!n_2\right){}^2
+3\left(n_1\!\cdot\!n_1\right)\left(n_2\!\cdot\!n_2\right)\right)\right)\right)\label{IRJ3Om3}\\ 
{\cal O}(m): &&
\frac{i m}{10 \pi } \Big{(}3 \left(n_2\!\cdot\!n_2\right)
\left(k\!\cdot\!n_2\right) \left(k\!\cdot\!n_1\right){}^3
+3 \left(n_1\!\cdot\!n_2\right) \left(3 \left(k\!\cdot\!n_2\right){}^2-
k^2 \left(n_2\!\cdot\!n_2\right)\right) \left(k\!\cdot\!n_1\right){}^2\0\\
&&+3 \left(k\!\cdot\!n_2\right) \left(\left(n_1\!\cdot\!n_1\right)
\left(k\!\cdot\!n_2\right){}^2-
k^2 \left(2 \left(n_1\!\cdot\!n_2\right){}^2+\left(n_1\!\cdot\!n_1\right)
\left(n_2\!\cdot\!n_2\right)\right)\right) \left(k\!\cdot\!n_1\right)\0\\
&& +
k^2 \left(n_1\!\cdot\!n_2\right) \left(k^2 \left(2
\left(n_1\!\cdot\!n_2\right){}^2+3 \left(n_1\!\cdot\!n_1\right)
\left(n_2\!\cdot\!n_2\right)\right)-
3 \left(n_1\!\cdot\!n_1\right) \left(k\!\cdot\!n_2\right){}^2\right)\Big{)}
\label{IRJ3Om}\\ 
{\cal O}(m^{-1}):&&-\frac{i k^6}{140 \pi m } \Pi^{(3)}_{\frac 32}(k,n_1,n_2) \label{IRJ3Om-1}
\ee
The coefficients of even powers in $m$ vanish, while the negative odd powers are
all proportional to the conserved structure (\ref{IRJ3Om-1}). 
The terms ${\cal O}(m^5),{\cal O}(m^3),{\cal O}(m)$ are local and non-conserved.

In the UV we have
\be 
{\cal O}(m^0):&&-\frac{k^5}{64}  \Pi^{(3)}_{\frac 32}(k,n_1,n_2)  \label{UVJ3Om0}\\
{\cal O}(m):&&\frac{im}{2 \pi  k^2}   \left(k\!\cdot\!\!n_1\right){}^3
\left(k\!\cdot\!\!n_2\right){}^3\label{UVJ3Om}\\ 
{\cal O}(m^2):&&  \frac{3m^2 k^3}{16}  \Pi^{(3)}_{\frac 32}(k,n_1,n_2)  \label{UVJ3Om2}\\ 
{\cal O}(m^3):&&\frac{2 im^3}{3 \pi  k^4} (k\!\cdot\! n_1) (k\!\cdot\! n_2)
\left(-5 \left(k\!\cdot\! n_1\right){}^2 \left(k\!\cdot\! n_2\right){}^2
+3 k^2 (n_1\!\cdot\! n_1) \left(k\!\cdot\! n_2\right){}^2\right.\0\\
&& {+}9 k^2 (n_1\!\cdot\! n_2)\left. (k\!\cdot\! n_1) (k\!\cdot\! n_2)+3 k^2
(n_2\!\cdot\! n_2) \left(k\!\cdot\! n_1\right){}^2\right) \label{UVJ3Om3}\\ 
{\cal O}(m^4):&&-\frac{3 m^4k}{4}  \Pi^{(3)}_{\frac 32}(k,n_1,n_2)  \label{UVJ3Om4}\\ 
{\cal O}(m^5):&&\frac{8 im^5}{5 \pi  k^6} \Big{(}6 k^4 \left(n_1\!\cdot\!
n_2\right){}^2 (k\!\cdot\! n_1) (k\!\cdot\! n_2)+3 k^4 (n_1\!\cdot\!
n_1)(n_1\!\cdot\! n_2)
 \left(k\!\cdot\! n_2\right){}^2\0\\
&&+3 k^4 (n_1\!\cdot\! n_1)( n_2\!\cdot\! n_2)( k\!\cdot\! n_1) (k\!\cdot\! n_2)
+3 k^4 (n_1\!\cdot\! n_2) (n_2\!\cdot\! n_2) \left(k\!\cdot\! n_1\right){}^2\0\\
&& -3 k^2 (n_1\!\cdot\! n_1) (k\!\cdot\! n_1) \left(k\!\cdot\! n_2\right){}^3 -9
k^2 (n_1\!\cdot\! n_2) \left(k\!\cdot\! n_1\right){}^2 \left(k\!\cdot\!
n_2\right){}^2
\0\\
&&-3 k^2 (n_2\!\cdot\! n_2) \left(k\!\cdot\! n_1\right){}^3 (k\!\cdot\! n_2)+5
\left(k\!\cdot\! n_1\right){}^3 \left(k\!\cdot\!
n_2\right){}^3\Big{)}\label{UVJ3Om5}\\ 
 {\cal O}(m^6):&&\frac{m^6}{  k}\Pi^{(3)}_{\frac 32}(k,n_1,n_2)   \label{UVJ3Om6} \0
\ee
The terms ${\cal O}(m^{2n})$ with $n\geq 4$ vanish. All terms are conserved,
except $ {\cal O}(m),{\cal O}(m^3),{\cal O}(m^5)$.
  
Proceeding as above we subtract from the non-conserved terms in the UV the
homogeneous local non-conserved terms in the IR and obtain conserved terms:
\be 
{\cal O}_{UV}(m) -{\cal O}_{IR}(m)\!\!&=& {-}\!\!\frac{im k^4}{5 \pi } 
 \Pi^{(3)}_{\frac 32}(k,n_1,n_2)\label{J3Om}\\
{\cal O}_{UV}(m^3) -{\cal O}_{IR}(m^3)\!\!&=&\!\!\frac{4im^3}{3 \pi }k^2 
\Pi^{(3)}_{\frac 32}(k,n_1,n_2) \label{J3Om3}\\
{\cal O}_{UV}(m^5) -{\cal O}_{IR}(m^5)\!\!&=& - \!\!  \frac{16im^5}{5 \pi}
\Pi^{(3)}_{\frac 32}(k,n_1,n_2)
\label{J3Om5}
\ee
Therefore {\bf up to local terms} {\it the effective action is a sum of infinite
many terms, all proportional to the same conserved structure 
with coefficients proportional to various monomials of $m$ and $k$.}They form a 
convergent series both in the IR and in the UV. In compact form:
\be
&&\frac{i}{480 \pi  k}\Bigg{(}960 m^6 \coth ^{-1}\left(\frac{2 m}{k}\right)-480
k m^5-720 k^2 m^4 \coth ^{-1}\left(\frac{2 m}{k}\right)
+320 \left(k^2\right){}^{3/2} m^3\0\\
&&+180 k^4 m^2 \coth ^{-1}\left(\frac{2 m}{k}\right)-66 k^4 k m
-15 k^6 \coth ^{-1}\left(\frac{2 m}{k}\right)\Bigg{)}\, \,\Pi_{\frac 32}^{(3)}(k,n_1,n_2)
\label{subtracted3dspinevenEA}
\ee
The term (\ref{J3Om3}) is local and gives rise to an eom, which is the nonlocal
version of the Fronsdal spin 3 equation of motion we have already met above.

\subsection{3d msm: higher spin currents}

This scheme repeats itself for higher spin currents. 
For spin 4 there are 4 non-conserved terms in the IR and 4 in the UV, while the
others are conserved or 0.
Subtracting the IR non-conserved terms from the corresponding UV ones all the
nonvanishing terms turn out to be proportional to the conserved structure:
\be
 \frac 13 \left(n_1\!\cdot\! \pi ^{(k)}\!\cdot\! n_2\right){}^4 
\!\!&+&\!\!\frac 18\left(n_1\!\cdot\! \pi ^{(k)}\!\cdot\! n_1\right){}^2
\left(n_2\!\cdot\! \pi ^{(k)}\!\cdot\! n_2\right){}^2 \0\\
&&+  (n_1\!\cdot\! \pi ^{(k)}\!\cdot\! n_1) \left(n_1\!\cdot\! \pi
^{(k)}\!\cdot\! n_2\right){}^2 (n_2\!\cdot\! \pi ^{(k)}\cdot n_2) 
\label{J4consstructure}
\ee
All terms with even powers of $m$ vanish, except $m^0,m^2,m^4, m^6,m^8$.

For spin 5 there are 5 non-conserved terms in the IR and 5 in the UV, while the
others are conserved or 0.
Subtracting the IR non-conserved terms from the corresponding UV ones all the
nonvanishing terms turn out to be proportional  
to the conserved structure:
\be 
   \left(n_1\!\cdot\! \pi ^{(k)}\!\cdot\! n_2\right){}^5
\!\!&+&\!\!\frac{15}{8}  \left(n_1\!\cdot\! \pi ^{(k)}\!\cdot\! n_1\right){}^2
(n_1\!\cdot\! \pi ^{(k)}\!\cdot\! n_2) \left(n_2\!\cdot\! \pi ^{(k)}\!\cdot\!
n_2\right){}^2\0\\
&&+ 5  (n_1\!\cdot\! \pi ^{(k)}\!\cdot\! n_1) \left(n_1\!\cdot\! \pi
^{(k)}\!\cdot\! n_2\right){}^3 (n_2\!\cdot\! \pi ^{(k)}\cdot
n_2)\label{J5consstructure}
\ee
All terms with even powers of $m$ vanish, except $m^0,m^2,m^4, m^6,m^8,m^{10}$.

{\bf Comment 1}. As we have seen above any conserved structure is connected to a
(non-local) higher spin field equation of motion. 
In particular eqs.(\ref{IRscalar1currentOm-1})
and (\ref{emOm}) are conserved structures which represent the linearized YM
and EH actions, respectively, the second one in a nonlocal version. 
Eq.(\ref{J3Om3}) is non-local and gives rise 
to a variant of the nonlocal Fronsdal equation discussed in sec.\ref{sec:appetizer}. 
It is clear that any two-point correlator structure can be uniquely related
to a given (linearized) equation of motion. The structure of the 2pt-functions conform 
to the general discussion in sec.\ref{sec:universal}. This will be confirmed by
the forthcoming analysis. 

It is remarkable that the conserved structures that appear in the above
expansions are
always the same for any fixed 2pt correlator. As we will see this is not the
case for
the effective field action originating from a fermion model.

{\bf Comment 2}. The nonvanishing even $m$ power terms are a finite number in
all cases. They come from the fact that the UV expansion of $\coth ^{-1} $
\be
\coth ^{-1}\left(\frac{2 m}{k}\right)= 
-\frac{i \pi }{2}+\frac{2 m}{k}+\frac{8 m^3}{3 k^3}+\frac{32 m^5}{5 k^5}+{\cal
O}\left(m^6\right)
\ee
contains the factor $-\frac{i \pi }{2}$. This is the reason why they are a
finite number and do not contain the factor $\frac i{\pi}$ like the others.
The factor $-\frac{i \pi }{2}$ comes from the logarithmic cut of $\coth ^{-1} $
and it is determined by the choice of the Riemann sheet. So it is scheme
dependent.

It is interesting to compare the ${\cal O}(m^0)$ results with the massless model case, obtained via (\ref{J2scalarm=0}). In the
massless case for spin 1 we get
\be
-\frac{1}{16}  k (n_1\!\cdot\! \pi ^{(k)}\!\cdot\! n_2)\label{spin1massless}
\ee
for spin 2
\be
\frac{k^3}{32}  \Pi^{(2)}_{\frac 12}(k,n_1,n_2) \label{spin2massless}
\ee
and for spin 3
\be
-\frac{ k^5}{64}   \Pi^{(3)}_{\frac 32}(k,n_1,n_2) \label{spin3massless}
\ee
These correlators are nonlocal and coincide with the   ${\cal O}_{UV}(m^0)$ terms evaluated
above\footnote{Appendix B contains a complete analysis of two-point functions for  massless scalar and fermion models.}. To be precise 
there is an indeterminacy in their sign due to the branch point at $k=0$
originated from the choice of sign of the square root $\sqrt{k^2}$. This indeterminacy
is present also in the $m\to 0$ limit of the massive model and it is related to the choice of Riemann sheet
mentioned above. As a consequence of it, in this paper we do not worry about  the sign in front of the Maxwell 
and EH kinetic terms that appear in the effective actions. We postpone to a future work the task of finding
a physically consistent prescription that eliminates this indeterminacy.

\section{3d fermion effective field theory action tomography}

We consider now the same analysis for the  massive fermion model (mfm). The starting
point are 
eqs.(\ref{actionA},\ref{actiong},\ref{Jmunuab},\ref{Tmnlr}) and the like for
higher spins (see also \cite{BCLPS}).

\subsection{3d mfm: spin 1 current}

This case is rather simple. It takes a very compact form  
\be
(n_1\!\cdot\! \tilde J(k) \!\cdot\! n_2)&=& \frac{i}{8 \pi  k} \Bigg{(}-\left(4
m^2 \coth ^{-1}\left(\frac{2 m}{k}\right)-2 k m
+k^2 \coth ^{-1}\left(\frac{2 m}{k}\right)\right) (n_1\!\cdot\! \pi
^{(k)}\!\cdot\! n_2)\0\\
&& + 4 i m \coth ^{-1}\left(\frac{2 m}{k}\right) \epsilon
\left(k\!\cdot\!n_1\!\cdot\!n_2\right)\Bigg{)}\label{spin1fermionEA}
\ee
and is conserved without any subtraction.

Expanding, the term
\be
{\cal O}(m^0): -\frac{\epsilon \left(k\!\cdot\!n_1\!\cdot\!n_2\right)}{4 \pi
}\label{IRfermionJ1Om0}
\ee
corresponds to the linearized CS action (here $\epsilon \left(k
\!\cdot\!n_1\!\cdot\!n_2\right)$
means $\epsilon_{\mu\nu\rho} k^\mu n_1^\nu n_2^\rho$), and the term 
\be 
{\cal O}(m^{-1}): { -}\frac{i}{12 \pi  m}\,  k^2 \,(n_1\!\cdot\! \pi ^{(k)}\!\cdot\!
n_2)  \label{IRfermionJ1Om-1}
\ee
in the IR corresponds to the linearized YM action.
 
\subsection{3d mfm: e.m. tensor - even part}

For the e.m. tensor we have in the IR (all formulas below have to be multiplied
by the factor ${ \frac 1{16}}$)
\be 
 {\cal O}(m^3):&&{ -2}\frac{im^3}{3 \pi } \left( \left(n_1\!\cdot\!n_2\right){}^2+
\left(n_1\!\cdot\!n_1\right)  \left(n_2\!\cdot\!n_2\right)\right) 
\label{IRfermionemOm3}\\
{\cal O}(m): &&{ -}\frac{im k^2}{6 \pi } \Pi^{(2)}_{-1}(k,n_1,n_2) 
  \label{IRfermionemOm0}\\
 {\cal O}(m^{-1}):&&  \frac {ik^4}{40 \pi m} \Pi^{(2)}_{-\frac 13}(k,n_1,n_2)   \label{IRfermionemOm-1}\\
 {\cal O}(m^{-3}):&&\frac {i k^6} {672 m^3\pi} \Pi^{(2)}_{-\frac 15}(k,n_1,n_2)    \label{IRfermionemOm-3}\\ 
....&&....\0
\ee
The even powers vanish.
The ${\cal O}(m^3)$ term is not conserved, while the other terms are all
conserved and proportional to {\it different combinations} of the two conserved
structures.

In the UV we have
\be 
{\cal O}(m^0):&& \frac {k^3}{32}  \Pi^{(2)}_{-\frac 12}(k,n_1,n_2)   \label{UVfermionemOm0}\\
{\cal O}(m): &&  0\label{UVfermionemOm}\\
{\cal O}(m^2):&& \frac{m^2}{8} k   (n_1\cdot \pi ^{(k)}\cdot n_1)( n_2\cdot
\pi ^{(k)}\cdot n_2) \label{UVfermionemOm2}\\
{\cal O}(m^3):&& { -2}\frac{im^3} {3 \pi  k^4} \left(\left(k^2 
\left(n_2\!\cdot\!n_2\right)-2  \left(k\!\cdot\!n_2\right){}^2\right)
\left(k\!\cdot\!n_1\right){}^2
+2 k^2 \left(n_1\!\cdot\!n_2\right)  \left(k\!\cdot\!n_2\right) 
\left(k\!\cdot\!n_1\right)\right.\0\\
&&+k^2 \left. \left(n_1\!\cdot\!n_1\right) 
\left(k\!\cdot\!n_2\right){}^2\right)  \label{UVfermionemOm3}\\
{\cal O}(m^4):&&{ -}\frac{m^4}{2 k}  \Pi^{(2)}_{\frac 12}(k,n_1,n_2) \label{UVfermionemOm4}\\
{\cal O}(m^5):&&{ -2}\frac {4i m^5}{5 \pi  k^2} \Pi^{(2)}_{\frac 13}(k,n_1,n_2)   \label{UVfermionemOm5}\\
{\cal O}(m^6): &&  0\0
 \ee
These are all conserved except  ${\cal O}(m^3)$.
But putting together the analogous non-conserved term in the UV and IR (that is
subtracting the local IR term from the (nonlocal) UV one) we recover
conservation:
\be 
{\cal O}_{UV}(m^3) -{\cal O}_{IR}(m^3)&=&{ 2} \frac{ im^3}{3 \pi } \Pi^{(2)}_{1}(k,n_1,n_2)\label{fermionemOm3}
\ee
Eq.(\ref{IRfermionemOm0}) is the linearized and {\it local} version of the EH
equation of motion (see sec.{\ref{sec:appetizer}). The other are non-local
versions of the same (except 
(\ref{UVfermionemOm2}). Actually, according to our general philosophy the term
${\cal O}_{IR}(m)$, which is divergent
in the IR limit, must be subtracted. It will therefore appear in the place of
the vanishing term (\ref{UVfermionemOm}) with
inverted sign.

Once again {\bf up to local terms} {\it the effective action is a sum of
infinite many terms, which form a convergent series both in the IR and in the
UV,
all of them proportional to various combinations of the conserved
structures 
with coefficients proportional to various monomials of $m$ and $k$.} In compact
form:
\be
&&{ -}\frac{i}{{ 96} \pi  k} \left({ 96} m^4 \coth ^{-1}\left(\frac{2
m}{k}\right)-{48} k m^3{ -4} k^3 m- 6 k^4 \coth ^{-1}\left(\frac{2
m}{k}\right)\right) 
\left(n_1\!\cdot\! \pi ^{(k)}\!\cdot\! n_2\right){}^2\0\\
&&{ -}\frac{i}{{ 96} \pi  k} \left({ 48} m^4 \coth ^{-1}\left(\frac{2 m}{k}\right)-{ 24} k m^3-{ 24} k^2
m^2 \coth ^{-1}\left(\frac{2 m}{k}\right){ +10} k^3 m\right.\0\\
&&\quad\quad\quad\quad\quad\quad\left. { +3} k^4 \coth ^{-1}\left(\frac{2
m}{k}\right)\right) (n_1\!\cdot\! \pi ^{(k)}\!\cdot\! n_1)( n_2\!\cdot\!
\pi
^{(k)}\!\cdot\! n_2). \label{spin2fermionevenEA}
\ee

\subsection{3d mfm: e.m. tensor - odd part}

In the IR (all formulas below have to be multiplied by the factor ${ \frac 1{16}}$)
\be 
 {\cal O}(m^3):&&0 \0\\
{\cal O}(m^2):&& -\frac{m^2}{\pi } \left(n_1\!\cdot\! n_2\right) \epsilon
\left(k\!\cdot\!n_1\!\cdot\!n_2\right)\label{IRfermionemoddOm2}\\
 {\cal O}(m^0): && \frac{k^2}{12 \pi } \epsilon
\left(k\!\cdot\!n_1\!\cdot\!n_2\right)
(n_1\!\cdot\! \pi ^{(k)}\!\cdot\! n_2)  \label{IRfermionemoddOm0}\\
 {\cal O}(m^{-2}): && \frac{k^4}{240 m^2 \pi } \epsilon
\left(k\!\cdot\!n_1\!\cdot\!n_2\right)
(n_1\!\cdot\! \pi ^{(k)}\!\cdot\! n_2)  \label{IRfermionemoddOm-2}\\
....&&....\0
\ee
the odd powers vanish.
The ${\cal O}(m^2)$ term is not conserved, while the other terms are all
conserved and proportional to  the unique odd conserved structure.

In the UV:
\be 
  {\cal O}(m^0): && 0\0\\
{\cal O}(m):&&-\frac{ik m}{8 } \epsilon \left(k\!\cdot\!n_1\!\cdot\!n_2\right)
(n_1\!\cdot\! \pi
^{(k)}\!\cdot\! n_2)  \label{UVfermionemoddOm1}\\
 {\cal O}(m^{2}):&&-\frac{m^2}{\pi  k^2}\left(k\!\cdot\!n_1\right)
\left(k\!\cdot\!n_2\right) \epsilon \left(k\!\cdot\!n_1\!\cdot\!n_2\right)
\label{UVfermionemoddOm2}\\
 {\cal O}(m^{3}): && { i } \frac {m^3} {2k}  \epsilon
\left(k\!\cdot\!n_1\!\cdot\!n_2\right)
(n_1\!\cdot\! \pi ^{(k)}\!\cdot\! n_2)  \label{UVfermionemoddOm3}\0\\
 {\cal O}(m^{4}):&& { -\frac {4 m^4}{3\pi k}  \epsilon
\left(k\!\cdot\!n_1\!\cdot\!n_2\right)(n_1\!\cdot\! \pi ^{(k)}\!\cdot\! n_2) }\0\\
....&&....\0
\ee
${\cal O}(m^{2})$ is not conserved, but
\be 
{\cal O}_{UV}(m^2) -{\cal O}_{IR}(m^2)&=&  \frac {m^2} {\pi}  \epsilon
\left(k\!\cdot\!n_1\!\cdot\!n_2\right) (n_1\!\cdot\! \pi ^{(k)}\!\cdot\! n_2) 
\label{UV-IRfermionemoddOm}  
\ee
is. In summary, after subtracting ${\cal O}_{IR}(m^2)$ the odd 2-pt correlator
is:
\be 
- \frac{m}{4 \pi  k}\left(4 m^2 \coth ^{-1}\left(\frac{2 m}{k}\right)-
2 k m-k^2 \coth ^{-1}\left(\frac{2 m}{k}\right)\right) \epsilon
\left(k\!\cdot\!n_1\!\cdot\!n_2\right) (n_1\!\cdot\! \pi ^{{k}},
n_2)\label{subtractedfermionoddEA}
\ee
The term (\ref{IRfermionemoddOm0}) and, in a scaling limit, also
(\ref{UVfermionemoddOm1}), give rise to the linearized CS action as discussed in
\cite{BCLPS}.

\subsection{3d mfm: spin 3, even part}

This was already discussed in \cite{BCLPS}, so 
we report here only the final results. One must subtract the local terms
${\cal O}(m^5),{\cal O}(m^3)$  in the IR, which are not
conserved.  After which the effective action becomes 
\be 
&&{ -}\frac{i}{216 \pi  k} \Bigg{(}192 m^6 \coth ^{-1}\left(\frac{2 m}{k}\right)-96
k m^5-48 k^2 m^4 \coth ^{-1}\left(\frac{2
m}{k}\right)\label{correctedspin3evenEA}\\
&&+16 k^3 m^3-12 k^4 m^2 \coth ^{-1}\left(\frac{2 m}{k}\right)
-6 k^5 m+3 k^6 \coth ^{-1}\left(\frac{2 m}{k}\right)\Bigg{)}  \left(n_1\!\cdot\!
\pi^{(k)}\!\cdot\! n_2\right) {}^3 \0\\
&& -\frac{i}{288 \pi  k} \Bigg{(}  384 m^6 \coth ^{-1}\left(\frac{2
m}{k}\right)-192 k m^5-128 k^2 m^4 \coth ^{-1}\left(\frac{2 m}{k}\right)
+48 k^3 m^3\0\\
&&+28 k^4 m^2 \coth ^{-1}\left(\frac{2 m}{k}\right)
+6 k^5 m-3 k^6 \coth ^{-1}\left(\frac{2 m}{k}\right) \Bigg{)}
(n_1\!\cdot\! \pi ^{(k)}\!\cdot\! n_1)( n_1\!\cdot\! \pi ^{(k)}\!\cdot\! n_2
)(n_2\!\cdot\! \pi ^{(k)}\!\cdot\! n_2)  \0
\ee
The ${\cal O}_{IR}(m)$ term is conserved and has to be subtracted from it.
The interpretation of these conserved structures in terms of massless Fronsdal
eom has been discussed above. At each order they are
different combinations of two conserved structures
\be 
\left(n_1\!\cdot\! \pi ^{(k)}\!\cdot\! n_2\right) {}^3 \quad\quad {\rm and} 
\quad\quad 
(n_1\!\cdot\! \pi ^{(k)}\!\cdot\! n_1)( n_1\!\cdot\! \pi ^{(k)}\!\cdot\! n_2
)(n_2\!\cdot\! \pi ^{(k)}\!\cdot\! n_2) \label{twostructures}
\ee
but it is actually easy to prove that all these combinations give rise to the
same eom (after taking the trace of the resulting equation and 
re-inserting it). The only condition is that the coefficient of the first
structure be nonvanishing.

\subsection{3d mfm: spin 3, odd part}

One must subtract the local terms ${\cal O}(m^4),{\cal O}(m^2)$ in the IR, which
are not conserved. 
After which the effective action becomes:
\be
&&-\frac{{1}}{216 \pi  k}\Bigg{(}96 m^5 \coth ^{-1}\left(\frac{2 m}{k}\right)-48 k
m^4-4 k^3 m^2-3 k^4 m \coth ^{-1}\left(\frac{2 m}{k}\right) \Bigg{)} \0\\
&&\cdot \epsilon \left(k\!\cdot\!n_1\!\cdot\!n_2\right) (n_1\!\cdot\! \pi
^{(k)}\!\cdot\! n_1)(
n_2\!\cdot\! \pi ^{(k)}\!\cdot\! n_2) \label{correctedspin3oddEA}\\
&&-\frac{1}{54 \pi  k}\Bigg{(}48 m^5 \coth ^{-1}\left(\frac{2 m}{k}\right)
-24 k m^4-24 k^2 m^3 \coth ^{-1}\left(\frac{2 m}{k}\right)+10 k^3 m^2\0\\
&&+3 k^4 m \coth ^{-1}\left(\frac{2 m}{k}\right) \Bigg{)} \epsilon
\left(k\!\cdot\!n_1\!\cdot\!n_2\right) \left(n_1\!\cdot\! \pi ^{(k)}\,
n_2\right){}^2 \0
\ee
The meaning of the term ${\cal O}(m)$ in the UV (in the scaling limit)
\be 
-i\frac {m}{|k|}\epsilon \left(k\!\cdot\!n_1\!\cdot\!n_2\right) k^4
\left(\frac{1}{36}   
\left(n_1\!\cdot\! \pi ^{(k)}\!\cdot\! n_2\right){}^2
-\frac{1}{144}  (n_1\!\cdot\! \pi ^{(k)}\!\cdot\! n_1)( n_2\!\cdot\! \pi
^{(k)}\!\cdot\! n_2)\right)\label{spin3oddUVOm}
\ee
and ${\cal O}(m^0)$ in the IR 
\be
\epsilon \left(k\!\cdot\!n_1\!\cdot\!n_2\right)\left(\frac{k^4}{240 \pi } 
(n_1\!\cdot\! \pi
^{(k)}\!\cdot\! n_1)( n_2\!\cdot\! \pi ^{(k)}\!\cdot\! n_2)-\frac{2 k^4}{135 \pi
}  
 \left(n_1\!\cdot\! \pi ^{(k)}\!\cdot\!
n_2\right){}^2\right)\label{spin3oddUVOm0}
\ee
have already been discussed in \cite{BCLPS}.

\section{Tomography in 5d}
\label{sec:5d}

There is no substantial difference between 3d and 5d. We start from the same
formulas as in 3d and change only the dimension. For obvious reasons of
readability we limit ourselves to the even parity part and the lowest spins,
although the generalization is at hand.

\subsection{5d scalar Model}

\subsubsection{5d msm: spin 1 current}

The analog of eq.(\ref{n1Jn2}) is 
\be
(n_1 \!\cdot\! \tilde J(k) \!\cdot\! n_2)&=&-\frac i{768 \pi^2 k^3}
\Bigg{(}k^2\left(n_1\!\cdot\!n_2\right) \left(40 k m^3+3 \left(k^2-4
m^2\right){}^2 \coth ^{-1}\left(\frac{2 m}{k}\right)-6 k^3 m\right)\0\\
&&+3 \left(8 k m^3-\left(k^2-4 m^2\right){}^2 \coth ^{-1}\left(\frac{2
m}{k}\right)+2 k^3
m\right)\left(k\!\cdot\!n_1\right)\left(k\!\cdot\!n_2\right)\Bigg{)}
\label{JJ5d}
\ee
This is not conserved, but the divergence is local. Expanding in powers of $m$
like in 3d, we get in the IR
\be
{\cal O}(m^3)&:&-\frac{im^3} {12 \pi ^2}\left(n_1\!\cdot\!
n_2\right)\label{5dscalarJIRm3}\\
{\cal O}(m)&:&\frac{im}{48 \pi ^2}  k^2 (n_1\!\cdot\! \pi ^{(k)}\!\cdot\!
n_2)\label{5dscalarJIRm}\\
{\cal O}(m^{-1})&:&-\frac{i}{960 \pi ^2m} k^4 (n_1\!\cdot\! \pi ^{(k)}\!\cdot\!
n_2)\label{5dscalarJIRm-1}\\
\ldots &:& \ldots\0
\ee
All terms corresponding to even powers of $m$ vanish. In the UV we have instead
\be 
{\cal O}(m^0): && -\frac 1{512 \pi}    k^3 (n_1\!\cdot\! \pi ^{(k)}\!\cdot\!
n_2)\label{5dscalarJUVm0}\\
{\cal O}(m):&& 0\0\\
{\cal O}(m^2):&& m^2 \frac{1} {64 \pi} k (n_1\!\cdot\! \pi ^{(k)}\!\cdot\! n_2
)\label{5dscalarJUVm2}\\
{\cal O}(m^3):&& -m^3 \frac {i}{12 \pi^2} \frac{  \left(k\!\cdot \!n_1 \right) 
\left(k\!\cdot \!n_2\right) }{k^2}
\label{5dscalarJUVm3}\\
{\cal O}(m^4):&& -m^4 \frac{1} {32 \pi k} (n_1\!\cdot\! \pi ^{(k)}\!\cdot\! n_2
)\label{5dscalarJUVm4}\\
....&&....\0
\ee
All even powers of $m$ $\geq 6$  vanish. All these terms are conserved except
${\cal
O}_{IR}(m^3)$ and ${\cal O}_{UV}(m^3)$. But once again ${\cal O}_{IR}(m^3)$ is
local and can be subtracted, and
\be
{\cal O}_{UV}(m^3)-{\cal O}_{IR}(m^3)= \frac{im^3}{12 \pi ^2}   (n_1\!\cdot\!
\pi^{(k)}\!\cdot\! n_2)\label{5dscalarUV-IR}
\ee
The term ${\cal O}(m)$ is conserved but divergent in the IR
limit. Therefore,
according to our recipe, it must be subtracted and will appear with opposite
sign in the UV list, where
the corresponding term is missing. This term yields the Maxwell (or linearized YM) action and EOM, with a coupling
${ \sim {m}}$.

\subsubsection{5d msm: spin 2 current}

For the full 2-pt function of the e.m. tensor is much too cumbersome see sec.\ref{sec:spins}.
As expected it is not conserved, as will be clear from the expansion in powers
of $m$, but the terms responsible
for the non-conservation are local. In the IR we have
\be
{\cal O}(m^5)&:&-\frac{i m^5}{15\pi ^2}\left(2  \left(n_1\!\cdot\!
n_2\right){}^2+ \left(n_1\!\cdot\!n_1\right) 
\left(n_2\!\cdot\!n_2\right)\right)\label{5dmetricJIRm5}\\
{\cal O}(m^3)&:& -\frac{i m^3} {36 \pi ^2}\left(\left(n_2\!\cdot\!n_2\right)
\left(k\!\cdot\!n_1\right){}^2+4\left(n_1\!\cdot\! n_2\right)
\left(k\!\cdot\!n_2\right) \left(k\!\cdot\!n_1\right)\right.\0\\
&&\left. -2 k^2 \left(n_1\!\cdot\! n_2\right){}^2- k^2
\left(n_1\!\cdot\!n_1\right) \left(n_2\!\cdot\!n_2\right)+
\left(n_1\!\cdot\!n_1\right)
\left(k\!\cdot\!n_2\right){}^2\right)\label{5dmetricJIRm3}\\
{\cal O}(m)&:&-\frac{im k^4}{120 \pi ^2}  \,\Pi_{\frac 12}^{(2)}(k,n_1,n_2)\label{5dmetricJIRm}\\
{\cal O}(m^{-1})&:& \frac{i k^6}{3360 \pi ^2m} \,\Pi_{\frac 12}^{(2)}(k,n_1,n_2)\label{5dmetricJIRm-1}\\
\ldots &:& \ldots\0
\ee
The terms corresponding to odd powers of $m$ vanish. In the UV we have
\be 
{\cal O}(m^0): && \frac{k^5}{1536 \pi }\,\Pi_{\frac 12}^{(2)}(k,n_1,n_2) \label{5dmetricJUVm0}\\
{\cal O}(m^2):&&-\frac {m^2k^3}{128 \pi } \,\Pi_{\frac 12}^{(2)}(k,n_1,n_2) \label{5dmetricJUVm2}\\
{\cal O}(m^3):&&-\frac{im^3}{12 \pi ^2 k^2} \left(k\!\!\cdot\!\! n_1\right){}^2
\left(k\!\cdot\!n_2\right){}^2  \label{5dmetricJUVm3}\\
{\cal O}(m^4):&&\frac{ m^4k}{32 \pi }\,\Pi_{\frac 12}^{(2)}(k,n_1,n_2) \label{5dmetricJUVm4}\\
{\cal O}(m^5):&&- \frac{im^5}{15 \pi ^2 k^4} \left( k^2 \left(n_1\!\cdot\!
n_1\right) \left(k\!\cdot\! n_2\right){}^2+4 k^2\left(n_1\!\cdot\! n_2\right)
\left(k\!\cdot\! n_1\right) \left(k\!\cdot\! n_2\right)\right.\0\\
&&\left.+ k^2 \left(n_2\!\cdot\! n_2\right) \left(k\!\cdot\! n_1\right){}^2-3
k^4 \left(k\!\cdot\! n_1\right){}^2 \left(k\!\cdot\! n_2\right){}^2\right)
\label{5dmetricJUVm5}\\
{\cal O}(m^6):&& - \frac {m^6}{24 \pi  k}  \,\Pi_{\frac 12}^{(2)}(k,n_1,n_2)   \label{5dmetricJUVm6}\\
....&&....\0
\ee
The terms ${\cal O}(m)$ and ${\cal O}(m^{2k})$ with $k\geq 4$ vanish. All the
nonvanishing terms are conserved except those of order 3 and 5. But the
non-conserved terms in the IR are local and
\be
{\cal O}_{UV}(m^5)-{\cal O}_{IR}(m^5)&=& \frac{2i m^5} {15 \pi ^2}  \,\Pi_{\frac 12}^{(2)}(k,n_1,n_2) \label{5dmetricUV-IRm5}\\
{\cal O}_{UV}(m^3)-{\cal O}_{IR}(m^3)&=& {-}\frac{i m^3k^2}{18 \pi
^2}  \,\Pi_{\frac 12}^{(2)}(k,n_1,n_2) \label{5dmetricUV-IRm3}\\....&&....\0
\ee
These are conserved. Eq.(\ref{5dmetricUV-IRm3}) gives rise to a (nonlocal)
version of the linearized EH action. Also in this case the term ${\cal
O}_{IR}(m)$ must be subtracted, 
although conserved, because it is divergent in the IR; as a consequence 
it will appear with opposite sign in the UV list, where
the corresponding term is missing. 

\subsubsection{5d msm: spin 3 current}

Once again the 2pt correlators of spin 3 currents can be calculated exactly,see sec.\ref{sec:spins},
but we will skip it here 
and go to the IR expansion. The terms ${\cal O}(m^7), {\cal O}(m^5),{\cal
O}(m^3)$ are not conserved, 
but local, while 
\be 
{\cal O}(m):&&\frac{i m k^6} {280 \pi ^2} \,\Pi_{\frac 32}^{(3)}(k,n_1,n_2) 
\label{5dspin3IRm}\\
{\cal O}(m^{-1}):&& {-}\!\!\-\frac{i k^8}{10080 \pi ^2m}  \,\Pi_{\frac 32}^{(3)}(k,n_1,n_2) \label{5dspin3IRm-1}\\
....&&....\0
\ee
Moreover ${\cal O}_{IR}(m^{n})=0$ for $n$ even.

Near the UV the nonvanishing terms are:
\be 
{\cal O}(m^0):&&\!\! -\frac{k^7}{4096\pi }  \,\Pi_{\frac 32}^{(3)}(k,n_1,n_2) \label{5dbspin3uvm0}\\
{\cal O}(m^2):&&\!\! \frac{m^2 k^5}{256 \pi }  \,\Pi_{\frac 32}^{(3)}(k,n_1,n_2) \label{5dbspin3uvm2}\\
{\cal O}(m^4):&&\!\!-\frac{3 m^4k^3}{128 \pi }  \,\Pi_{\frac 32}^{(3)}(k,n_1,n_2) \label{5dbspin3uvm4}\\
{\cal O}(m^6):&&\!\!\frac{k m^6}{16 \pi } \,\Pi_{\frac 32}^{(3)}(k,n_1,n_2) \label{5dbspin3uvm6}\\
{\cal O}(m^8):&&\!\!-\frac {m^8}{16 \pi  k}  \,\Pi_{\frac 32}^{(3)}(k,n_1,n_2) \label{5dbspin3uvm8}\\
\ee
while ${\cal O}(m^{2n})=0$ for $n\geq 5$. As for the odd $m$ power terms they
are conserved for $n\geq 9$:
\be
{\cal O}(m^9):&&\!\! -\frac{32 im^9}{315 \pi ^2 k^2}  \,\Pi_{\frac 32}^{(3)}(k,n_1,n_2) \label{5dbspin3UVm9}\\
....&&....\0
\ee
while ${\cal O}(m)=0$ and ${\cal O}(m^3),{\cal O}(m^5),{\cal O}(m^7)$ are
non-local and non-conserved. But once again
\be
{\cal O}_{UV}(m^7)\!&-&\!{\cal O}_{IR}(m^7)= \frac{8 im^7}{35 \pi ^2} \,\Pi_{\frac 32}^{(3)}(k,n_1,n_2)\label{5dbspin3UV-IRm7}\\
{\cal O}_{UV}(m^5)\!&-&\!{\cal O}_{IR}(m^5)= {-}\frac{2 im^5k^2}{15 \pi ^2}\,\Pi_{\frac 32}^{(3)}(k,n_1,n_2) \label{5dbspin3UV-IRm5}\\
{\cal O}_{UV}(m^3)\!&-&\!{\cal O}_{IR}(m^3)= \frac{ im^3k^4}{30 \pi ^2}\,\Pi_{\frac 32}^{(3)}(k,n_1,n_2) \label{5dbspin3UV-IRm3})
\ee
The term (\ref{5dbspin3UV-IRm5}) corresponds to the spin 3 Fronsdal EOM.
As we see from these examples the scheme for 5d is similar to 3d. Once again the
term ${\cal O}_{IR}(m)$ must be subtracted, 
although conserved, because it is divergent in the IR; as a consequence 
it will appear with opposite sign in the UV list, where
the corresponding term is missing. 

\subsection{5d fermion model}

\subsubsection{5d mfm: spin 1 current}

The analog of eq.(\ref{spin1fermionEA}) (for the even part) is 
\be(n_1\!\cdot\! \tilde J(k) \!\cdot\! n_2)&=& -\frac{i}{128 \pi ^2 k  }
\left(-16  m^4 \coth ^{-1}\left(\frac{2 m}{k}\right)+
8 k   m^3-8 k^2  m^2 \coth ^{-1}\left(\frac{2 m}{k}\right)\right. \0\\
&&-6 k^3  m\left.
+3 k^4  \coth ^{-1}\left(\frac{2 m}{k}\right)\right)( n_1\!\cdot\! \pi
^{(k)}\!\cdot\! n_2)  \label{JJ5dfernmion}
\ee
which is conserved. 

Expanding in powers of $m$
like in 3d, all coefficients have of course the same conserved structure. In the
IR all even $m$-power coefficient vanish 
and, for instance, 
\be
{\cal O}(m)&:&\frac{im}{12 \pi ^2}  k^2 (n_1\!\cdot\! \pi ^{(k)}\!\cdot\!
n_2)\label{5dfscalarJIRm}
\ee
which (with reversed sign) corresponds to the Maxwell action. In the UV we have instead,  
\be 
{\cal O}(m^0): &&- \frac 3{256 \pi}    k^3 (n_1\!\cdot\! \pi ^{(k)}\!\cdot\!
n_2)\label{5dfJUVm0}\\
{\cal O}(m^2):&& m^2 \frac{1} {32 \pi}k (n_1\!\cdot\! \pi ^{(k)}\!\cdot\! n_2
)\label{5dfJUVm2}\\
{\cal O}(m^4):&& m^4 \frac{1} {16 \pi k} (n_1\!\cdot\! \pi ^{(k)}\!\cdot\! n_2
)\label{5dfJUVm4}\\
{\cal O}(m^5):&& m^5 \frac{4i} {15 \pi^{ 2} k^2} (n_1\!\cdot\! \pi ^{(k)}\!\cdot\!
n_2
)\label{5dfJUVm5}\\
{\cal O}(m^7):&& m^7 \frac{32i} {105 \pi^{ 2} k^{ 4}} (n_1\!\cdot\! \pi ^{(k)}\!\cdot\!
n_2
)\label{5dfJUVm7}\\
....&&....\0
\ee
while ${\cal O}(m)={\cal O}(m^3)={\cal O}(m^n)=0$ for even $n\geq 6$. According
to our recipe the term ${\cal O}(m)$
must be subtracted and will appear in the UV list with opposite sign

\subsubsection{5d mfm: e.m. tensor} 

In this subsection every result must be multiplied by a factor of $\frac 1{16}$.

In the IR the even $m$-power coefficients vanish. The nonvanishing ones are
\be
{\cal O}(m^5):&&\frac{2 im^5}{15 \pi ^2} \left(\left(n_1\!\cdot\!
n_2\right){}^2+(n_1\!\cdot\!n_1) (n_2\!\cdot\!n_2)\right) \label{5dfmetricm5}\\
{\cal O}(m^3):&&\frac{ik^2m^3} {18 \pi ^2} \Pi^{(2)}_{-1}(k,n_1,n_2)
\label{5dfmetricm3}\\
{\cal O}(m):&&-\frac{i k^4m}{40 \pi ^2}  \Pi^{(2)}_{-\frac 13}(k,n_1,n_2)
\label{5dfmetricm}\\
{\cal O}(m^{-1}):&&\frac{i k^6}{672 \pi ^2m} \Pi^{(2)}_{-\frac 15}(k,n_1,n_2) \label{5dfmetricm-1}\\
....&&....\0
\ee
Except ${\cal O}(m^5)$ they are all conserved. 
In the UV we find  $ {\cal O}(m)={\cal O}(m^3)={\cal O}(m^{2n})=0$, for even $n\geq
4$, and
\be 
{\cal O}(m^0): &&  
 \frac {k^5}{384 \pi}  \,\Pi_{-\frac 14}^{(2)}(k,n_1,n_2)\label{5dfmetricUVm0}\\
{\cal O}(m^2):&& -m^2 \frac {k^3}{64 \pi}  \,\Pi_{-\frac 12}^{(2)}(k,n_1,n_2)\label{5dfmetricUVm2}\\
{\cal O}(m^4):&& -m^4  \frac {k}{32 \pi} ( n_1\!\cdot\! \pi^{(k)}\!\cdot\! n_1)(
n_2\!\cdot\! \pi^{(k)}\!\cdot\! n_2) 
\label{5dfmetricUVm4}\\  
{\cal O}(m^6):&& \frac {m^6}{12 \pi k}  \,\Pi_{\frac 12}^{(2)}(k,n_1,n_2)  \label{5dfmetricUVm6}\\
{\cal O}(m^7):&& \frac {8i m^7 }{35 \pi k^2} \,\Pi_{\frac 13}^{(2)}(k,n_1,n_2)
\label{5dfmetricUVm7}\\
....&&....\0
\ee
${\cal O}(m^5)$ is nonlocal and non-conserved, but
\be
{\cal O}_{UV}(m^5)-{\cal O}_{IR}(m^5)&=&-\frac {2im^5}{15\pi^2}\,\Pi_{1}^{(2)}(k,n_1,n_2)
\label{5dfmetricUV-IRm5}
\ee
The remaining terms are conserved. In particular ${\cal O}_{UV}(m^2)$
corresponds to the linearized EH action.
The terms ${\cal O}_{IR}(m),{\cal O}_{IR}(m^3)$ are conserved but divergent in
the IR limit. So they must be
subtracted and will appear in the UV list with opposite sign.

\subsubsection{5d mfm: spin 3 current}

We give a brief account because this case varies with respect to the scalar
model only in one respect: the various conserved terms 
in the $m$ expansion do not have always the same conserved structure like in the
latter case.
In the IR the even power terms vanish, while the odd power terms ${\cal O}(m^n)$
are nonvanishing for $n\leq 7$.
Moreover ${\cal O}(m^7),{\cal O}(m^5) $ are not conserved, while all the others
are. For instance
\be 
{\cal O}(m^3):&& {-}\frac{8 im^3 k^4}{405 \pi ^2} \,\Pi_{-1}^{(3)}(k,n_1,n_2)\label{5dfspin3IRm3}\\
{\cal O}(m):&& \frac{4imk^6}{945 \pi ^2} \,\Pi_{-\frac {37}{64}}^{(3)}(k,n_1,n_2) \label{5dfspin3IRm}\\
....&&....\0
\ee
In the UV ${\cal O}(m)={\cal O}(m^3)={\cal O}(m^n)=0 $ for even $n\geq 10$,
while 
${\cal O}(m^0),{\cal O}(m^2), {\cal O}(m^4)$,  ${\cal O}(m^6), {\cal O}(m^8)$
are nonvanishing and conserved. For instance
\be 
{\cal O}(m^8):&& \frac{m^8}{ 18 \pi k}\,\Pi_{\frac 32}^{(3)}(k,n_1,n_2) \label{5dfspin3UVm8}
\ee
The odd powers ${\cal O}(m^n)$ are nonvanishing for $n>0$ and conserved except
for $n=5,7$. But again 
{\small
\be
{\cal O}_{UV}(m^5)-{\cal O}_{IR}(m^5)&=& \frac {im^5 k^2}{15\pi^2} \left(
(n_1\!\cdot\! \pi ^{(k)}\!\cdot\! n_1)( n_1\!\cdot\! \pi ^{(k)}\!\cdot\! n_2)(
n_2\!\cdot\! 
\pi ^{(k)}\!\cdot\! n_2) \right)\label{5dfspin3UV-IRm5}\\
{\cal O}_{UV}(m^7)-{\cal O}_{IR}(m^7)&=& {-}\frac{128im^7}{945 \pi ^2}\,\Pi_{\frac {23}{16}}^{(3)}(k,n_1,n_2)  \label{5dfspin3UV-IRm7}
\ee}
It is curious that the fermionic model in 5d does not reproduce exactly the spin
3 Fronsdal operator. 
In fact the term (\ref{5dfspin3UV-IRm5}) has the right form but lacks the
essential
$k^2 \left(n_1\!\cdot\!  \pi ^{(k)}\!\cdot\! n_2\right){}^3$ part. This has to
be considered a combinatorial
coincidence. The terms ${\cal O}_{IR}(m),{\cal O}_{IR}(m^3)$ are conserved but
divergent in the IR limit. So they must be
subtracted and will appear in the UV list with opposite sign.

\vskip 1cm
{\bf Comment}. The structure of the 2pt functions in 5d essentially repeats the
scheme of 3d. The $m$-power expansions  
both in the IR and in the UV are similar: in the IR there are non conserved {\it
local} terms, while in the UV there
are non-conserved {\it nonlocal} terms. Subtracting the former from the latter
one obtains 
conserved structures (and a finite IR limit).
All the other terms are conserved and have analogous types of structures in both
the fermionic and the scalar model.

\section{Tomography in 4d} 
\label{sec:4d}

Even dimensional models present an additional problem concerning their
regularization. 
For odd $d$ works by itself as a complete regulator in carrying out the
integrals generated by the Feynman diagrams.  
This is not anymore true for even $d$. The way out is well-known, we will set
$d=4+\varepsilon$. Another difference we 
will come across with, which is related to this, is the appearance of $log$
terms in the form factors. 
We will again expand the two-point functions in powers of $m$ near the IR and UV
limits. 

In almost all the two-point correlators and, therefore, 
in all the one-loop effective actions, we will find non-conserved terms and
terms that diverge
in the IR $m\to\infty$, like in the odd dimensional case, but we will find also
$\varepsilon$-divergent terms.
Our general attitude is to recover both conservation and finiteness in the IR.
This
is possible because all the nonconserved and all divergent terms in the IR, as
well as all $\varepsilon$-divergent terms,
are local. We will therefore subtract all the terms that diverge
in the IR and in $\varepsilon$. They include, in particular, all the
nonconserved ones. 

There remains however an ambiguity. Beside divergent and/or nonconserved terms, in the case
of $m^0$ we meet also finite contributions, both in the IR and in the UV. Also for these
terms we subtract the IR from the UV contribution, on the assumption that it is this difference
that contains the physical information.

\subsection{4d Scalar Model}

The basic formulas are again
(\ref{scalarlag},\ref{scalarcurrent},\ref{scalarcurrents},\ref{Vsst})
and (\ref{2ptscalar}) together with the analogous ones for higher spins, with
$d=4+\varepsilon$.

\subsubsection{4d msm: spin 1 current}

The full formula for the 2pt correlator is expressed in terms of hypergeometric
functions and  parameter derivatives thereof,
and we dispense with writing it down explicitly here, see however sec.\ref{sec:spins}. We will focus on the
power of $m$ expansions. As just mentioned, we have 
to consider also $\log (m)$ and $ \frac 1\varepsilon$ factors. In the IR we find
\be
{\cal O}(m^2):&&-\frac{im^2}{8 \pi ^2   }   
\Big{(}  \gamma -1-\log (4 \pi )+2    \log (m)+\frac 2{\varepsilon}\Big{)} 
\left(n_1\!\cdot\!n_2\right) \label{4dbscalarIRm2}\\
{\cal O}(m):&& 0\0\\
{\cal O}(\log(m)):&&\frac{i \log(m)}{24 \pi ^2}  k^2 (n_1\!\cdot\! \pi
^{k}\!\cdot\! n_2) \label{4dbscalarIRlogm}\\
{\cal O}(m^0):&& \frac{i k^2}{48 \pi ^2   }  \Big{(}\gamma   -\log (4 \pi
)+\frac 2{\varepsilon}\Big{)} 
(n_1\!\cdot\! \pi ^{(k)}\!\cdot\! n_2) \label{4dbscalarIRm0}\\
{\cal O}(m^{-1}):&& 0\0\\
{\cal O}(m^{-2}):&&-\frac{i k^4}{480 \pi ^2m^2}  (n_1\!\cdot\! \pi
^{(k)}\!\cdot\! n_2) \label{4dbscalarIRm-2}\\
....&&....\0
\ee
These coefficients are conserved except ${\cal O}(m^2)$. All the odd powers of
$m$ vanish.

In the UV we find:
\be
{\cal O}(m^0):&&-i\frac{k^2} {144 \pi ^2 }
 \Big{(}8    -3  \gamma    
-   \log \left(\frac{1}{64 \pi ^3}\Big{)}
-3  \log \left(-k^2\right)-\frac 6 \varepsilon   \right) (n_1\!\cdot\! \pi
^{(k)}\!\cdot\! n_2)\label{4dbscalarUVm0}\\
{\cal O}(m):&&0\\
{\cal O}(m^2):&&-\frac{im^2}{24 \pi ^2   k^2}\Big{(}  \left(-3 \log
\left(-	\frac{k^2}{m^2} \right)  
+3  \right) (k\!\cdot\! n_1)( k\!\cdot\! n_2)\0\\
&&  +k^2 (n_1\!\cdot\! n_2) \left(3    
(-2+\gamma  -\log (4 \pi ))+3  \log \left(-k^2\right)+\frac 6 \varepsilon  
\right)\Big{)}\label{4dbscalarUVm2}\\
{\cal O}(m^3):&&0\\
{\cal O}(m^4):&&-i\frac{m^4}{16 \pi ^2 k^2}\Big{(}-2 \log \left(-	\frac{k^2}{m^2}\right) -3\Big{)} 
(n_1\!\cdot\! \pi ^{(k)}\!\cdot\! n_2) \label{4dbscalarUVm4}\\
....&&....\0
\ee
All odd powers of $m$ vanish. The even powers are conserved except
(\ref{4dbscalarUVm2}). Subtracting from the
latter the analogous (local) non-conserved term in the IR we find a conserved
term
\be
{\cal O}_{UV}(m^2)-{\cal O}_{IR}(m^2)&=& -\frac{im^2}{8 \pi ^2}\Big{(}2  \log
\left(-\frac {k^2}{m^2}\right) -1 \Big{)} 
(n_1\!\cdot\! \pi ^{(k)}\!\cdot\! n_2)\label{4dbscalarUV-IRm2}
\ee
The ${\cal O}(\log(m))$ term is divergent in the IR, and the ${\cal O}(m^0)$ is
divergent in the $\varepsilon\to 0$ limit.
Luckily they are local and can be subtracted with the following result:
\be
{\cal O}_{UV}(m^0)-{\cal O}_{IR}(m^0)-{\cal
O}_{IR}(\log(m))&=&-  \frac{i k^2}{144 \pi ^2} \left(-3 \log \left(- \frac {k^2}{m^2}\right)
+8 \right) (n_1\!\cdot\! \pi ^{(k)}\!\cdot\! n_2)\0\\
&&\label{4dbscalarUV-IRm0}
\ee
This term corresponds to the linearized Maxwell action with an energy dependent
coupling.

\subsubsection{4d msm: spin 2 current}

In the IR the odd powers of $m$ vanish. The nonvanishing even powers are
\be
{\cal O}(m^4):&&-\frac{im^4}{16 \pi ^2   }
 \left(2 \left(n_1\!\cdot\! n_2\right){}^2+(n_1\!\cdot\! n_1) (n_2\!\cdot\!
n_2)\right) ( 2 \gamma -3 -2 \log (4\pi )+4  \log (m)+\frac 4{\varepsilon})\0\\
\label{4dbTIRm4}\\
{\cal O}(m^2):&&-\frac{im^2}{24 \pi ^2  } \Big{(}(n_2\!\cdot\! n_2)
\left(k\!\cdot\! n_1\right){}^2+4 (n_1\!\cdot\! n_2)( k\!\cdot\! n_2)(
k\!\cdot\! n_1)\label{4dbTIRm2}\\
&&\quad\quad -k^2 \left(2 \left(n_1\!\cdot\! n_2\right){}^2  {+}(n_1\!\cdot\! n_1)
(n_2\!\cdot\! n_2)\right)
+(n_1\!\cdot\! n_1) \left(k\!\cdot\! n_2\right){}^2\Big{)} \0\\
&&\quad\quad\quad\quad\cdot ( \gamma -1-\log (4 \pi )+2    \log (m)+\frac
2{\varepsilon})\0\\
{\cal O}(\log(m)):&&-\frac{i\log(m)}{60 \pi ^2} k^4 \Pi^{(2)}_{\frac 12}(k,n_1,n_2)\label{4dbTIRlogm}\\
{\cal O}(m^0):&& {-}\frac{ik^4  }{ {120} \pi ^2}\Pi^{(2)}_{\frac 12}(k,n_1,n_2) {(  \gamma - \log (4\pi )+\frac 2{\varepsilon})}  \label{4dbTIRm0}\\
{\cal O}(m^{-2}):&&\frac{ik^6}{1680 m^2\pi ^2}\Pi^{(2)}_{\frac 12}(k,n_1,n_2)  \label{4dbTIRm-2}\\
....&&....\0
\ee
The first two terms are not conserved, the logarithmic term is conserved but
divergent in the IR, the $m^0$ term is divergent in the limit $\varepsilon\to
0$. They all must be subtracted. The remaining terms are conserved. 

In the UV all the odd powers of $m$ vanish. The nonvanishing even powers are
\be
{\cal O}(m^0): && \frac{ik^4}{ 1800 \pi ^2   } 
 \left(46   -15  \gamma    +15    \log (4 \pi )-15 \log
\left(-k^2\right)-\frac {30}{\varepsilon}  \right)\label{4dbTUVm0}\\ 
&&\cdot\, \Pi^{(2)}_{\frac 12}(k,n_1,n_2)  \0\\
{\cal O}(m^2): && -\frac{im^2}{72 \pi ^2}\Bigg{(}\left(\frac 6{\varepsilon}   -8+3
\gamma +\log \left(\frac{1}{64 \pi ^3}\right)+3   \log \left(-k^2\right)
\right)\label{4dbTUVm2}\\
&&\Big{(}(n_2\!\cdot\! n_2) \left(k\!\cdot\! n_1\right){}^2+4 (n_1\!\cdot\! n_2)
(k\!\cdot\! n_2)( k\!\cdot\! n_1)+(n_1\!\cdot\! n_1) \left(k\!\cdot\!
n_2\right){}^2\0\\
&& -k^2 \left(2 \left(n_1\!\cdot\! n_2\right){}^2+(n_1\!\cdot\! n_1)(
n_2\!\cdot\! n_2)\right)\Big{)}\0\\
&&-\frac{3}{k^2} \left(-3  \log \left(-k^2\right)+6  \log (m)+5 \right)
\left(k\!\cdot\! n_1\right){}^2 \left(k\!\cdot\! n_2\right){}^2\Bigg{)}\0
\ee
and 
\be
&&{\cal O}(m^4):\quad -\frac{im^4}{16 \pi ^2   k^4} \Bigg{(}  k^2 \left(2 \log
\left(-	\frac{k^2}{m^2}\right)-1
 \right)\label{4dbTUVm4}\\
&&\cdot \Big{(}(n_2\!\cdot\! n_2) \left(k\!\cdot\! n_1\right){}^2
+4 (n_1\!\cdot\! n_2) (k\!\cdot\! n_2) (k\!\cdot\! n_1)+(n_1\!\cdot\! n_1)
\left(k\!\cdot\! n_2\right){}^2 
-3     \left(k\!\cdot\! n_1\right){}^2 \left(k\!\cdot\!
n_2\right){}^2\Big{)}\0\\
&&+2 k^4 \Big{(}2 \left(n_1\!\cdot\! n_2\right){}^2
+(n_1\!\cdot\! n_1)( n_2\!\cdot\! n_2)\Big{)} \left( 
-2+\gamma  -\log (4 \pi )-   \log \left(-k^2\right)+\frac
2{\varepsilon}\right) \Bigg{)}\0\\
&&{\cal O}(m^6):\frac{im^6}{36 \pi ^2 k^2}\left(6  \log\left(-	\frac{k^2}{m^2}\right)
+11 \right)\, \Pi^{(2)}_{\frac 12}(k,n_1,n_2)  \label{4dbTUVm6} \\
&&\ldots \quad\quad: \quad\quad\dots\0
\ee
${\cal O}(m^0)$ and all terms with even $m$ power larger than 4 are conserved,
while ${\cal O}(m^2)$ and ${\cal O}(m^4)$
are not. According to our prescription we have to subtract not only ${\cal
O}_{IR}(m^2)$ and ${\cal O}_{IR}(m^4)$, but also
${\cal O}_{IR}(m^0)$ and ${\cal O}_{IR}(\log(m))$. We obtain
\be 
{\cal O}_{UV}(m^4)-{\cal O}_{IR}(m^4)&=& -\frac {im^4}{8\pi^2} \left( {+}2  \log
\left(-	\frac{k^2}{m^2} \right) -    {1}\right)\, \Pi^{(2)}_{\frac 12}(k,n_1,n_2)  \label{4dbTUV-IRm4}\\ 
{\cal O}_{UV}(m^2)-{\cal O}_{IR}(m^2)&=& \frac {im^2}{36\pi^2}k^2 \left(3  \log
\left(-\frac {k^2}{m^2} \right) {-5}  \right) \,\Pi^{(2)}_{\frac 12}(k,n_1,n_2)  \label{4dbTUV-IRm2} 
\ee
and
\be
{\cal O}_{UV}(m^0)-{\cal O}_{IR}(m^0)\!&-&\!{\cal O}_{IR}(\log(m))=\frac
{i}{1800\pi^2}k^4(-15  \log \left(-\frac {k^2}{m^2} \right)+46 )\0\\
&&\cdot\,  \Pi^{(2)}_{\frac 12}(k,n_1,n_2) \label{4dbTUV-IRm0}
\ee
They are all conserved. (\ref{4dbTUV-IRm2}) contains a nonlocal linearized
version of the EH eom.

\subsubsection{4d msm: spin 3 current}

The scheme is the same as above. In the IR the odd power of $m$ vanish. The even
powers $m^{2n}$ with $n\leq 0$ are conserved
together with the term proportional to $\log(m)$. The terms ${\cal
O}_{IR}(m^2),{\cal O}_{IR}(m^6),{\cal O}_{IR}(m^6)$ are not conserved.
Of course ${\cal O}(\log[m])$ diverges in the IR, while the term  ${\cal
O}_{IR}(m^0)$ diverges for $\varepsilon \to 0$.
According to our prescription all these terms, which are local, have to be
subtracted from the effective action. 
The result is as follows:
\be
{\cal O}(m^{-2}):&& -\frac {ik^8}{5040 m^2\pi
^2}\,\Pi_{\frac 32}^{(3)}(k,n_1,n_2)\label{4dbspin3IRm-2}\\
{\cal O}(m^{-4}):&& -\frac {ik^{10}}{110880 m^4\pi ^2}
\,\Pi_{\frac 32}^{(3)}(k,n_1,n_2)\label{4dbspin3IRm-4}\\
....&&....\0
\ee
In the UV the odd $m$ power terms vanish. The even power of order $2,4,6$ are
not conserved, but
\be 
{\cal O}_{UV}(m^0)-{\cal O}_{IR}(m^0) \!&-&\! {\cal
O}_{IR}(\log(m))\label{4dbspin3UV-IRm0}\\
&=&-\frac {ik^6}{29400 \pi ^2} \,\Pi_{\frac 32}^{(3)}(k,n_1,n_2)
\left(-105 \log \left(-\frac {k^2}{m^2}\right) +352\right )\0
\ee
and
\be
{\cal O}_{UV}(m^2)-{\cal O}_{IR}(m^2)&=&\frac {im^2k^4}{300 \pi ^2}
\,\Pi_{\frac 32}^{(3)}(k,n_1,n_2) \left(
( 31  -15 \log \left(-\frac {k^2}{m^2}\right)\right)\label{4dbspin3UV-IRm2}\\
{\cal O}_{UV}(m^4)-{\cal O}_{IR}(m^4)&=&- \frac{i m^4 k^2}{24 \pi ^2}
\,\Pi_{\frac 32}^{(3)}(k,n_1,n_2)\left( 
( 7 -6\log \left(\frac {k^2}{m^2} \right)\right)\label{4dbspin3UV-IRm4}\\
{\cal O}_{UV}(m^6)-{\cal O}_{IR}(m^6)&=&\frac{i m^6}{12 \pi ^2}
\,\Pi_{\frac 32}^{(3)}(k,n_1,n_2) \left(
( 1 -6\log \left(-\frac {k^2}{m^2}\right)\right)\label{4dbspin3UV-IRm6}\\
{\cal O}_{UV}(m^8)&=& \frac{i m^8}{48k^2 \pi ^2} \,\Pi_{\frac 32}^{(3)}(k,n_1,n_2) \left(
( 25 -12\log \left(-\frac {k^2}{m^2}\right)\right)\label{4dbspin3UVm8}\\
....&&....\0
\ee
are all conserved. Eq.(\ref{4dbspin3UV-IRm4}) is related to a nonlocal version
of the spin 3 Fronsdal equation.

\subsection{4d Fermion Model}

We consider now the same analysis for the fermion massive model. We start again
from
eqs.(\ref{actionA},\ref{actiong},\ref{Jmunuab},\ref{Tmnlr}) and the like for
higher spins.

\subsubsection{4d mfm: Spin 1 current}  

The full formula for the 2pt correlator is similar to the scalar case and
expressed in terms of parameter derivatives of 
hypergeometric functions, see sec.\ref{sec:spins}. A full expression in terms of simple functions can be found in Appendix \ref{sec:4dfullamplitudes}. The $m$-power
expansion in the IR is as follows
\be
{\cal O}(m^2):&&\frac{im^2}{4 \pi ^2 \varepsilon }     
\left(n_1\!\cdot\!n_2\right) \label{4dfermionIRm2}\\
{\cal O}(\log(m)):&&\frac{i \log(m)}{6 \pi ^2}  k^2 (n_1\!\cdot\! \pi
^{k}\!\cdot\! n_2) \label{4dfermionIRlogm}\\
{\cal O}(m^0):&&  -\frac{i}{24 \pi ^2  } \Big{(}k^2 (  -2 \gamma +1
+\log (16\pi^2)-\frac 4\varepsilon)\left(n_1\!\cdot\! n_2\right) \0\\
&&+2 (\gamma   -\log (4 \pi )
+\frac
2\varepsilon)\left(k,n_1\right)\left(k,n_2\right)\Big{)}\label{4dfermionIRm0}\\
{\cal O}(m^{-2}):&&-\frac{i k^4}{60 \pi ^2m^2}  (n_1\!\cdot\! \pi
^{(k)}\!\cdot\! n_2) \label{4dfermionIRm-2}\\
....&&....\0
\ee
All odd powers of $m$ vanish. The above terms are all conserved except
(\ref{4dfermionIRm2}) and (\ref{4dfermionIRm0}).
${\cal O}(m^2)$ and ${\cal O}(\log(m))$ are divergent in the IR and ${\cal
O}(m^0)$ is divergent in $\varepsilon$.

In the UV all odd powers of $m$ vanish, while
\be
{\cal O}(m^0):&&\frac{i} {24 \pi ^2  }
 \Bigg{(}  \left(\frac 4\varepsilon   -\frac {13} 3 +2 \gamma   +2i \pi   
-\log \left( 16 \pi ^2 \right)  \right) k^2 (n_1\!\cdot\! n_2) \0\\
&&-\frac 23 \Big{(}\frac 6 \varepsilon -5+3\gamma +3 i\pi - \log\left(
16\pi^2\right)\Big{)}
(n_1\!\cdot\!k)(k\!\cdot\! n_2)\Bigg{)}\label{4dfermionUVm0}\\
 {\cal O}(m^2):&&-\frac{im^2}{4 \pi ^2   k^2}\Big{(} 
-2 (k\!\cdot\! n_1)( k\!\cdot\! n_2)+ k^2 (n_1\!\cdot\!n_2)\Big{)}  
\label{4dfermionUVm2}\\
{\cal O}(m^4):&&-i\frac{m^4}{4 \pi ^2 k^2}\Big{(}2\log \left(-	\frac{k^2}{m^2} \right)+1 \Big{)} 
(n_1\!\cdot\! \pi ^{(k)}\!\cdot\! n_2) \label{4dfermionUVm4}\\
....&&....\0
\ee
All the terms  are conserved except the first two. But, subtracting from them
the corresponding local terms 
in the IR we get {\small
\be
{\cal O}_{UV}(m^0)-{\cal O}_{IR}(m^0)-{\cal O}_{IR}(\log(m))& =&\frac{i}{36 \pi
^2} \left(3 \log \left(-\frac {k^2}{m^2}\right)-5 
 \right)\, k^2\,  (n_1\!\cdot\! \pi ^{(k)}\!\cdot\! n_2)
\label{4dbfermionUV-IRm0} \\
{\cal O}_{UV}(m^2)-{\cal O}_{IR}(m^2)&=& -\frac{im^2}{2\pi ^2} (n_1\!\cdot\! \pi
^{(k)}\!\cdot\! n_2)\label{4dbfermionUV-IRm2}
\ee}
Clearly (\ref{4dbfermionUV-IRm0}) reproduces the Maxwell action.

\subsubsection{4d mfm: e.m. tensor}

A full expression in terms of simple functions can be found in Appendix \ref{sec:4dfullamplitudes}. In this subsection every result must be multiplied by a factor of $\frac 1{16}$.
In the IR the odd powers of $m$ vanish. The nonvanishing even powers are {\small
\be
{\cal O}(m^4):&&\frac{im^4}{8 \pi ^2   }
 \Big{(}\left(n_1\!\cdot\! n_2\right){}^2\left( 2 \gamma -1 -2 \log (4\pi )+4 
\log (m)+\frac 4{\varepsilon}\right)\label{4dfTIRm4}\\
&&+(n_1\!\cdot\! n_1) (n_2\!\cdot\! n_2)  \left( 2 \gamma -3 -2 \log (4\pi )+4 
\log (m)+\frac 4{\varepsilon}\right)\Big{)}	\0\\
{\cal O}(m^2):&&\frac{im^2}{12 \pi ^2  } \Bigg{(}\Big{(}(n_2\!\cdot\! n_2)
\left(k\!\cdot\! n_1\right){}^2
+(n_2\!\cdot\! n_2) ( k\!\cdot\! n_1)^2-k^2   (n_1\!\cdot\! n_1) (n_2\!\cdot\!
n_2)\Big{)} \0\\ 
&&\cdot\left(  \gamma -1-\log (4 \pi )+2   \log (m)+\frac
2{\varepsilon}\right)\label{4dfTIRm2}\\
&&+(n_1\!\cdot\! n_2) \left(k\!\cdot\! n_1\right)\left(k\!\cdot\! n_2\right)
\left(3-2 \gamma +\log (4 \pi )-4   \log (m)-\frac
4{\varepsilon}\right)\Bigg{)}\0\\
{\cal O}(\log(m)):&&-\frac{i\log(m)}{20 \pi ^2} k^4  \Pi^{(2)}_{-\frac 13} (k,n_1,n_2)\label{4dfTIRlogm}
\ee
and
\be
{\cal O}(m^0):&& -\frac{i}{120 \pi ^2} \Bigg{(}k^4 \left( \frac{6}{\varepsilon
}+3 \gamma -1 -3 \log (4\pi ) \right)
 \left(n_1\!\cdot\! n_2\right){}^2\label{4dfTIRm0}\\
&&+k^2 \left(-\frac{12}{\varepsilon }-6 \gamma 
+1+ 6 \log (4\pi )\right) (n_1\!\cdot\! n_2) (k\!\cdot\! n_1)( k\!\cdot\! n_2)
+\left(\frac{2}{\varepsilon }+
\gamma -\log (4 \pi )\right) \0\\
&&\cdot\Big{(}-k^4\left(n_1\!\cdot\! n_1 \right) (n_2\!\cdot\! n_2)
+k^2 (n_1\!\cdot\! n_1) \left(k\!\cdot\! n_2\right){}^2+\left(k\!\cdot\!
n_1\right){}^2 \left(2 \left(k\!\cdot\! n_2\right){}^2
+k^2 (n_2\!\cdot\! n_2)\right)\Big{)}\Bigg{)} \0\\
{\cal O}(m^{-2}):&&\frac{i}{336 m^2\pi ^2} k^6\Pi^{(2)}_{-\frac 15} (k,n_1,n_2) \label{4dfTIRm-2}\\
....&&....\0
\ee}

The first two terms are not conserved, the logarithmic term is conserved but
divergent in the IR, 
the $m^0$ term is not conserved and 
divergent in the limit $\varepsilon\to 0$. They all must be subtracted. The
remaining terms are conserved. 

In the UV all the odd powers of $m$ vanish. The nonvanishing even powers
are{\small
\be
{\cal O}(m^0): && \frac{i}{1800 \pi ^2} \Bigg{(}\left(\frac{30}{\varepsilon }+15
\log \left(-k^2\right)-46+15 \gamma 
 -15 \log (4 \pi )\right)\label{4dfTUVm0}\\ 
&&\cdot \Big{(}k^4 \left((n_1\!\cdot\! n_1) (n_2\!\cdot\! n_2)-3
\left(n_1\!\cdot\! n_2\right){}^2\right)
-k^2 (n_1\!\cdot\! n_1) \left(k\!\cdot\! n_2\right){}^2-k^2 (n_2\!\cdot\! n_2)
\left(k\!\cdot\! n_1\right){}^2\Big{)}\0\\
&&+3 k^2 (n_1\!\cdot\! n_2) (k\!\cdot\! n_1) (k\!\cdot\!
n_2)\left(\frac{60}{\varepsilon }+30\log \left(-k^2\right)-77
+30 \gamma  -30 \log (4 \pi )\right)\0\\
&&-2 \left(k\!\cdot\! n_1\right){}^2 
 \left(k\!\cdot\! n_2\right){}^2\Big{(}\frac{30}{\varepsilon }+15 \log
\left(-k^2\right) 
-31+15 \gamma   -15 \log (4 \pi )\Big{)}
 \Bigg{)}\0\\ 
{\cal O}(m^2): && -\frac{m^2}{36 \pi ^2 k^2}i \Bigg{(}(-6 \left(k\!\cdot\!
n_1\right){}^2 \left(k\!\cdot\! n_2\right){}^2 \label{4dfTUVm2}\\
&&-  k^2 (n_1\!\cdot\! n_2)( k\!\cdot\! n_1) (k\!\cdot\! n_2)\left(
-\frac{12}{\varepsilon }-6 \log \left(-k^2\right)
+7-6 \gamma   +6 \log (4 \pi )\right)\0\\
&& + \Big{(}k^4 \left(-\left(n_1\!\cdot\! n_1\right)\right) (n_2\!\cdot\!
n_2)+k^2 (n_1\!\cdot\! n_1)
 \left(k\!\cdot\! n_2\right){}^2+k^2 (n_2\!\cdot\! n_2) \left(k\!\cdot\!
n_1\right){}^2\Big{)}\0\\
&&\cdot\left( \frac{6}{\varepsilon }+3\log \left(-k^2\right)-8+3 \gamma  
-3 \log (4 \pi )\right)\0\\
&& -  k^4 \left(n_1\!\cdot\! n_2\right){}^2 \left(\frac{6}{\varepsilon }
+3 \log \left(-k^2\right)-3 \log \left(4 \pi  \right)-5+3 \gamma \right)
\Bigg{)}\0
\ee} 
and {\small	
\be
&&{\cal O}(m^4): \frac{im^4}{16 \pi ^2   k^4} \Bigg{(}  k^2 \left(2\log
\left(-\frac {k^2}{m^2}\right)-1
 \right)\label{4dfTUVm4}\\
&&\cdot\Big{(}(n_2\!\cdot\! n_2) \left(k\!\cdot\! n_1\right){}^2
+4 (n_1\!\cdot\! n_2) (k\!\cdot\! n_2) (k\!\cdot\! n_1)+(n_1\!\cdot\! n_1)
\left(k\!\cdot\! n_2\right){}^2 
-3     \left(k\!\cdot\! n_1\right){}^2 \left(k\!\cdot\!
n_2\right){}^2\Big{)}\0\\
&&-2 k^4 \Big{(}2 \left(n_1\!\cdot\! n_2\right){}^2
+(n_1\!\cdot\! n_1)( n_2\!\cdot\! n_2)\Big{)} \left( 
-2+\gamma  -\log (4 \pi )+  \log \left(-\frac {k^2}{m^2}\right)+\frac
2{\varepsilon}\right) \Bigg{)}\0\\
&&{\cal O}(m^6):-\frac{im^6}{12 \pi ^2 k^2}\left(6 \log\left(-\frac {k^2}{m^2} \right)
+7\right) \left(n_1\!\cdot\! \pi ^{(k)}\!\cdot\! n_2\right){}^2
\label{4dfTUVm6}\\
&&+\frac{im^6}{36 \pi ^2 k^2}\left(6  \log\left(-\frac {k^2}{m^2} \right) +11
  \pi \right)   (n_1\!\cdot\! \pi ^{(k)}\!\cdot\! n_1)
( n_2\!\cdot\! \pi ^{(k)}\!\cdot\! n_2) \0\\
&&\ldots \quad\quad: \quad\quad\dots\0
\ee}
All terms with even $m$ power larger than 4, as well as ${\cal O}(log(m))$, are
conserved,
while ${\cal O}(m^0),{\cal O}(m^2)$ and ${\cal O}(m^4)$ are not. According to
our prescription 
we have to subtract not only ${\cal O}_{IR}(m^0),{\cal O}_{IR}(m^2)$ and ${\cal
O}_{IR}(m^4)$, but also
${\cal O}_{IR}(\log(m))$. We obtain  
\be 
{\cal O}_{UV}(m^4)-{\cal O}_{IR}(m^4)&=& \frac {im^4}{8\pi^2} \left( {+}2 
\log\left(-	\frac{k^2}{m^2} \right) -5
  \right) \left(n_1\!\cdot\! \pi ^{(k)}\!\cdot\! n_2\right){}^2
\label{4dfTUV-IRm4}\\ 
&&+\frac {im^4}{8\pi^2}  \left(2 \log\left(-\frac {k^2}{m^2} \right) -1
 \right) \left(    (n_1\!\cdot\! \pi ^{(k)}\!\cdot\! n_1)
( n_2\!\cdot\! \pi ^{(k)}\!\cdot\! n_2) \right)\0\\
{\cal O}_{UV}(m^2)-{\cal O}_{IR}(m^2)&=&  \frac {im^2k^2}{36\pi^2} \left(3 
\log\left( -\frac {k^2}{m^2}\right)  {+}1
 \right) \left(n_1\!\cdot\! \pi ^{(k)}\!\cdot\! n_2\right){}^2
\label{4dfTUV-IRm2}\\ 
&&-\frac {im^2k^2}{36\pi^2}  \left(3  \log\left(-	\frac{k^2}{m^2} \right) -5
  \right)      (n_1\!\cdot\! \pi ^{(k)}\!\cdot\! n_1)
( n_2\!\cdot\! \pi ^{(k)}\!\cdot\! n_2)   \0
\ee
and
\be
{\cal O}_{UV}(m^0)\!&-&\!{\cal O}_{IR}(m^0)-{\cal O}_{IR}(\log(m))=\frac
{i}{1800\pi^2}k^4\label{4dfTUV-IRm0}\\
&&\cdot \Bigg{(}9
\left(-5 \log \left(-\frac {k^2}{m^2} \right)+12  \right)
   \left(n_1\!\cdot\! \pi ^{(k)}\!\cdot\! n_2\right){}^2\0\\
&& - \left(-15  \log \left(-\frac {k^2}{m^2} \right)+46  \right)(n_1\!\cdot\! \pi
^{(k)}\!\cdot\! n_1)
( n_2\!\cdot\! \pi ^{(k)}\!\cdot\! n_2) \Bigg{)}\0
\ee
They are all conserved. (\ref{4dfTUV-IRm2}) contains a nonlocal linearized
version of the EH eom.

\subsubsection{4d mfm: spin 3 current}

A full expression of the correlator in terms of simple functions can be found in Appendix \ref{sec:4dfullamplitudes}.
The scheme is the same as above. In the IR the odd power of $m$ vanish. The even
powers $m^{2n}$ with $n< 0$ are conserved
together with the term proportional to $\log(m)$. The terms 
${\cal O}_{IR}(m^0),{\cal O}_{IR}(m^2),{\cal O}_{IR}(m^6)$ and ${\cal
O}_{IR}(m^6)$ are not conserved.
Of course ${\cal O}(\log[m])$ diverges in the IR, while the term  ${\cal
O}_{IR}(m^0)$ diverges for $\varepsilon \to 0$.
According to our prescription all these terms, which are local, have to be
subtracted from the effective action. 
The result is as follows.  
\be
{\cal O}(m^{-2}):&& -\frac {ik^8}{2835 m^2\pi ^2}\, \Pi^{(3)}_{-\frac 3
{16}}(k,n_1,n_2)\label{4dfspin3IRm-2}\\
{\cal O}(m^{-4}):&& -\frac {2 ik^{10}}{93555 m^4\pi ^2}\, \Pi^{(3)}_{\frac
{49}{128}}(k,n_1,n_2)\label{4dfspin3IRm-4}\\
....&&....\0
\ee
In the UV the odd $m$ power terms vanish. The even power of order $2,4,6$ are
not conserved, but again {\small
\be  
&&{\cal O}_{UV}(m^0)-{\cal O}_{IR}(m^0) -{\cal
O}_{IR}(\log(m))\label{4dfspin3UV-IRm0}\\
&=&-\frac {2ik^6}{99225 \pi ^2}\left(-210  \log \left(-\frac {k^2}{m^2}\right) +599\right) 
\left(n_1\!\cdot\! \pi ^{(k)}\!\cdot\! n_2\right){}^3\0\\
&+&\!\frac {ik^6}{1587600 \pi^2}\left(-3885 \log \left(-\frac {k^2}{m^2}\right) +13339
\right) 
(n_1\cdot \pi ^{(k)}\cdot n_2)( n_1\cdot \pi ^{(k)}\cdot n_1)( n_2\cdot \pi
^{(k)}\cdot n_2)\0
\ee
and 
\be
&&{\cal O}_{UV}(m^2)-{\cal O}_{IR}(m^2)= -\frac {4im^2k^4 }{2025 \pi ^2}\left(15
\log \left(-\frac {k^2}{m^2}\right) -16 \right) 
\left(n_1\!\cdot\! \pi ^{(k)}\!\cdot\! n_2\right){}^3\label{4dfspin3UV-IRm2}\\
&&+\!\frac {im^2k^4}{ 16200\pi^2}\left(480  \log \left(-\frac {k^2}{m^2}\right) - {857}
\right) 
(n_1\cdot \pi ^{(k)}\cdot n_2)( n_1\cdot \pi ^{(k)}\cdot n_1)( n_2\cdot \pi
^{(k)}\cdot n_2)\0  \\
&&{\cal O}_{UV}(m^4)-{\cal O}_{IR}(m^4)= \frac {4im^4k^2 }{27 \pi ^2}  
\left(n_1\!\cdot\! \pi ^{(k)}\!\cdot\! n_2\right){}^3\label{4dfspin3UV-IRm4}\\
&&-\!\frac {im^4k^2}{ 144\pi^2}\left(18 \log \left(-\frac {k^2}{m^2}\right) {-}23 \right) 
(n_1\cdot \pi ^{(k)}\cdot n_2)( n_1\cdot \pi ^{(k)}\cdot n_1)( n_2\cdot \pi
^{(k)}\cdot n_2)\0  \\
&&{\cal O}_{UV}(m^6)-{\cal O}_{IR}(m^6)= \frac {4im^6 }{81\pi ^2}\left(6 
\log \left(-\frac {k^2}{m^2}\right) -7\right) 
\left(n_1\!\cdot\! \pi ^{(k)}\!\cdot\! n_2\right){}^3\label{4dfspin3UV-IRm6}\\
&&+\!\frac {im^6}{ 162\pi^2}\left(69 \log \left(-\frac {k^2}{m^2}\right) -70 \right) 
(n_1\cdot \pi ^{(k)}\cdot n_2)( n_1\cdot \pi ^{(k)}\cdot n_1)( n_2\cdot \pi
^{(k)}\cdot n_2)\0  \\
&&{\cal O}_{UV}(m^8)= -\frac {2im^8 }{81\pi ^2k^2}\left(12 \log \left(-\frac {k^2}{m^2} \right) +19 \right) 
\left(n_1\!\cdot\! \pi ^{(k)}\!\cdot\! n_2\right){}^3\label{4dfspin3UVm8}\\
&&-\!\frac {im^2k^4}{2592\pi^2 k^2}\left( 1356 \log \left( -\frac {k^2}{m^2}\right) +1637
 \right) 
(n_1\cdot \pi ^{(k)}\cdot n_2)( n_1\cdot \pi ^{(k)}\cdot n_1)( n_2\cdot \pi
^{(k)}\cdot n_2)\0  \\
&&....\quad\quad ....\0
\ee}
are all conserved. Eq.(\ref{4dfspin3UV-IRm4}) is related to a nonlocal version
of the spin 3 Fronsdal equation.

\section{Tomography in 6d}
\label{sec:6d}

\subsection{6d Scalar Model}

The basic formulas are again
(\ref{scalarlag},\ref{scalarcurrent},\ref{scalarcurrents},\ref{Vsst}
and (\ref{2ptscalar}) together with the analogous ones for higher spins, with
$d=6+\varepsilon$. For the full two-point correlator formulas see next section. Here we limit ourselves to
IR and UV expansions. 

\subsubsection{6d msm: spin 1 current}

Like in 4d, we have 
to consider also $\log (m)$ and $ \frac 1\varepsilon$ factors. In the IR the nonvanishing terms are
\be
{\cal O}(m^4):&&\frac{im^4}{128 \pi ^3   }   
\Big{(}2  \gamma -3-2\log (4 \pi )+4    \log (m)+\frac 4{\varepsilon}\Big{)} 
\left(n_1\!\cdot\!n_2\right) \label{6dbscalarIRm4}\\
{\cal O}(m^2):&&\frac{im^2k^2}{192 \pi ^3   }   
\Big{(} \gamma -1-\log (4 \pi )+2    \log (m)+\frac 2{\varepsilon}\Big{)} 
 (n_1\!\cdot\! \pi^{k}\!\cdot\! n_2)  \label{6dbscalarIRm2}\\
{\cal O}(\log(m)):&&\frac{i \log(m)}{960 \pi ^3}  k^4 (n_1\!\cdot\! \pi
^{k}\!\cdot\! n_2) \label{6dbscalarIRlogm}\\
{\cal O}(m^0):&&\frac{i k^4}{1920 \pi ^3   }  \Big{(} \gamma   -\log (4 \pi
)+\frac 2{\varepsilon}\Big{)} 
(n_1\!\cdot\! \pi ^{(k)}\!\cdot\! n_2) \label{6dbscalarIRm0}\\
{\cal O}(m^{-2}):&&-\frac{i k^6}{26880 \pi ^3m^2}  (n_1\!\cdot\! \pi
^{(k)}\!\cdot\! n_2) \label{6dbscalarIRm-2}\\
....&&....\0
\ee
These coefficients are conserved except ${\cal O}(m^4)$. All the odd powers of
$m$ vanish.

In the UV we find:
\be
{\cal O}(m^0):&&\frac{ik^4} {28800 \pi ^3 }
 \Big{(}-46    +15 \gamma     
-15   \log \left(4 \pi\right)
+ 15 \log \left(-k^2 \right)+\frac {30} \varepsilon \Big{)} (n_1\!\cdot\! \pi^{(k)}\!\cdot\! n_2)\label{6dbscalarUVm0}\\
 {\cal O}(m^2):&&\frac{im^2k^2}{576 \pi ^3 }\Big{(}  \left(-3\log(-k^2)
  -3\gamma+8+3\log (4 \pi )  -\frac 6 \varepsilon  
\right)\Big{)}(n_1\!\cdot\! \pi^{(k)}\!\cdot\! n_2)\label{6dbscalarUVm2}\\
{\cal O}(m^4):&& \frac{im^4}{ {128} \pi ^3   k^2}\Big{(} \left(-2 \log
\left(-\frac {k^2}{m^2} \right) 
+1 \right) (k\!\cdot\! n_1)( k\!\cdot\! n_2)\0\\
&&  + {2} k^2 (n_1\!\cdot\! n_2) \left(   
-2+\gamma   -\log (4 \pi )+  \log \left(-k^2\right)+\frac 2 \varepsilon  
\right)\Big{)}\label{6dbscalarUVm4}\\
....&&....\0
\ee
All odd powers of $m$ vanish. The even powers are conserved except
(\ref{4dbscalarUVm2}). Subtracting from the
latter the analogous (local) non-conserved term in the IR we find a conserved
term
\be
{\cal O}_{UV}(m^4)-{\cal O}_{IR}(m^4)&=&- \frac{im^4}{128 \pi ^3}\Big{(}-2 \log
\left(-\frac {k^2}{m^2} \right) +1 \Big{)} 
(n_1\!\cdot\! \pi ^{(k)}\!\cdot\! n_2)\label{6dbscalarUV-IRm4}
\ee
The ${\cal O}(m^2),{\cal O}(\log(m))$ terms are divergent in the IR, and the ${\cal O}(m^0)$ is
divergent in the $\varepsilon\to 0$ limit, but they are local and can be subtracted with the following result:{\small
\be
{\cal O}_{UV}(m^2)-{\cal O}_{IR}(m^2)&=&\frac{im^2k^2}{576 \pi ^3 }\Big{(}  \left(-3 \log
\left(-\frac {k^2}{m^2}\right)+5 \right)\Big{)}(n_1\!\cdot\! \pi
^{(k)}\!\cdot\! n_2)\label{6dbscalarUV-IRm2}\\
{\cal O}_{UV}(m^0)-{\cal O}_{IR}(m^0)-{\cal
O}_{IR}(\log(m)) 
&=& - \frac{i k^4}{28800 \pi ^3} \left(-15 \log \left(-\frac {k^2}{m^2} \right)
+46  \right) (n_1\!\cdot\! \pi ^{(k)}\!\cdot\! n_2)\0\\
\label{6dbscalarUV-IRm0}
\ee}
(\ref{6dbscalarUV-IRm2}) corresponds to the linearized Maxwell action with an energy dependent
coupling.

\subsubsection{6d msm: spin 2 current}

In the IR the odd powers of $m$ vanish. The nonvanishing even powers are{\small
\be
{\cal O}(m^6):&&\frac{im^6}{576 \pi ^3   }
 \left(2 \left(n_1\!\cdot\! n_2\right){}^2+(n_1\!\cdot\! n_1) (n_2\!\cdot\!
n_2)\right) ( 6 \gamma -11 -6 \log (4\pi )+12 \log (m)+\frac {12}{\varepsilon})\0\\
\label{6dbTIRm6}\\
{\cal O}(m^4):&&\frac{im^4}{384\pi ^3  } \Big{(}(k\!\cdot\! n_2){}^2
\left(n_1\!\cdot\! n_1\right)+4 (n_1\!\cdot\! n_2)( k\!\cdot\! n_2)(
k\!\cdot\! n_1)+ \left(n_2,n_2\right) \left(k\!\cdot\!n_1\right){}^2\0\\
&&-k^2 \left(2 \left(n_1\!\cdot\!n_2\right){}^2+\left(n_1\!\cdot\!n_1\right) \left(n_2\!\cdot\!n_2\right)\right)\Big{)} 
 (2 \gamma -3-2\log (4 \pi )+4    \log (m)+\frac
4{\varepsilon})\label{6dbTIRm4}\\
{\cal O}(m^2):&&\frac{im^2k^4}{{480} \pi ^3  }\left(\gamma  -1 -\log (4 \pi )+2 \varepsilon  \log (m)+\frac 2{\varepsilon}\right)
\Pi^{(2)}_{\frac 12} (k,n_1,n_2) \label{6dbTIRm2}\\
{\cal O}(\log(m)):&&-\frac{i\log(m)}{3360 \pi ^3} k^6\Pi^{(2)}_{\frac 12} (k,n_1,n_2)  \label{6dbTIRlogm}\\
{\cal O}(m^0):&&-\frac{i }{6720 \pi ^3} k^6\left(\gamma    -\log (4 \pi )  +\frac 2{\varepsilon}\right)  
\Pi^{(2)}_{\frac 12} (k,n_1,n_2) \label{6dbTIRm0}\\
 ....&&....\0
\ee}
The first two terms are not conserved, the logarithmic term is conserved but
divergent in the IR, the $m^0$ term is divergent in the limit $\varepsilon\to
0$. They all must be subtracted (including the finite ${\cal O}(m^0)$ part). The remaining terms are conserved. 

In the UV all the odd powers of $m$ vanish. The nonvanishing even powers are
\be
{\cal O}(m^0): &&\frac{ik^6}{ 705600 \pi ^3   } 
 \left(352  -105  \gamma    +105    \log (4 \pi )-105  \log
\left(-k^2\right)-\frac { {210}}{\varepsilon}  \right)\label{6dbTUVm0}\\ 
&&\cdot \Pi^{(2)}_{\frac 12} (k,n_1,n_2)  \0\\
{\cal O}(m^2): &&-\frac{im^2k^4}{7200 \pi ^3}\Big{(}-\frac {30}{\varepsilon}  +46-15
\gamma  +15\log \left(4 \pi \right)-15  \log \left(-k^2\right)\Big{)}\label{6dbTUVm2} \\
&&\cdot  \Pi^{(2)}_{\frac 12} (k,n_1,n_2)\0
\ee
and 
\be
&&{\cal O}(m^4):\quad \frac{im^4}{1152 \pi ^3 k^2} \Bigg{(}  2 k^2\left(3\log
\left(-k^2\right)-8+3\gamma -3\log(4\pi)
 +\frac 6{\varepsilon} \right)\Big{(}(n_2\!\cdot\! n_2) \left(k\!\cdot\! n_1\right){}^2\0\\
&&+4 (n_1\!\cdot\! n_2) (k\!\cdot\! n_2) (k\!\cdot\! n_1)+(n_1\!\cdot\! n_1)
\left(k\!\cdot\! n_2\right){}^2 
-k^2\left(2 \left(n_1\!\cdot\! n_2\right){}^2+ \left(n_1\!\cdot\! n_1\right) \left(n_2\!\cdot\! n_2\right)\right)\Big{)}\label{6dbTUVm4}\\
&&+3 \left(-6\log \left(-\frac {k^2}{m^2} \right)+7  \right)\left(k\!\cdot\! n_1\right){}^2 \left(k\!\cdot\!
n_2\right){}^2\Bigg{)}\0\\
&&{\cal O}(m^6):-\frac{im^6}{576 \pi ^3 k^4}\left(6  \log\left(-\frac {k^2}{m^2} \right)
-1 \right)\label{6dbTUVm6}\\
&& \Big{(} \left((n_1\!\cdot\! n_1)\left(k\!\cdot\! n_2\right){}^2 +4 (n_1\!\cdot\! n_2) (k\!\cdot\! n_2) (k\!\cdot\! n_1) 
+ (k\!\cdot\! n_1){}^2\left(n_2\!\cdot\! n_2\right)\right)k^2 -3  \left(k\!\cdot\! n_1\right){}^2 \left(k\!\cdot\! n_2\right){}^2\big{)}\0\\
&& - 6 k^4\bigg{(}-2+\gamma  -\log(4\pi) +\log(-k^2)+\frac 2 {\varepsilon}\bigg{)}\Big{(} 2 \left(n_1\!\cdot\! n_2\right){}^2
+ \left(n_1\!\cdot\! n_1\right) \left(n_2\!\cdot\! n_2\right)\Big{)}\0\\
&&{\cal O}(m^8): \frac{im^8}{1152 \pi ^3k^6}\Bigg{(}-25-12 \log \left(-\frac {k^2}{m^2} \right)\Bigg{)} \Pi^{(2)}_{\frac 12} (k,n_1,n_2)\label{6dbTUVm8} \\
 ....&&....\0
\ee
${\cal O}(m^0)$ and ${\cal O}(m^2)$ as well as all terms with even $m$ power larger than 6 are conserved,
while ${\cal O}(m^4)$ and $ {\cal O}(m^6)$
are not. According to our prescription we have to subtract not only ${\cal
O}_{IR}(m^4)$ and ${\cal O}_{IR}(m^6)$, but also
${\cal O}_{IR}(m^0), {\cal O}_{IR}(\log(m))$ and ${\cal O}_{IR}(m^2)$. We obtain{\small
\be 
{\cal O}_{UV}(m^6)-{\cal O}_{IR}(m^6)&=&-\frac {im^6}{288\pi^3} \left(-6 \log
\left(-\frac {k^2}{m^2} \right) +1 \right) \Pi^{(2)}_{\frac 12} (k,n_1,n_2)\label{6dbTUV-IRm6}\\ 
 {\cal O}_{UV}(m^4)-{\cal O}_{IR}(m^4)&=&\frac {im^4}{576\pi^3} k^2\left(-6 \log
\left(-\frac {k^2}{m^2} \right) +7 \right) \Pi^{(2)}_{\frac 12} (k,n_1,n_2)\label{6dbTUV-IRm4}\\ 
 {\cal O}_{UV}(m^2)-{\cal O}_{IR}(m^2)&=&\frac {im^2}{7200 \pi^3}k^4 \left(-15  \log
\left(-\frac {k^2}{m^2} \right) +31 \right) \Pi^{(2)}_{\frac 12} (k,n_1,n_2)\label{6dbTUV-IRm2}
 \ee}
and
\be
{\cal O}_{UV}(m^0)-{\cal O}_{IR}(m^0)\!&-&\!{\cal O}_{IR}(\log(m))=\frac
{i}{705600\pi^3}k^6\left(-105  \log \left(-\frac {k^2}{m^2} \right)+352  \right)\0\\
&&\cdot\Pi^{(2)}_{\frac 12} (k,n_1,n_2) \label{6dbTUV-IRm0}
\ee
They are all conserved. (\ref{6dbTUV-IRm4}) contains a nonlocal linearized
version of the EH eom.

\subsubsection{6d msm: spin 3 current}

The scheme is the same as above. In the IR the odd power of $m$ vanish. The even
powers $m^{2n}$ with $n\leq 2$ are conserved
together with the term proportional to $\log(m)$. The terms ${\cal
O}_{IR}(m^4),{\cal O}_{IR}(m^6),{\cal O}_{IR}(m^8)$ are not conserved.
Of course ${\cal O}(\log(m)),{\cal O}_{IR}(m^2)$ diverge in the IR, while the term  ${\cal
O}_{IR}(m^0)$ diverges for $\varepsilon \to 0$.
According to our prescription all these terms, which are local, have to be subtracted from the effective action. 
 Using again $ \Pi^{(3)}(k,n_1,n_2)$, see (\ref{Pi3}),  the result is as follows. 
\be
{\cal O}(m^{-2}):&& -\frac {ik^{10}}{443520 m^2\pi
^3}\,\Pi_{\frac 32}^{(3)}(k,n_1,n_2)\label{6dbspin3IRm-2}\\
{\cal O}(m^{-4}):&& -\frac {ik^{12}}{11531520 m^4\pi ^3}
\,\Pi_{\frac 32}^{(3)}(k,n_1,n_2)\label{6dbspin3IRm-4}\\
....&&....\0
\ee
In the UV the odd $m$ power terms vanish. The even power of order $4,6,8$ are
not conserved, but
\be 
{\cal O}_{UV}(m^0)-{\cal O}_{IR}(m^0) \!&-&\!{\cal
O}_{IR}(\log(m))\label{6dbspin3UV-IRm0}\\
&=&\frac {ik^8}{6350400 \pi ^3} \,\Pi_{\frac 32}^{(3)}(k,n_1,n_2)
\left(-315\log \left(-\frac {k^2}{m^2}\right) +1126 \right )\0
\ee
and
\be
{\cal O}_{UV}(m^2)-{\cal O}_{IR}(m^2)&=&  \frac {im^2k^6}{117600 \pi ^3}
\,\Pi_{\frac 32}^{(3)}(k,n_1,n_2) \left(
(247 -105   \log \left(-\frac {k^2}{m^2} \right)\right)\label{6dbspin3UV-IRm2}\\
{\cal O}_{UV}(m^4)-{\cal O}_{IR}(m^4)&=& -\frac{i m^4 k^4}{4800 \pi ^3}
\,\Pi_{\frac 32}^{(3)}(k,n_1,n_2)\left( 
( 47  {-} 30 \log \left(-\frac {k^2}{m^2} \right)\right)\label{6dbspin3UV-IRm4}\\
{\cal O}_{UV}(m^6)-{\cal O}_{IR}(m^6)&=& -\frac{i m^6k^2}{288 \pi ^3}
\,\Pi_{\frac 32}^{(3)}(k,n_1,n_2) \left(
( -5+6 \log \left(-\frac {k^2}{m^2} \right)\right)\label{6dbspin3UV-IRm6}\\
{\cal O}_{UV}(m^8)&=& -\frac{i m^8}{384 \pi ^3} \,\Pi_{\frac 32}^{(3)}(k,n_1,n_2) \left(
(-1 -12  \log \left(-\frac {k^2}{m^2} \right)\right)\label{6dbspin3UV-IRm8}\\
....&&....\0
\ee
are all conserved. Eq.(\ref{4dbspin3UV-IRm6}) is related to a nonlocal version
of the spin 3 Fronsdal equation.

\subsection{6d Fermion Model}

We consider now the same analysis for the fermion massive model. We start again
from eqs.(\ref{actionA},\ref{actiong},\ref{Jmunuab},\ref{Tmnlr}) and the like for
higher spins and set $d=6+\varepsilon$.

\subsubsection{6d mfm: spin 1 current}

We will limit ourselves to the power of $m$ expansions in the IR. The terms proportional 
to $m^4,m^2, m^0$ and $\log(m)$ are local, nonconserved and/or 
divergent. Thus they must be subtracted. Therefore the first nonvanishing term in the IR is:
\be
{\cal O}(m^{-2}):&&-\frac{i k^6}{1120 \pi ^3m^2}  (n_1\!\cdot\! \pi
^{(k)}\!\cdot\! n_2) \label{6dfJIRm-2}\\
....&&....\0
\ee
In the UV, after subtracting the local terms we find:
{\small
\be
{\cal O}_{UV}(m^0)-{\cal O}_{IR}(m^0)\!&-&\!{\cal
O}_{IR}(\log(m)) \label{6dfJUV-IRm0}\\
&=& - \frac{i k^4}{ {3600} \pi ^3} \left(-30 \log \left(-\frac {k^2}{m^2} \right)
+77 \right) (n_1\!\cdot\! \pi ^{(k)}\!\cdot\! n_2)\0\\
{\cal O}_{UV}(m^2)-{\cal O}_{IR}(m^2)&=&\frac{im^2k^2}{ {72} \pi ^3 }\Big{(}  \left(-3 \log
\left(-\frac {k^2}{m^2} \right)+2 \right)\Big{)}(n_1\!\cdot\! \pi ^{(k)}\!\cdot\! n_2)\label{6dfJUV-IRm2}
\ee }
and 
\be
{\cal O}_{UV}(m^4)-{\cal O}_{IR}(m^4)&=& \frac{im^4}{ {8} \pi ^3}
(n_1\!\cdot\! \pi ^{(k)}\!\cdot\! n_2)\label{6dfJUV-IRm4}\\
{\cal O}_{UV}(m^6)&=&-\frac{im^6}{ {72} \pi ^3k^2}\Big{(}  \left(-6 \log
\left( -\frac {k^2}{m^2}\right)-5 \right)\Big{)}
(n_1\!\cdot\! \pi ^{(k)}\!\cdot\! n_2)\label{6dfJUVm6}\\
....&&....\0
\ee

(\ref{6dfJUV-IRm2}) corresponds to the linearized Maxwell action with an energy dependent coupling.

\subsubsection{6d mfm: spin 2 current}

In this subsection all results must be multiplied by a factor of $\frac 1{16}$
In the IR the odd powers of $m$ vanish. The terms proportional 
to $m^6,m^4,m^2, m^0$ and $\log(m)$ are local, nonconserved and/or 
divergent. Thus they must be subtracted.
\be
 {\cal O}(m^{-2}):&&\frac{i k^8}{8640 \pi ^3 m^2}  \left( \left(n_1\!\cdot\! \pi ^{(k)}\!\cdot\! n_2\right){}^2 -\frac 17
(n_1\!\cdot\! \pi ^{(k)}\!\cdot\! n_1)
( n_2\!\cdot\! \pi ^{(k)}\!\cdot\! n_2) \right)  \label{6dfscalarIRm-2}\\
 ....&&....\0
\ee}
 
In the UV all the odd powers of $m$ vanish. After subtracting the above local terms we have
\be
{\cal O}_{UV}(m^0)-{\cal O}_{IR}(m^0)\!& -&\!{\cal O}_{IR}(\log(m))\label{6dfTUV-IRm0}\\
&=&\frac
{i}{352800\pi^3}k^6\Bigg{(} 25\left(-21\log \left( -\frac {k^2}{m^2}\right)+62 \right) \left(n_1\!\cdot\! \pi ^{(k)}\!\cdot\! n_2\right){}^2 
\0\\
&&-\left(- 105  \log \left(-\frac {k^2}{m^2} \right)+352  \right)(n_1\!\cdot\! \pi ^{(k)}\!\cdot\! n_1)
( n_2\!\cdot\! \pi ^{(k)}\!\cdot\! n_2)  \Bigg{)} \0\\
{\cal O}_{UV}(m^2)-{\cal O}_{IR}(m^2)&=& -\frac {im^2}{3600 \pi^3}k^4\Bigg{(}9 \left(- 5 \log
\left(-\frac {k^2}{m^2} \right) +7 \right)\left(n_1\!\cdot\! \pi ^{(k)}\!\cdot\! n_2\right){}^2\label{6dfTUV-IRm2}\\ 
&&+ \left(15 \log
\left(-\frac {k^2}{m^2} \right) -31 \right) 
(n_1\!\cdot\! \pi ^{(k)}\!\cdot\! n_1)
( n_2\!\cdot\! \pi ^{(k)}\!\cdot\! n_2) \Bigg{)}\0
\ee
and
\be 
{\cal O}_{UV}(m^4)-{\cal O}_{IR}(m^4)&=& -\frac {im^4}{288\pi^3}k^2\Bigg{(}\left(6 \log
\left(-\frac {k^2}{m^2} \right) +5\right) \left(n_1\!\cdot\! \pi ^{(k)}\!\cdot\! n_2\right){}^2\label{6dfTUV-IRm4}\\ 
&&+\left(-6 \log
\left(-\frac {k^2}{m^2} \right) +7 \right)
(n_1\!\cdot\! \pi ^{(k)}\!\cdot\! n_1)
( n_2\!\cdot\! \pi ^{(k)}\!\cdot\! n_2) \Bigg{)} \0\\
{\cal O}_{UV}(m^6)-{\cal O}_{IR}(m^6)&=&-\frac {im^6}{144\pi^3}\Bigg{(} \left(6 \log
\left(-\frac {k^2}{m^2}\right) -13\right)\left(n_1\!\cdot\! \pi ^{(k)}\!\cdot\! n_2\right){}^2\label{6dfTUV-IRm6}\\ 
&&+ \left(6 \log
\left( -\frac {k^2}{m^2}\right) -1\right)
(n_1\!\cdot\! \pi ^{(k)}\!\cdot\! n_1)
( n_2\!\cdot\! \pi ^{(k)}\!\cdot\! n_2) \Bigg{)} \0\\  
{\cal O}_{UV}(m^8)&=&\frac {im^8}{576\pi^3k^2}\Bigg{(} 3 \left(12 \log
\left(-\frac {k^2}{m^2}\right) +17\right)\left(n_1\!\cdot\! \pi ^{(k)}\!\cdot\! n_2\right){}^2\label{6dfTUVm8}\\ 
&&+ \left(12\log\left(-\frac {k^2}{m^2}\right) +25 \right) 
(n_1\!\cdot\! \pi ^{(k)}\!\cdot\! n_1)
( n_2\!\cdot\! \pi ^{(k)}\!\cdot\! n_2)  \Bigg{)} \0
\ee

They are all conserved. (\ref{6dfTUV-IRm4}) contains a nonlocal linearized version of the EH eom.

\subsubsection{6d mfm: spin 3 current}

The scheme is the same as above. In the IR the odd power of $m$ vanish. The even
powers $m^{2n}$ with $n\leq 2$ are conserved
together with the term proportional to $\log(m)$. The terms ${\cal
O}_{IR}(m^4),{\cal O}_{IR}(m^6),{\cal O}_{IR}(m^8)$ are not conserved.
Of course ${\cal O}(\log[m]),{\cal O}_{IR}(m^2)$ diverge in the IR, while the term  ${\cal
O}_{IR}(m^0)$ diverges for $\varepsilon \to 0$.
According to our prescription all these terms, which are local, have to be subtracted from the effective action. 

Using again $ \Pi^{(3)}(k,n_1,n_2)$, see (\ref{Pi3}), the result is  
\be
{\cal O}(m^{-2}):&&- \frac {ik^{10}}{93555 m^2\pi
^3}\,\Pi_{\frac{49}{128}}^{(3)}(k,n_1,n_2)\label{6dfspin3IRm-2}\\
....&&....\0
\ee
In the UV the odd $m$ powers vanish. The even powers of order $4,6,8$ are
not conserved. After subtracting the local terms in the IR one gets
\be 
{\cal O}_{UV}(m^0)\!&-&\! {\cal O}_{IR}(m^0) -{\cal
O}_{IR}(\log(m))\label{6dfspin3UV-IRm0}\\
&=&-\frac {ik^8}{57153600 \pi ^3}\Bigg{(}
32\left(-315 \log \left(-\frac {k^2}{m^2} \right) +1021  \right )\left(n_1\!\cdot\! \pi ^{(k)}\!\cdot\! n_2\right){}^3\0\\
&-&\!3\left(- 630 \log \left(-\frac {k^2}{m^2} \right) +3617  \right ) 
\left(n_1\!\cdot\! \pi ^{(k)}\!\cdot\! n_2\right)(n_1\!\cdot\! \pi ^{(k)}\!\cdot\! n_1)
( n_2\!\cdot\! \pi ^{(k)}\!\cdot\! n_2)  \Bigg{)}\0
\ee
and{\small
\be
{\cal O}_{UV}(m^2)\!&-&\!{\cal O}_{IR}(m^2) = \frac {im^2k^6}{3175200\pi ^3}
\Bigg{(}
32\left(- {210}\log \left(-\frac {k^2}{m^2} \right) +389  \right )\left(n_1\!\cdot\! \pi ^{(k)}\!\cdot\! n_2\right){}^3\0  \\
&& +\left( 3885\log \left( -\frac {k^2}{m^2}\right) -9454 \right ) 
\left(n_1\!\cdot\! \pi ^{(k)}\!\cdot\! n_2\right)(n_1\!\cdot\! \pi ^{(k)}\!\cdot\! n_1)
( n_2\!\cdot\! \pi ^{(k)}\!\cdot\! n_2)  \Bigg{)}\label{6dfspin3UV-IRm2}\\
{\cal O}_{UV}(m^4)\!&-&\!{\cal O}_{IR}(m^4) = \frac{i m^4 k^4}{64800 \pi ^3}\Bigg{(}
 16\left(- 17 +30  \log \left(-\frac {k^2}{m^2} \right)\right)\left(n_1\!\cdot\! \pi ^{(k)}\!\cdot\! n_2\right){}^3\label{6dfspin3UV-IRm4}\\
&& +\left(- 480 \log \left(-\frac {k^2}{m^2} \right) +617  \right ) 
\left(n_1\!\cdot\! \pi ^{(k)}\!\cdot\! n_2\right)(n_1\!\cdot\! \pi ^{(k)}\!\cdot\! n_1)
( n_2\!\cdot\! \pi ^{(k)}\!\cdot\! n_2)  \Bigg{)}\0\\
{\cal O}_{UV}(m^6)\!&-&\!{\cal O}_{IR}(m^6) = -\frac{i m^6k^2}{864 \pi ^3}\Bigg{(}\frac { {64}}3 
\left(n_1\!\cdot\! \pi ^{(k)}\!\cdot\! n_2\right){}^3 \label{6dfspin3UV-IRm6}\\ 
 && -\left( 18\log \left(-\frac {k^2}{m^2} \right) -17  \right ) 
\left(n_1\!\cdot\! \pi ^{(k)}\!\cdot\! n_2\right)(n_1\!\cdot\! \pi ^{(k)}\!\cdot\! n_1)
( n_2\!\cdot\! \pi ^{(k)}\!\cdot\! n_2)  \Bigg{)}\0\\
{\cal O}_{UV}(m^8)\!&-&\!{\cal O}_{IR}(m^8) = \frac{i m^8}{5184 \pi ^3} \Bigg{(} 16 \left(
(11 -12\log \left(-\frac {k^2}{m^2}\right)\right)\label{6dfspin3UV-IRm8}\\
 && +\left(211 -276 \log \left(-\frac {k^2}{m^2}\right) \right ) 
\left(n_1\!\cdot\! \pi ^{(k)}\!\cdot\! n_2\right)(n_1\!\cdot\! \pi ^{(k)}\!\cdot\! n_1)
( n_2\!\cdot\! \pi ^{(k)}\!\cdot\! n_2)  \Bigg{)}\0\\
....&&....\0
\ee}
are all conserved. Eq.(\ref{6dfspin3UV-IRm6}) is related to a nonlocal version
of the spin 3 Fronsdal equation.

\section{Spin $s$ current two-point correlators in any dimension}
\label{sec:spins}

In this section we derive general formulas for the two-point correlators for spin $s=1,2$ and 3 in any dimension. The procedure is slightly different from the one used so far. In the previous sections we
fixed the dimension of space-time, that is we set $d=3,4+\varepsilon,5, 6+\varepsilon$ in the scalar integrals (see sec.\ref{sec:method}). In this section we leave the parameter $d$ free and 
evaluate specific cases at the end. The two procedures often lead to different intermediate results. Of course the results of physical interest must coincide.

In the following we focus on the massive case, while the massless computations are deferred to Appendix \ref{sec:massless}.

\subsection{Fermion model}

In this subsection we compute the even part of two point correlator for a fermion in $d$ dimensions for spin $s=1,2,3$
 \be 
 \tilde J_{\mu_1 \ldots \mu_s\nu_1\ldots\nu_s}(k)& =& 
-\int\frac
{d^3p}{(2\pi)^3}\, {\rm Tr} 
\left( \frac {i} {\slashed{p}-m} \gamma_\sigma 
\frac {{ i}} {\slashed{p}-\slashed{k}-m} \gamma_\tau \right)
V^\sigma_{\mu_1\ldots\mu_s}V^\tau_{\nu_1\ldots\nu_s}\label{Jmunu0}
\ee
where the Feynman vertices are
\be
V^\sigma_{\mu_1\ldots\mu_s} = { i}  \quad \delta^\sigma_\mu\sum_{j=0}^{\left\lfloor 
  \frac{s-1}{2}\right\rfloor} 
  \frac{i^{{ 2} s-2 j-{ 2}}(k_\mu-2 p_\mu)^{s-2 j-1} (2 (p\!\cdot\!(p-k) - m^2) \eta_{\mu\mu}-4 (p-k)_\mu p_\mu)^j}{(2 j+1)! (n-2 j-1)!}\0\\
	\label{Vmumumu}
\ee

\subsubsection{Fermion model - massive case}

Let us compute the even part of two point correlator for a massive fermion in $d$ dimensions for spin 1. The one-loop contribution is
\be \label{JmunuF}
\tilde J_{\mu\nu}(k) &=& - \int \frac{d^dp}{(2\pi)^d} {\rm Tr}
\left( {\gamma_\nu} \frac 1{\slashed{p} -m } {\gamma_\mu}   
\frac 1{\slashed{p}-\slashed{k} -m }\right) \0\\
&=&  -{\rm Tr}(1)
\int\frac{d^dp}{(2\pi)^d}\frac {p_\nu(p-k)_\mu
 + p_\mu(p-k)_\nu -p\!\cdot\!(p-k)\eta_{\mu\nu}+m^2 \eta_{\mu\nu}}{(p^2-m^2)((p-k)^2-m^2)}
\ee
${\rm Tr}(1)$ is the trace of the identity operator on the vector space on which the Dirac matrices act. Since we are working with the lowest dimensional complex spinors in each dimension we have ${\rm Tr}(1)=2^{\lfloor\frac{d}{2}\rfloor}$. In the odd dimensional cases $d$ can be simply replaced
by the values $3,5,...$, so this factor is an overall numerical factor. In the even dimensional cases we have to replace $d$ by $2+\varepsilon,4+\varepsilon,...$, so the same factor contains an $\varepsilon$ dependence in addition to the overall numerical factor. The $\varepsilon$ dependence cannot change the divergent pole part in dimensional regularization but will only affect the finite local part. However
when subtracting the infinite and finite IR terms from the effective action this dependence disappears.
So we will ignore it.

{\bf Warning}. In evaluating the scalar integral $\tilde{\cal I}^{(1)}(k)$, which is our basic quantity, the results below have been obtained by choosing a reference value 4 for ${\rm Tr}(1)$. This value is appropriate only for $d=4,5$, but must be corrected for the other dimensions: for $d=2,3$ the results must be divided by 2, for $d=6,7$ they must be multiplied by 2, and so on.

By Davydychev tensor reduction procedure it is possible to rewrite such an amplitude in terms of scalar integrals as
\be
\tilde J_{\mu\nu}(k) &=& \eta_{\mu\nu}\tilde {\cal I}^{(1)}(k)+k_\mu k_\nu \tilde{\cal I}^{(2)}(k)
=-4 \Bigg{(}\eta_{\mu\nu} \bigg{(}m^2 \tilde I^{(2)}(d,1,1;k,m) \label{Jmunured}\\
&& -2 \pi  \left(2 k^2 \left(\tilde I^{(2)}(d+2,2,1;k,m)+8 \pi \tilde I^{(2)}(d+4,3,1;k,m) \right)\right.  \0\\
&& \quad\quad \left.-(d-2)\tilde I^{(2)}(d+2,1,1;k,m)\right)\bigg{)}\0\\
&&+8 \pi  k_\mu k_\nu \left(\tilde I^{(2)}(d+2,2,1;k,m) +8 \pi  I^{(2)}(d+4,3,1;k,m) \right)\Bigg{)}\0 
\ee
Because of dimensional reasons the superficial degree of divergence of  $\tilde{\cal I}^{(1)}$ and  
$\tilde{\cal I}^{(2)}$ are always such that 
${\rm deg}(\tilde{\cal I}^{(1)})={\rm deg}(\tilde{\cal I}^{(2)})+2=d-2$. 
One can check the Ward identity
\be
k^2\tilde {\cal I}^{(2)}(k)+\tilde{\cal I}^{(1)}(k)=0,\label{WII1I2}
\ee
which implies we can rewrite the amplitude \eqref{JmunuF} as
\be
\tilde J_{\mu\nu}(k) =\left( \eta_{\mu\nu}-\frac{k_\mu k_\nu}{k^2}\right)\tilde{\cal I}^{(1)}(k),\label{Jmunuproj}
\ee
This is in fact the main advantage of the regularization we are using in this section: the two point function is conserved without any subtraction. This will not be true for higher spin currents (see below).

The explicit form of $\tilde{\cal I}^{(1)}(k)$ is
\be
&&\tilde {\cal I}^{(1)}(k)=-\frac{ 2^{1-d} \, i\,e^{-\frac{1}{2} i \pi  d} \pi ^{-\frac{d}{2}} \left(-m^2\right)^{d/2 -1} \Gamma \left(2-\frac{d}{2}\right)}{(d-2)}\label{calI1}\\
&&\cdot\left(8 \frac{k^2}{4 m^2} \, _2F_1\left(2,1-\frac{d}{2};\frac{3}{2};\frac{k^2}{4 m^2}\right)-4\left(d \frac{k^2}{4m^2}+1\right) \, _2F_1\left(1,1-\frac{d}{2};\frac{3}{2};\frac{k^2}{4 m^2}\right)+4\right)\0\\
 &=&\frac{ 2^{3-d}\, i\, e^{-\frac{1}{2} i \pi  d} \pi ^{-d/2} \left(-m^2\right)^{\frac{d-2}{2}}
   \Gamma \left(2-\frac{d}{2}\right) \left(((d-2) z+1) \,
   _2F_1\left(1,-\frac{d}{2};\frac{1}{2};z\right)+2 z-1\right)}{(d-2) d (z-1) z}\0 
\ee
where $z=\frac{k^2}{4m^2}$.

For  $d=2$ there is no pole.  For  even $d>2$ the relevant $\varepsilon$-expansion is given by
\be
&&\tilde {\cal I}^{(1)}(k) =\frac 1{\varepsilon} \frac{ 2^{4-d} i^{d+1} \pi ^{-d/2} m^{d-2} \left(((d-2) z+1) \,
   _2F_1\left(1,-\frac{d}{2};\frac{1}{2};z\right)+2 z-1\right)}{(d-2) d (z-1) z 
    \Gamma \left(\frac{d}{2}-1\right)}\label{calI1eps}\\
&&-\frac{  2^{3-d}i^{d+1}  \pi ^{-d/2} m^{d-2}}{(d-2)^2 d^2 (z-1)
   z \Gamma \left(\frac{d}{2}-1\right)}\0\\
&&\cdot\Bigg{(}-2 (d-2) d \bigg{(}z \,
   _2F_1\left(1,-\frac{d}{2};\frac{1}{2};z\right)-\frac{1}{2} ((d-2) z+1)
   \, _2F_1^{(0,1,0,0)}\left(1,-\frac{d}{2},\frac{1}{2},z\right)\bigg{)} \0\\
&&+
   \left(((d-2) z+1) \, _2F_1\left(1,-\frac{d}{2};\frac{1}{2};z\right)+2 z-1\right)\0\\
	&&\quad\quad\cdot  \left(d^2 \log (4 \pi )-2 (d-2) d \log (m)-2 d (\log (4 \pi )-2)-4\right)\0\\
&&+
(d-2) d
   \left(H_{\frac{d}{2}-2}-\gamma \right) \left(((d-2) z+1) \,
   _2F_1\left(1,-\frac{d}{2};\frac{1}{2};z\right)+2 z-1\right)\Bigg{)}
\ee
where $_2F_1^{(0,1,0,0)}\left(1,-\frac{d}{2},\frac{1}{2},z\right)$ means the derivative of 
 $_2F_1^{(0,1,0,0)}\left(1, \beta,\frac{1}{2},z\right)$ with respect to $\beta$ at $\beta= - \frac d2$,
and $H_{\frac{d}{2}-2}$ are the harmonic numbers.

\subsubsection{d even}

Let us examine first the even $d$ case. For even $d$ the hypergeometric functions boil down to finite order polynomials according to the formulas
\be
\, _2F_1\left(2,1-\frac{d}{2};\frac{3}{2};z\right)&=&\sum _{n=0}^{\frac{d}{2}-1} \frac{(2)_n \left(1-\frac{d}{2}\right)_n}{n! \left(\frac{3}{2}\right)_n}z^n \0\\
 \, _2F_1\left(1,1-\frac{d}{2};\frac{3}{2};z\right)&=&\sum _{n=0}^{\frac{d}{2}-1} \frac{(1)_n \left(1-\frac{d}{2}\right)_n}{n! \left(\frac{3}{2}\right)_n} z^n\0
\ee
so that one can easily check the $1/\varepsilon$ part is just
\be
d=4&&\frac{2 i m^2 z}{3 \pi ^2},\label{calI1div}\\
d=6&&\frac{i m^4 z (4 z-5)}{30 \pi ^3},\0\\
d=8&&\frac{i m^6 z \left(24 z^2-56 z+35\right)}{1680 \pi ^4},\0\\
d=10&&\frac{i m^8 z \left(64 z^3-216 z^2+252 z-105\right)}{60480 \pi ^5},\0\\
d=12&&\frac{i m^{10} z \left(640 z^4-2816 z^3+4752 z^2-3696 z+1155\right)}{10644480 \pi ^6},\0\\
&\ldots&\0
\ee
So in general the divergent part (for $\varepsilon \to 0$) is a polynomial in $z=\frac{k^2}{4m^2}$ of degree $d/2-1$, where the constant term is missing because the front factor of highest dimension is $m^{d-2}$. According to Weinberg's theorem this  corresponds to the degree of divergence $d-4$ of the two point functions, which is therefore lower than the expected one $d-2$ because of gauge invariance. 
The above divergent terms are local and appear both in the IR and the UV limit, as we have seen many times above. Some of them are divergent for $m\to\infty$ and must be subtracted to guarantee finiteness
of the IR limit (or decoupling of infinite mass modes). Others are of order $m^0$. The reason why 
we subtract them from the effective action is, according to our attitude, because the physical information is contained in the difference between the IR and UV limits (rather than in their absolute value).

As for the finite part we cannot give a closed formula for generic $d$, but  thanks to the formulae
\be
\, _2F_1\left(1,-\frac{d}{2};\frac{1}{2};z\right)&=&\sum _{n=0}^{d/2} \frac{(1)_n \left(-\frac{d}{2}\right)_n}{\left(\frac{1}{2}\right)_n}\frac{z^n}{n! }\label{F21id1} \\
&=&\sum _{n=0}^{d/2} \frac{\Gamma(-\frac{d}2+n)}{\Gamma(-\frac{d}2)}\frac{\Gamma(\frac12)}{\Gamma(\frac12 + n)}z^n\0\\
\, _2F_1^{(0,1,0,0)}\left(1,-\frac{d}{2};\frac{1}{2};z\right)&=&-2\sum _{n=0}^{\infty}\frac{\partial}{\partial d} \frac{\Gamma(-\frac{d}2+n)}{\Gamma(-\frac{d}2)}\frac{\Gamma(\frac12)}{\Gamma(\frac12 + n)}z^n\0\\
&=&-\sum _{n=0}^{\infty}\frac{\sqrt{\pi } \Gamma \left(n-\frac{d}{2}\right) \left(\psi
   ^{(0)}\left(-\frac{d}{2}\right)-\psi ^{(0)}\left(n-\frac{d}{2}\right)\right)}{\Gamma
   \left(-\frac{d}{2}\right) \Gamma \left(n+\frac{1}{2}\right)}z^n\0
\ee
we can recognize the IR behavior is analytic in $z$, which is to be expected as $m$ acts as an IR regulator. More explicitly, we get the following behaviors
\be
d=4&:&\frac{i m^2 z ( \log (\frac{m^2}{4\pi})+\gamma)}{3 \pi ^2}-\frac{4 i m^2 z^2}{15 \pi ^2}-\frac{4 i m^2 z^3}{35 \pi ^2}-\frac{64 i m^2 z^4}{945 \pi ^2}+{\cal O}(z^5),\label{calI1IR}\\
d=6&:&\frac{i m^4 z (- \log (\frac{m^2}{4\pi})- \gamma +1)}{12 \pi ^3}+\frac{i m^4 z^2 ( \log (\frac{m^2}{4\pi})+ \gamma )}{15 \pi ^3}\0\\
&&-\frac{i m^4 z^3}{35 \pi ^3}-\frac{8 i m^4 z^4}{945 \pi ^3}+{\cal O}(z^5),\0\\
d=8&:&\frac{i m^6 z (2 \log (\frac{m^2}{4\pi})+2 \gamma -3)}{192 \pi ^4}+\frac{i m^6 z^2 (-\log (\frac{m^2}{4\pi})- \gamma +1)}{60 \pi ^4}\0\\
&&+\frac{i m^6 z^3 (\log (\frac{m^2}{4\pi})+ \gamma)}{140 \pi ^4}-\frac{2 i m^6 z^4}{945 \pi ^4}+{\cal O}(z^5),\0\\
d=10&:&\frac{i m^8 z (-6 \log (\frac{m^2}{4\pi})-6 \gamma +11)}{6912 \pi ^5}+\frac{i m^8 z^2 (2 \log (\frac{m^2}{4\pi})+2 \gamma -3)}{960 \pi ^5}\0\\
&&+\frac{i m^8 z^3 (-\log (\frac{m^2}{4\pi})- \gamma +1)}{560 \pi ^5}+\frac{i m^8 z^4 ( \log (\frac{m^2}{4\pi})+ \gamma )}{1890 \pi ^5}+{\cal O}(z^5),\0\\
  \ldots&&\ldots\0
\ee
So the dominating term is $\sim m^{d-2}(A^{(1)} \log m +B^{(1)})z\equiv m^{d-4}(A^{(1)} \log m +B^{(1)})k^2 $, whereas the term with highest power of momentum and dimensionless constant is $\sim m^{d-2}(A^{(d/2-1)} \log m +B^{(d/2-1)})z^{d/2-1}\equiv (A^{(d/2-1)} \log m +B^{(d/2-1)})k^{d-2}$. So, in coordinate space the following terms are dominating for large $m$
\be
\sim (A^{(d/2-1)} \log m +B^{(d/2-1)})F^{\mu\nu}\Box^{d/2-2}F_{\mu\nu},\ldots, m^{d-4}(A^{(1)} \log m +B^{(1)})F^{\mu\nu}F_{\mu\nu}\,,\0
\ee
whereas all the others are suppressed by negative powers of $m$. For $d>4$ those terms would give a non-decoupling of IR dynamics from high-energy physics, but we can notice they are the same as the local counterterms appearing in the divergent part.\footnote{This case corresponds, in ordinary interacting gauge theories, to the fact that the vertex for the spin $1$ current is a non-renormalizable interaction for $d>4$ and in fact higher derivatives operators are generated. $d=4$ is the special case when just the two-derivative operator with dimensionless front constant is generated and it corresponds to the fact that spin $1$ vertex in $d=4$ is a renormalizable interaction. For $d<4$ the spin $1$ vertex is super-renormalizable.} So they have to be subtracted, as we have done many times before. In (\ref{calI1IR}) there are also terms of order $m^0$. They have to be subtracted from the effective action for the
same reason explained above: the physical meaning is contained in the difference between the UV and the IR.

In  $d=2$ no pole shows up. In fact we have
\be
\frac{i \left(   \, _2F_1^{(0,1,0,0)}\left(1,-1,\frac{1}{2},z\right)+2 z (2
   z-1)\right)}{2 \pi  (z-1) z}\,,\0
\ee
whose IR expansion is
\be
-\frac{4 i z}{3 \pi }-\frac{16 i z^2}{15 \pi }-\frac{32 i z^3}{35 \pi }-\frac{256 i
   z^4}{315 \pi }-\frac{512 i z^5}{693 \pi }-\frac{2048 i z^6}{3003 \pi
   }+O\left(z^7\right)\,,\label{2dIRexp}
\ee
meaning all the local terms are suppressed by negative powers of the mass. There is no need to remove them by finite subtraction in order to have decoupling.

The asymptotic behavior of the finite part in the UV ($z\to \infty$) is determined by the formulae
\be
\, _2F_1\left(1,-\frac{d}{2};\frac{1}{2};z\right)&=&\frac{(1)_{\frac{d}{2}} (-z)^{\frac{d}{2}}}{\left(\frac{1}{2}\right)_{\frac{d}{2}}} \, _2F_1\left(-\frac{d}{2},\frac{1-d}{2};-\frac{d}{2};\frac{1}{z}\right)\label{F21id2} \\
&=&\Gamma\left(\frac12\right)\frac{\Gamma\left(\frac{d+2}2\right)}{\Gamma\left(\frac{d+1}2\right)}(-z)^{\frac{d}{2}} \, _2F_1\left(-\frac{d}{2},\frac{1-d}{2};-\frac{d}{2};\frac{1}{z}\right)\0\\
&=&(-z)^{d/2} \sum _{n=0}^{\frac{d}{2}} \frac{ \left(\cos \left(\frac{\pi  d}{2}\right) \Gamma \left(\frac{d}{2}+1\right) \Gamma \left(-\frac{d}{2}+n+\frac{1}{2}\right)\right)}{\sqrt{\pi } \Gamma (n+1)}\left(\frac{1}{z}\right)^n\0
\ee
and
\be
\, _2F_1^{(0,1,0,0)}\left(1,-\frac{d}{2};\frac{1}{2};z\right)&=&-2\frac{\partial}{\partial d}\, _2F_1\left(1,-\frac{d}{2};\frac{1}{2};z\right)\label{F21id3} \\
 &=&-z^{d/2} \log (-z)\sum _{n=0}^{\frac{d}{2}} \frac{ \Gamma \left(\frac{d}{2}+1\right) \Gamma \left(-\frac{d}{2}+n+\frac{1}{2}\right)}{\sqrt{\pi } \Gamma (n+1)}\left(\frac{1}{z}\right)^n\0\\
&&+z^{d/2}\sum _{n=0}^{\infty}\frac{\Gamma \left(\frac{d}{2}+1\right) \Gamma \left(-\frac{d}{2}+n+\frac{1}{2}\right)\left(H_{-\frac{d}{2}+n-\frac{1}{2}}-H_{\frac{d}{2}}\right)}{\sqrt{\pi } \Gamma (n+1)} \left(\frac{1}{z}\right)^n\0
\ee

We notice, as before, that the asymptotic expansion of $\, _2F_1^{(0,1,0,0)}\left(1,-\frac{d}{2};\frac{1}{2};z\right)$ contains $\log (-z)$, which can have an imaginary part even for real $-z$. We should keep in mind that the expansion is valid under the assumption $| \arg (-z) | <\pi $, which means $-z$, if real, must be assumed to be positive. This is for example the case for Euclidean momenta which are such that $\frac{k_E^2}{4m^2}=-\frac{k_M^2}{4m^2}=-z$. When the metric is Lorentzian a cut appears at $z=1$ and we have to rely on the analytic continuation by choosing a definite Riemann sheet: it is not surprising that in this case an imaginary part of the correlator appear.  
Dimension by dimension the UV expansions for the finite part are:
\be
d=4: &&  \frac{i m^2 z (3 \log (\frac{m^2}\pi)+3 \log (-z)+3 \gamma -5)}{9 \pi ^2}-\frac{i m^2 (2 \log (-4 z)+1)}{16 \pi ^2 z}\label{UVexp}\\
&&-\frac{i m^2}{2 \pi ^2}+O\left(\left(\frac{1}{z}\right)^2\right),\0\\
d=6:&&  \frac{i m^4 z^2 (30\log (\frac{m^2}\pi)+30 \log (-z)+30 \gamma -77)}{450 \pi ^3}+\frac{i m^4 (6 \log (-4 z)+5)}{576 \pi ^3 z}\0\\
&&-\frac{i m^4 z (3\log (\frac{m^2}\pi))+3 \log (-z)+3 \gamma -5)}{36 \pi ^3}+\frac{i m^4}{16 \pi ^3}+O\left(\left(\frac{1}{z}\right)^2\right),\0\\
d=8:&&  \frac{i m^6 z^3 (105 \log (\frac{m^2}\pi)+105 \log (-z)+105 \gamma -317)}{14700 \pi ^4}\0\\
&&-\frac{i m^6 z^2 (30 \log (\frac{m^2}\pi)+30 \log (-z)+30 \gamma -77)}{1800 \pi ^4}\0\\
&&+\frac{i m^6 z (3 \log (\frac{m^2}\pi)+3 \log (-z)+3 \gamma -5)}{288 \pi ^4}-\frac{i m^6 (12 \log (-4 z)+13)}{18432 \pi ^4 z}\0\\
&&-\frac{i m^6}{192 \pi ^4}+O\left(\left(\frac{1}{z}\right)^2\right),\0\\
 d=10:&&  \frac{i m^8 z^4 (1260\log (\frac{m^2}\pi)+1260 \log (-z)+1260 \gamma -4189)}{2381400 \pi ^5}\0\\
 &&-\frac{i m^8 z^3 (105 \log (\frac{m^2}\pi))+105 \log (-z)+105 \gamma -317)}{58800 \pi ^5}\0\\
 &&+\frac{i m^8 z^2 (30 \log (\frac{m^2}\pi)+30 \log (-z)+30 \gamma -77)}{14400 \pi ^5}\0\\
 &&-\frac{i m^8 z (3 \log (\frac{m^2}\pi)+3 \log (-z)+3 \gamma -5)}{3456 \pi ^5}+\frac{i m^8}{3072 \pi ^5}\0\\
&&+\frac{i m^8 (60 \log (-4 z)+77)}{1843200 \pi ^5 z}+O\left(\left(\frac{1}{z}\right)^2\right),\0\\
%  d=12&& \frac{i m^{10} z^5 (6930 \log (m)+3465 \log (-z)+3465 \gamma -12323-3465 \log (\pi %))}{115259760 \pi ^6}-\frac{i m^{10} z^4 (2520 \log (m)+1260 \log (-z)+1260 \gamma -4189-1260 \log %(\pi ))}{9525600 \pi ^6}+\frac{i m^{10} z^3 (210 \log (m)+105 \log (-z)+105 \gamma -317-105 \log (\pi %))}{470400 \pi ^6}-\frac{i m^{10} z^2 (60 \log (m)+30 \log (-z)+30 \gamma -77-30 \log (\pi ))}{172800 %\pi ^6}+\frac{i m^{10} z (6 \log (m)+3 \log (-z)+3 \gamma -5-3 \log (\pi ))}{55296 \pi ^6}-\frac{i %m^{10} (20 \log (-4 z)+29)}{14745600 \pi ^6 z}-\frac{i m^{10}}{61440 \pi %^6}+O\left(\left(\frac{1}{z}\right)^2\right)\,,\0\\
\ldots && \ldots  \0
\ee
As it is to be expected the leading behavior is $\sim m^{d-2}z^{d/2-1}\log (-z)$ corresponding to the UV behavior of the divergent part.  The presence of logarithms is to be interpreted as the consequence of running of parameters. In ordinary (interacting) gauge theories, once these logarithms are reabsorbed in the running parameters, the remaining polynomial behavior can itself be subtracted by proper counterterms leading to a well-behaved amplitude in the UV.

\subsubsection{d odd}

Let us discuss now the odd dimensional case.
In odd dimensions there is no divergent part in the $\epsilon$-expansion. 
For the IR we get the expansion
\be
d=3&&\sqrt{-m^2}\left(-\frac{2 z}{3 \pi }-\frac{4 z^2}{15 \pi }-\frac{6 z^3}{35 \pi }-\frac{8 z^4}{63
   \pi }-\frac{10 z^5}{99 \pi }+O\left(z^6\right)\right),\0\\
d=5&&\left(-m^2\right)^{3/2}\left(-\frac{z}{3 \pi ^2}+\frac{2 z^2}{15 \pi
   ^2}+\frac{z^3}{35 \pi ^2}+\frac{4 z^4}{315 \pi ^2}+\frac{5 z^5}{693 \pi
   ^2}+O\left(z^6\right)\right),\0\\
 d=7&&\left(-m^2\right)^{5/2}\left(-\frac{z}{18 \pi ^3}+\frac{z^2}{15 \pi ^3}-\frac{z^3}{70 \pi
   ^3}-\frac{2 z^4}{945 \pi ^3}-\frac{z^5}{1386 \pi ^3}+O\left(z^6\right)\right),\0\\
d=9&& \left(-m^2\right)^{7/2
   }\left(-\frac{z}{180
   \pi ^4}+\frac{z^2}{90 \pi ^4}-\frac{z^3}{140 \pi ^4}+\frac{z^4}{945 \pi
   ^4}+\frac{z^5}{8316 \pi ^4}+O\left(z^6\right)\right),\0\\
d=11&&\left(-m^2\right)^{9/2}\left(-\frac{z}{2520 \pi ^5}+\frac{z^2}{900
   \pi ^5}-\frac{z^3}{840 \pi ^5}+\frac{z^4}{1890 \pi ^5}-\frac{z^5}{16632 \pi
   ^5}+O\left(z^6\right)\right)\0\\
 \ldots&&\ldots\label{doddIRexp}
\ee
Again the dominating term is $\sim m^{d-2}z^2$.
The case $d=3$ is the one in which no term with positive power of $m$ shows up.

For the UV behavior we get
\be
d=3&&\sqrt{-m^2}\left(\frac{i \sqrt{z}}{4}+\frac{1}{4} i \sqrt{\frac{1}{z}}+O\left(\left(\frac{1}{z}\right)^{3/2}\right)\right),\0\\
d=5&&\left(-m^2\right)^{3/2}\left( -\frac{3 i z^{3/2}}{32 \pi }+\frac{i \sqrt{z}}{16 \pi }+\frac{i \sqrt{\frac{1}{z}}}{32 \pi
   }-\frac{1}{15 \pi ^2 z}+O\left(\left(\frac{1}{z}\right)^{3/2}\right)\right),\0\\
d=7&&\left(-m^2\right)^{5/2}\left(\frac{5 i z^{5/2}}{384 \pi
   ^2}-\frac{3 i z^{3/2}}{128 \pi ^2}+\frac{i \sqrt{z}}{128 \pi ^2}+\frac{i
   \sqrt{\frac{1}{z}}}{384 \pi ^2}-\frac{1}{210 \pi ^3 z}+O\left(\left(\frac{1}{z}\right)^{13/2}\right)\right),\0\\
d=9&&\left(-m^2\right)^{7/2
   }\Bigg{(}-\frac{7
   i z^{7/2}}{6144 \pi ^3}+\frac{5 i z^{5/2}}{1536 \pi ^3}-\frac{3 i z^{3/2}}{1024 \pi
   ^3}+\frac{i \sqrt{z}}{1536 \pi ^3}+\frac{i \sqrt{\frac{1}{z}}}{6144 \pi
   ^3}\0\\
	&&-\frac{1}{3780 \pi ^4 z}+O\left(\left(\frac{1}{z}\right)^{13/2}\right)\Bigg{)},\0\\
d=11&&\left(-m^2\right)^{9/2}\left(\frac{3 i z^{9/2}}{40960 \pi
   ^4}-\frac{7 i z^{7/2}}{24576 \pi ^4}+\frac{5 i z^{5/2}}{12288 \pi ^4}-\frac{i
   z^{3/2}}{4096 \pi ^4}+\frac{i \sqrt{z}}{24576 \pi ^4}\right.\0\\
   &&\left.+\frac{i
   \sqrt{\frac{1}{z}}}{122880 \pi ^4}-\frac{1}{83160 \pi ^5 z}+O\left(\left(\frac{1}{z}\right)^{13/2}\right)\right)\,,\0\\
  \ldots&&\ldots\label{doddUVexp}
\ee

\subsubsection{Spin 2}
 
For spin $s=2$ the two-point correlator in any dimension $d$ is
\be 
\tilde J_{\mu\mu\nu\nu}(k) &=& \frac{ 2^{2-d+\left\lfloor \frac{d}{2}\right\rfloor}\, i\,e^{-\frac{1}{2} i \pi  d} \pi ^{-d/2}
   \left(-m^2\right)^{d/2} \Gamma \left(1-\frac{d}{2}\right)}{d (d+1) k^2}\0\\
 &&  \left[ \left(-\left((d-1)
   k^2+8 m^2\right) \,
   {_2F_1}\left[1,-\frac{d}{2};\frac{1}{2};\frac{k^2}{4
   m^2}\right]-(d+1) k^2+8
   m^2\right)\pi_{\mu\nu}^2 \right. \0\\
&& \left.- \left(\left(4
   m^2-k^2\right) \,
   {_2F_1}\left[1,-\frac{d}{2};\frac{1}{2};\frac{k^2}{4
   m^2}\right]+(d+1) k^2-4 m^2\right)\pi_{\mu\mu}\pi_{\nu\nu}\right]\0\\
   &&+\frac{ 2^{2-d+\left\lfloor \frac{d}{2}\right\rfloor}\,i\, \pi ^{-d/2} m^d \Gamma
   \left(1-\frac{d}{2}\right)
   \left(\eta^2_{\mu\nu}+\eta_{\mu\mu}\eta_{\nu\nu}\right)}{d} \0
\ee 
We can check the Ward identity
\be
k^\mu\tilde J_{\mu\mu\nu\nu}(k)&=& - 2^{2-d+\left\lfloor \frac{d}{2}\right\rfloor}\,i\, \pi ^{-d/2} m^d \Gamma \left(-\frac{d}{2}\right)
   \left(k_{\nu}\eta_{\mu\nu}+k_{\mu}\eta_{\nu\nu}\right)
\ee
For spin $s=2$ the two-point correlator is transverse up to local counterterms.

\subsubsection{Spin 3}

For spin $s=3$ the two-point correlator in any dimension $d$ can be written as
\be
\tilde J_{\mu\mu\mu\nu\nu\nu}&=&-\frac{ 2^{1-d+\left\lfloor \frac{d}{2}\right\rfloor}\, i\,
  e^{-\frac{1}{2} i \pi  d} \pi ^{-d/2} \left(-m^2\right)^{d/2} (d+2)\Gamma \left(2-\frac{d}{2}\right)}{9 d (d+3) \left(d^4-5
   d^2+4\right) k^{4}} \pi_{\mu\nu}\0\\
   && \left[32 \,\left(-\left(k^2-4 m^2\right){}^2 \left(d k^2+12 m^2\right) \,
   {_2F_1}\left[1,-\frac{d}{2};-\frac{1}{2};\frac{k^2}{4 m^2}\right]\right.\right. \0\\
   &&\left.\left. +4 m^2 \left(48 m^4+8 (2 d-3) k^2 m^2+(3-2 d (d+1)) k^4\right)\right)\,\pi_{\mu\nu}^2  \right.\0\\
   && \left. - \,\left(4 k^4 m^2
   \left((d (9 d-44)+219) \, {_2F_1}\left[1,-\frac{d}{2};-\frac{1}{2};\frac{k^2}{4 m^2}\right]+d (95 d+68)-219\right) \right. \right. \0\\
   && \left. \left. +k^6 \left((d ((d-6) d+3)-54) \,
   {_2F_1}\left[1,-\frac{d}{2};-\frac{1}{2};\frac{k^2}{4 m^2}\right]-18 (d-1) (d+1) (d+3)\right) \right. \right. \0\\
   &&\left. \left. +9216 m^6 \left(\, {_2F_1}\left[1,-\frac{d}{2};-\frac{1}{2};\frac{k^2}{4
   m^2}\right]-1\right)\right.\right. \0\\
   &&\left.\left. +128 k^2 m^4 \left((5 d-39) \, {_2F_1}\left[1,-\frac{d}{2};-\frac{1}{2};\frac{k^2}{4 m^2}\right]-23 d+39\right)\right)\,\pi_{\mu\mu}\pi_{\nu\nu}\right]\0\\
   && -\frac{ 2^{1-d+\left\lfloor \frac{d}{2}\right\rfloor} \, i\,e^{-\frac{1}{2} i \pi  d} \pi ^{-d/2} \left(-m^2\right)^{d/2} \Gamma \left(2-\frac{d}{2}\right)}{9 d \left(d^2-4\right)}\0\\
   &&  \left[\eta_{\mu\nu}
   \left[\eta_{\nu\nu} \left(2 \eta_{\mu\mu}\left(208 m^2-9 (d+2) k^2\right)+21 (d+2) k_{\mu}k_{\mu}\right) +256 m^2\eta_{\mu\nu}\eta_{\mu\nu}\right]\right.\0\\
   &&\left. +21 (d+2) \eta_{\mu\mu} \eta_{\mu\nu} k_{\nu}k_{\nu}+18 (d+2)
   \eta_{\mu\mu} \eta_{\nu\nu} k_{\mu}k_{\nu}\right]
\ee
For spin $s=3$ the two-point correlator {is conserved up to local counterterms 
\be
k^\mu\tilde J_{\mu\mu\mu\nu\nu\nu}&=&-\frac{ 2^{1-d+\left\lfloor \frac{d}{2}\right\rfloor} \, i\,e^{-\frac{1}{2} i \pi  d} \pi ^{-d/2} \left(-m^2\right)^{d/2} \Gamma \left(2-\frac{d}{2}\right)}{9 d \left(d^2-4\right)}\0\\
&& \left[k_{\nu}
   \left(\eta_{\nu\nu}\left(57 (d+2) k_{\mu}k_{\mu}+416 m^2 \eta_{\mu\mu}\right) +768 m^2\eta_{\mu\nu}\eta_{\mu\nu}\right) \right. \0\\
 &&  \left. +2 \eta_{\mu\nu} \eta_{\nu\nu}k_{\mu} \left(3 (d+2) k^2+416 m^2\right)+21
   (d+2) \eta_{\mu\mu} k_{\nu}k_{\nu}k_{\nu}\right. \0\\
 &&  \left. +42 (d+2) \eta_{\mu\nu}k_{\nu}k_{\nu}\right]
\ee
%%%%%

\subsection{Scalar model}

Let us compute the even part of two point correlator for a scalar in $d$ dimensions for spin $s$
 \be 
 \tilde J_{\mu_1 \ldots \mu_s\nu_1\ldots\nu_s}(k)& =& 
\int\frac
{d^dp}{(2\pi)^d}\,
\frac i{p^2-m^2} V_{\mu_1\ldots\mu_s} 
\frac i{(p-k)^2-m^2} V_{\nu_1\ldots\nu_s}
\ee
where the vertex for an incoming scalar with momentum $p$ and outgoing scalar with momentum $p'$ and an outgoing spin-$s$ field with momentum $k$
is
\be
V_{\mu_1\ldots\mu_s} = { i } (p+p')_{\mu_1} \ldots (p+p')_{\mu_s} \delta^{(d)}(p-p'-k) \label{Vsst1}
\ee

\subsubsection{Scalar model - massive case}

Let us compute the two point correlator for the massive scalar in any dimension $d$ for spin $s=1$
\be
\tilde J_{\mu\nu}(k)&=&\frac{ 2^{2-d}\, i\, e^{-\frac{1}{2} i \pi  d} \pi ^{-d/2}
   \left(-m^2\right)^{d/2} \Gamma \left(2-\frac{d}{2}\right)}{(d-2) k^2 m^2}\0\\
   &&
   \left(\left(k^2\eta_{\mu\nu}-k_\mu k_\nu\right) \,\,
   {_2F_1}\left[1,1-\frac{d}{2};\frac{3}{2};\frac{k^2}{4
   m^2}\right]+k_\mu k_\nu\right)
\ee
$\tilde J_{\mu\nu}$ is not conserved
\be
k^\mu \tilde J_{\mu\nu}(k)=\frac{ 2^{-d} \, i\, e^{-\frac{1}{2} i \pi  d} \pi ^{-d/2}
   \left(-m^2\right)^{d/2} d\,\Gamma \left(-\frac{d}{2}\right)}{m^2}k_\nu
\ee

To make $\tilde J_{\mu\nu}(k)$ conserved let us add a local counterterm with an arbitrary constant $a$
\be
\tilde J_{\mu\nu}(k)&=&\frac{ 2^{2-d} \, i\,
e^{-\frac{1}{2} i \pi  d} \pi ^{-d/2}\left(-m^2\right)^{d/2} \Gamma \left(2-\frac{d}{2}\right)}{(d-2) k^2 m^2}\0\\
&& \left(a k^2\eta_{\mu\nu}+k_\mu k_\nu+\left(k^2\eta_{\mu\nu}-k_\mu k_\nu\right) \,
{_2F_1}\left[1,1-\frac{d}{2};\frac{3}{2};\frac{k^2}{4m^2}\right]\right)
\ee
We get conservation for $a=-1$. The conserved 2pt is
\be 
\tilde J_{\mu\nu}(k) &=& \frac{ 2^{2-d}\, i\, e^{-\frac{1}{2} i \pi  d} \pi ^{-d/2}
   \left(-m^2\right)^{d/2} \Gamma \left(2-\frac{d}{2}\right)}{(d-2)  m^2} \pi_{\mu\nu}\0\\
   && \left(\,
   {_2F_1}\left[1,1-\frac{d}{2};\frac{3}{2};\frac{k^2}{4
   m^2}\right]-1\right)
\ee

For massive scalar for spin $s=2$ we get

\be
\tilde J_{\mu\mu\nu\nu}(k)&=& 2^{2-d}\, i\, e^{-\frac{1}{2} i \pi  d} \pi ^{-d/2} \left(-m^2\right)^{d/2} \Gamma
   \left(-\frac{d}{2}\right) \left(2\pi_{\mu\nu}^2+\pi_{\mu\mu}\pi_{\nu\nu}\right)\0\\
   && \left(\,
   {_2F_1}\left[1,-\frac{d}{2};\frac{3}{2};\frac{k^2}{4 m^2}\right]-1\right)\0\\
   && {+ } \frac{ 2^{-d} \, i\,e^{-\frac{1}{2} i \pi  d} \pi ^{-d/2} (-m^2)^{d/2}  \Gamma
   \left(-\frac{d}{2}\right)}{k^2 m^2}\0\\
   && \left(d\, k_\mu k_\mu k_\nu k_\nu+4 k^2 m^2 \left(2
   \eta_{\mu\nu}\eta_{\mu\nu}+\eta_{\mu\mu}\eta_{\nu\nu}\right)\right)
\ee

$\tilde J_{\mu\mu\nu\nu}(k)$ is not conserved
\be
k^{\mu}\tilde J_{\mu\mu\nu\nu}(k)&=&\frac{ 2^{1-d} \, i\,e^{-\frac{1}{2} i \pi  d} \pi ^{-d/2} (-m^2)^{d/2}  \Gamma\left(-\frac{d}{2}\right)}{m^2}\0\\
&& \left(k_{\mu} \left(d\,
   k_{\nu} k_{\nu}+4 m^2 \eta_{\nu\nu}\right)+8 m^2
   \eta_{\mu\nu} k_{\nu}\right)
\ee

For spin $s=3$ we have

\be
\tilde J_{\mu\mu\mu\nu\nu\nu}(k)&=&-3 \cdot 2^{1-d}\, i\, e^{-\frac{1}{2} i \pi  d} \pi ^{-d/2} \left(-m^2\right)^{d/2} \Gamma
   \left(-\frac{d}{2}-1\right) \pi_{\mu\nu} \left(2\pi_{\mu\nu}^2+3\pi_{\mu\mu}\pi_{\nu\nu}\right)\0\\
   && \left(\,
   8m^2\left({_2F_1}\left[2,-\frac{d}{2}-1;\frac{3}{2};\frac{k^2}{4 m^2}\right]-1\right)+(d+2)k^2{_2F_1}\left[1,-\frac{d}{2};\frac{3}{2};\frac{k^2}{4 m^2}\right]\right)\0\\
   &&-\frac{i 2^{-d-1} e^{-\frac{1}{2} i \pi  d} \pi ^{-d/2} (-m^2)^{\frac{d}2} \Gamma
 \left(-\frac{d}{2}-1\right)}{k^4 m^2}\0\\
   && \left[(d+2)k^2\, k_\mu k_\nu(k_\mu k_\mu(d\,k_\nu k_\nu+12m^2\eta_{\nu\nu})+12m^2\eta_{\mu\mu} k_\nu k_\nu+36m^2\eta_{\mu\nu }k_\mu k_\nu)\right.\0\\
&&\left.-20(d+2)m^2 k_\mu k_\mu k_\mu k_\nu k_\nu k_\nu+48k^4 m^4\eta_{\mu\nu}(2\eta_{\mu\nu}\eta_{\mu\nu}+3\eta_{\mu\mu}\eta_{\nu\nu}))\right]
\ee

$\tilde J_{\mu\mu\mu\nu\nu\nu}(k)$ is not conserved
\be
k^{\mu}\tilde J_{\mu\mu\nu\nu}(k)&=&\frac{3\cdot 2^{-d-1} \, i\,e^{-\frac{1}{2} i \pi  d} \pi ^{-d/2} (-m^2)^{\frac{d}2} \Gamma\left(-\frac{d}{2}-1\right)}{m^2}\0\\
&& \left[24m^2\eta_{\mu\nu} k_\mu((d+2)k_\nu k_\nu+4m^2\eta_{\nu\nu})\right.\0\\
&&\left. +4m^2 k_\nu(\eta_{\mu\mu}((d+2)k_\nu k_\nu+12m^2 \eta_{\nu\nu})+24m^2\eta_{\mu\nu}\eta_{\mu\nu})\right.\0\\
&&\left. +(d+2)k_\mu k_\mu k_\nu(d \,k_\nu k_\nu+12m^2\eta_{\nu\nu})\right]
\ee

\subsubsection{Concluding remark}

In this section we have produced two-point correlator formulas for spin 1,2 and 3 in any dimension. In all the cases where it is possible to make a comparison between the results obtained in this section and the previous ones (spin 1 in d=3,4,5,6) the results coincide\footnote{In making the comparison one should not  forget
to correct the results of section 11.1.2 and 11.1.3 for the ${\rm Tr}(1)$ factor as explained before eq.(\ref{Jmunured} ), i.e.
by dividing the $d=2,3$ results by 2, multiplying the $d=6,7$ ones by 2, etc.}. However they do {\it only} if we subtract the infinite and finite IR terms from the effective action. In other words this confirms that only the difference between the UV and the IR can have a physical meaning.

\section{Conclusion}
\label{Conclusion}

We have seen a large number of examples that the one-loop effective action of a free
massive model coupled to external sources contains complete information about 
the (classical) equations of motion of the sources. In this paper we have considered only the
two-point functions and so the relevant information involve the linearized equations of motion.
Moreover we have considered only completely symmetric bosonic external sources. Within these limitations
we have produced overwhelming evidence that our previous statement is correct. We have considered
both a free scalar model and a free fermion model in different dimensions, and shown that in all cases
the two-point functions of conserved currents are built out of the differential operators
which define the linearized (Fronsdal) equations of motion of the fields that couple to the currents. 
There is no doubt that such free field theories know about the dynamics of the fields that can couple 
to them (via a conserved current)\footnote{The limitation to free field theories does not seem to be essential provided the currents
are conserved, but of course explicit calculations are far more complicated in the case of interacting theories.}.

At this stage a specification is in order. Our intent in this paper was to show the
universal appearance of non-local Fronsdal (as well as Maxwell and EH) linearized eom in the one-loop effective actions of a free scalar and boson field coupled to external currents, while postponing other subtler questions to future research. In particular we did not tackle the problem raised by \cite{Francia1,Francia2}, concerning the form of the Fronsdal equation that guarantees the right propagator for the relevant higher spin field. In order to do that one must first of all specify to what equations one refers to, for we have seen that in the IR and UV limits
of the OLEA's the conserved structures very often are different, and different from the various tomographic sections, although for spin $s$ they are all characterized by the presence of the leading (\ref{A0}) term (the scalar model is in this sense a particular, though less interesting, case, because the conserved structures are always the same for given spin). This part of our research is work in progress, see \cite{BCDGLS1}.

The results of this paper opens a new research territory. Beside the just mentioned problem, we would like to  know whether the above results
extend to other external sources, fermionic fields as well as not completely symmetric fields. The next question is interaction, which requires analyzing three-point functions. In this context interactions have been
considered for the simplest cases (spin 1 and 2) in 3d in \cite{BCLPS}. From three-point functions one expects 
to find information about the consistency of the (field or fields) interaction with the source field symmetry. For instance,
for spin 1 with gauge symmetry, for spin 2 with diffeomorphisms. For higher spins we do not know, in general, neither the interaction 
nor the full form of the symmetry transformations. But knowing the three-point functions may be the key to
constructing both. There are anyhow some exceptions to our ignorance in this field (higher spin theories in 3d, or Vasiliev's higher spin 
theory in $AdS_4$, or string field theory). One can hopefully use this knowledge to test the approach suggested here. 

If our conjecture is correct, that is if the analysis of three-point correlators in theories coupled to external sources
confirms their consistency with the dynamics of the latter, as we believe, an obvious question comes next: what does this mean?
The correspondence between one free field theory and higher (or low) spin theories is not a type of duality we are familiar with, like AdS/CFT.
First of all it concerns models in the same dimension. Secondly, from one free theory we retrieve knowledge about (infinite) many theories.
So the correspondence would be one to (infinite) many. And this is clearly not satisfactory. The results of this paper points 
rather toward the possibility of
a correspondence between theories with infinite many fields. If, say, a starting free (or interacting) theory knows about the
dynamics of other fields, why shouldn't the latter be included with the initial one in a unique theory?  Arguing this way one is  led
to a (for the time being, generic) concept of {\it involutive theory}: a theory is {\it involutive} if it includes all the fields it is able to couple with (in the OLEA) while preserving a fundamental symmetry.

 {A good playground to test this concept could be string field theory (SFT)}.
Such a theory is formulated in terms of a basic string field $\Phi$. The latter, in the field theory regime, is a superposition of Fock space states, each with a coefficient given by a suitable ordinary spacetime field. Restricting ourselves to bosonic
SFT, the action formulated by Witten is well known, and is given by the formula below with $\Psi$ replaced by $\Phi$. Analyzing
it in the spirit of this paper amounts to studying the theory
\be
S= \frac 1{2g_o} \int \left( \Phi \ast Q \Phi + \frac 23  \Phi \ast \Phi \ast \Psi\right)\label{SFT}
\ee 
where the first piece is the free SFT and the second is the simplest interaction with the source term ($\Psi$ is the source string field).
The first piece is invariant under the BRST transformation $\delta \Phi=Q\Lambda$. The second term carries the invariance under
 $\delta \Psi=Q\Lambda$ provided that $\Phi$ is on shell, i.e. $Q\Phi=0$. 
 {This mimics what we have done previously for simple field theory models.}

\bigskip

%%%%%%%%%%%%%%%%%%%%%%%%%
\noindent
{\bf \large Acknowledgements}%%%
%%%%%%%%%%%%%%%%%%%%%%%%%

\bigskip

\noindent
We would like to thank Ivica Smoli{\'c} for his collaboration 
in the early stage of this research.
One of us (L.B.) would like to thank Dario Francia and Mikhail Vasiliev
for useful discussions. The research has been supported by the Croatian
Science Foundation under the project No.~8946 and by the University of
Rijeka under the research support No.~13.12.1.4.05.

\vspace{15pt}

\appendix
\section*{Appendix}

\section{Proof that a conserved structure can be written in terms of products
of $\pi$ alone}
\label{sec:proof}

By induction in steps of 2. In the lowest case (spin 1), the most general
Lorentz covariant (dimensionless) conserved even structure can be written 
in terms of $\eta_{\mu\nu}$ and $\frac {k_\mu k_\nu}{k^2}$.
Imposing conservation the result is $\sim \eta_{\mu\nu}-\frac {k_\mu
k_\nu}{k^2}=\,\pi_{\mu\nu}$. In the same way
one can prove the property for the case $s=2$. Now we suppose that
the proposition is true for $s$. So it is true for the combination   
$T^{(s)}(k\!\cdot\!n_1^s\!\cdot\!n_2^s)=\tilde E^{(s)}(k)=  \sum_{l=0}^{[s/2]}
a_l \tilde
A^{(s)}_l$ (see above), 
meaning that $k^\mu \frac {\partial}{\partial n_1^\mu} T^{(s)}=0$.
In order to construct $T^{(s+1)}$ we can multiply  $T^{(s)}$ 
by $(n_1n_2)$ or $\frac {(n_1k)(n_2k)}{k^2}$ or multiplying  $T^{(s-1)}$ by
$(n_1n_1)  (n_2n_2), \frac {(n_1k)(n_1k)}{k^2}(n_2n_2), (n_1n_1)\frac
{(n_2k)(n_2k)}{k^2}$ 
or by $ \frac {(n_1k)(n_1k)}{k^2} \frac {(n_2k)(n_2k)}{k^2}$,
because the construction is in steps or 2.
So we can have only
\be
T^{(s+1)}&=&a_1 (n_1n_2)T^{(s)}+ a_2  (n_1n_1)  (n_2n_2) T^{(s-1)} +b_1
\frac{(n_1k)(n_2k)}{k^2}T^{(s)}\0\\
&&+ b_2 \frac {(n_1k)(n_1k)}{k^2}(n_2n_2) T^{(s-1)}+b_3 (n_1n_1)\frac
{(n_2k)(n_2k)}{k^2} T^{(s-1)}\0\\
&&+b_4 \frac {(n_1k)(n_1k)}{k^2} \frac {(n_2k)(n_2k)}{k^2}
T^{(s-1)}\label{combi}
\ee
Now applying $k^\mu \frac {\partial}{\partial n_1^\mu}$ to this expression we
find that conservation 
requires $a_1=-b_1, a_2=-b_2=-b_3=b_4$. So that (\ref{combi})
becomes
\be
T^{(s+1)}= a (n_1 , \pi^{(k)}\!\cdot \!n_2) T^{(s)}+ b (n_1,
\pi^{(k)}\!\cdot \!n_1)  (n_2, \pi^{(k)}\!\cdot \!n_2)
T^{(s-1)}\label{Ts+1}
\ee
with arbitrary $a$ and $b$.

\section{Massless models}
\label{sec:massless}

In this Appendix we consider the massless case both for the scalar and the fermion models, i.e. we
set $m=0$ in their action, and derive the
relevant two-point functions for several tensorial currents in any dimension. These results are based 
on the scalar integral (\ref{J2scalarm=0}).

The results we report below have to be compared with the results obtained in the section 6-9, precisely with
the ${\cal O}_{UV}(m^0)-{\cal O}_{IR}(m^0)-{\cal O}_{IR}(\log(m))$ terms therein. It can be easily checked that
for odd $d$ the results coincide exactly, as far as the even parity part of the correlators is concerned, while the massless model approach is unable to reproduce the odd parity part (at least perturbatively). In the even dimensional case the
results of the two approaches do not, in general, coincide. Only the 
terms proportional to $\log\left(- k^2\right)$ are the same in the two approaches. These differences are due to the lack of IR regularization in the massless model approach.

For conciseness in this Appendix we use a simplified notation, taken from the literature on higher spin fields:
the same repeated subscript, say $\mu\ldots \mu$ repeated $s$ times, stand for $s$ completely symmetrized labels.

In this appendix we also construct traceless two-point correlators.
As a matter of fact, in this paper we are only marginally interested in zero trace currents. But the tracelessness condition may be relevant for further developments.

\subsection{Massless fermion model}

Let us start with spin 1 case for  massless fermions. The two point correlator is
 \be \label{J1F1}
\tilde J_{\mu\nu}(k) &=& - \int \frac{d^dp}{(2\pi)^d} {\rm Tr}
\left(\gamma_\nu\, \frac {{ 1 }}{\slashed{p}}\,\gamma_\mu   
\frac {{ 1 }}{\slashed{p}-\slashed{k} }\right) \0\\
&=& {- }{\rm Tr}(1)
\int\frac{d^dp}{(2\pi)^d}\frac {p_\nu(p-k)_\mu
 + p_\mu(p-k)_\nu -p\!\cdot\!(p-k)\eta_{\mu\nu}}{p^2(p-k)^2}
\ee
where ${\rm Tr}(1)=2^{\left\lfloor \frac{d}{2}\right\rfloor}$. Using Davydychev methods, the two point correlator for spin s=1 and any dimension $d$ reads
\be 
\tilde J_{\mu\nu}={- }\frac{2^{2-2d+\left\lfloor \frac{d}{2}\right\rfloor}  \pi ^{\frac 32 -\frac{d}{2}} \left(k^2\right)^{\frac d2-1} (d-2)}{\left(-1+e^{i \pi  d}\right) \Gamma (\frac{d+1}2)} \pi_{\mu\nu} \label{J1F2}
\ee 
where $\pi_{\mu\nu} $ is the usual projector.

In a similar way for spin 2 in the massless case we get
\be 
\tilde J_{\mu\mu\nu\nu}=\frac{2^{1-2 d+\left\lfloor \frac{d}{2}\right\rfloor} \pi ^{\frac{3}{2}-\frac{d}{2}}
   \left(k^2\right)^{\frac d2}(d-1) }{\left(-1+e^{i \pi d}\right) \Gamma \left(\frac{d+3}{2}\right)}\left(\pi^2_{\mu\nu}-\frac{1}{d-1}\pi_{\mu\mu}\pi_{\nu\nu}\right)\label{J2F1}
\ee
The two-point correlator in $s=2$ case is traceless for any $d$.

For spin 3 we obtain
\be
\tilde J_{\mu\mu\mu\nu\nu\nu}&=&-\frac{2^{-2-2d+\left\lfloor \frac{d}{2}\right\rfloor} \pi ^{\frac{3}{2}-\frac{d}{2}} \left(k^2\right)^{\frac d2+ 1} }{9 \left(-1+e^{i \pi  d}\right) \Gamma \left(\frac{d+5}{2}\right)} \pi_{\mu\nu}\0\\
&& \left(32 \,d\, \pi_{\mu\nu}^2+(d ((d-6)d+3)-54) \pi_{\mu\mu}\pi_{\nu\nu}\right)\label{J3F1}
\ee
We can check if this expression is traceless
\be
\eta^{\mu\mu}\tilde J_{\mu\mu\mu\nu\nu\nu}&=&-\frac{2^{-1-2 d+\left\lfloor \frac{d}{2}\right\rfloor} (d-3)^2 \left(d^2+d-6\right) \pi ^{\frac{3}{2}-\frac{d}{2}} \left(k^2\right){}^{\frac{d}{2}+1}}{9
   \left(-1+e^{i \pi  d}\right) \Gamma \left(\frac{d+5}{2}\right)} \pi_{\mu\nu}\pi_{\nu\nu}\label{J3F2}
\ee   
For dimension $d=3$, the two-point correlator $J_{\mu\mu\mu\nu\nu\nu}$ is traceless. To obtain tracelessness for other dimensions we can add local counterterms to the two point function
\be
\tilde J_{\mu\mu\mu\nu\nu\nu}&=&-\frac{ 2^{-2-2d+\left\lfloor \frac{d}{2}\right\rfloor} \pi ^{\frac{3}{2}-\frac{d}{2}} \left(k^2\right)^{\frac d2 +1} }{9 \left(-1+e^{i \pi  d}\right) \Gamma \left(\frac{d+5}{2}\right)} \pi_{\mu\nu}\0\\
&& \left((32 \,d+a_1)\, \pi_{\mu\nu}^2+(d ((d-6)d+3)-54+a_2) \pi_{\mu\mu}\pi_{\nu\nu}\right)\label{J3F3}
\ee
The counterterms are local only for even $d$. In this case it is easy to verify that
for the full expression to be traceless we must have
\be
a_1=\frac{1}{3} \left(-a_2 (d+1)-\left(d^2+d-6\right) (d-3)^2\right)\0
\ee
The traceless 2pt function is therefore
\be
\tilde J_{\mu\mu\mu\nu\nu\nu}&=&-\frac{2^{-2-2d+\left\lfloor \frac{d}{2}\right\rfloor} \pi ^{\frac{3}{2}-\frac{d}{2}} \left(k^2\right)^{\frac d2+1}(1+d)(-54+3d-6d^2+d^3+a_2)}{9 \left(-1+e^{i \pi  d}\right) \Gamma \left(\frac{d+5}{2}\right)}\0\\
&& \pi_{\mu\nu} \left( \pi_{\mu\nu}^2-\frac{3}{(d+1)}\pi_{\mu\mu}\pi_{\nu\nu}\right)\label{J3F4}
\ee

For spin $s=4$ the two-point correlator is
\be
\tilde J_{\mu\mu\mu\mu\nu\nu\nu\nu}&=&\frac{2^{-3-2 d+\left\lfloor \frac{d}{2}\right\rfloor} \pi ^{\frac{3}{2}-\frac{d}{2}} \left(k^2\right)^{\frac d2+2}}{9 \left(-1+e^{i \pi  d}\right)  \Gamma \left(\frac{d+7}{2}\right)} 
\left((24 (d+1)) \pi_{\mu\nu}^4\right. \0\\
&&\left. +((d-7) (d(d+2)+9))\pi_{\mu\nu}^2 \pi_{\mu\mu} \pi_{\nu\nu}+((9-(d-4) d))
   \pi_{\mu\mu}^2 \pi_{\nu\nu}^2\right)\label{J4F1}
\ee
Again, unless $d=3$, the expression for $\tilde J_{\mu\mu\mu\mu\nu\nu\nu\nu}$ is not traceless
For even $d>3$ we can add local counterterms
\be
\tilde J_{\mu\mu\mu\mu\nu\nu\nu\nu}&=&\frac{2^{-3-2 d+\left\lfloor \frac{d}{2}\right\rfloor} \pi ^{\frac{3}{2}-\frac{d}{2}} \left(k^2\right)^{\frac d2+2}}{9 \left(-1+e^{i \pi  d}\right) \Gamma \left(\frac{d+7}{2}\right)} 
\left((24 (d+1)+a_1) \pi_{\mu\nu}^4\right.\label{J4F2}\\
&&\left. +((d-7) (d(d+2)+9)+a_2)\pi_{\mu\nu}^2 \pi_{\mu\mu} \pi_{\nu\nu}+((9-(d-4) d)+a_3)
   \pi_{\mu\mu}^2 \pi_{\nu\nu}^2\right)\0
\ee
and we obtain  a traceless 2pt correlator for
\be
a_1=-(1/3) (1 + d) ((-3 + d)^2 (5 + d) - a_3(3 + d) ),\quad\quad
a_2=(d-3)^2 (d+5)-2 a_3 (d+1)\0
\ee
The traceless two point correlator is now
\be
\tilde J_{\mu\mu\mu\mu\nu\nu\nu\nu}&=&-\frac{2^{-3-2 d+\left\lfloor \frac{d}{2}\right\rfloor} \pi ^{\frac{3}{2}-\frac{d}{2}} \left(-a_3+d^2-4
   d-9\right) \left(k^2\right)^{\frac d2+2}(d+1) ( d+3)}{27
   \left(-1+e^{i \pi  d}\right) \Gamma
   \left(\frac{d+7}{2}\right)}\0\\
&& \left(
   \pi _{\mu \nu }^4-\frac{6}{(d+3)} \pi _{\mu \mu } \pi _{\mu \nu }^2 \pi
   _{\nu \nu }+\frac{3}{(d+1) ( d+3)} \pi _{\mu \mu }^2 \pi _{\nu \nu }^2\right)\label{J4F3}
\ee

In general, conserved and traceless 2pt function for spin $s$ in any (even) dimension is proportional to
\be
\tilde J_{\mu_1\dots\mu_s\nu_1\dots\nu_s} &\sim& \sum_{l=0}^{\left\lfloor \frac{s}{2}\right\rfloor} \frac {(-1)^l}{2^{2l}l!} \frac {s!}{(s-2l)!} \frac{\Gamma\left(s+\frac{d-3}2-l\right)}{\Gamma\left(s+\frac {d-3}2\right)}\pi_{\mu\nu}^{s-2l} (\pi_{\mu\mu}\pi_{\nu\nu})^l\label{JsF1}
\ee
We can write the sum as
\be
\tilde J_{\mu_1\dots\mu_s\nu_1\dots\nu_s} &\sim& \,{_2F_1}\left(\frac{1-s}{2},-\frac{s}{2};\frac{5-d-2s}{2};\frac{\pi_{\mu\mu}\pi_{\nu\nu}}{\pi_{\mu\nu}^2}\right)\pi_{\mu\nu}^s\label{JsF2}
\ee

\subsection{Massless scalar model }

Let us compute the  two point correlator for a scalar in $d$ dimensions for spin $s=1$. Using Davydychev methods we get:
\be
\tilde J_{\mu\nu}&=&-\frac{2^{3-2d} \,\pi ^{\frac 32-\frac d2}\,
   \left(k^2\right)^{\frac{d}{2}-1}}{(-1+e^{i \pi d})\,\Gamma (\frac{d+1}{2})}\,\pi_{\mu\nu}  \label{J1S1}
\ee
The expression for $\tilde J_{\mu\nu}$ is traceless and conserved.

For spin 2 we get
\be
\tilde J_{\mu\mu\nu\nu}&=&\frac{2^{2-2d} \,\pi ^{\frac 32-\frac d2}\, \left(k^2\right)^{\frac d2}}{(-1+e^{i \pi d})\,\Gamma (\frac{d+3}{2})} \left(2\pi_{\mu\nu}^2+\pi_{\mu\mu}\pi_{\nu\nu}\right)   \label{J2S1}
\ee
which is conserved, but not traceless:
\be
\eta^{\mu\mu}\tilde J_{\mu\mu\nu\nu}&=&\frac{2^{3-2d} \,\pi ^{\frac 32-\frac d2}\, \left(k^2\right)^{\frac d2}}{(-1+e^{i \pi d})\,\Gamma (\frac{d+3}{2})} \left(d+1\right) \pi_{\nu\nu}  \label{J2S2}
\ee
Let us consider the counterterms $\tilde J_{\mu\mu\nu\nu}$ 
\be
\tilde J_{\mu\mu\nu\nu}&=&\frac{2^{2-2d} \,\pi ^{\frac 32-\frac d2}\, \left(k^2\right)^{\frac d2}}{(-1+e^{i \pi d})\,\Gamma (\frac{d+3}{2})} \left((2+a_1)\pi_{\mu\nu}^2+(1+a_2)\pi_{\mu\mu}\pi_{\nu\nu}\right)  \0\\
   && \label{J2S3}
\ee
which is local for even $d$. Adding it to $\tilde J_{\mu\mu\nu\nu}$, the  trace becomes 
\be
\eta^{\mu\mu}\tilde J_{\mu\mu\nu\nu}&=&\frac{2^{3-2d} \,\pi ^{\frac 32-\frac d2}\, \left(k^2\right)^{\frac d2}}{(-1+e^{i \pi d})\,\Gamma (\frac{d+3}{2})} \left(d+1+a_1+a_2(d-1)\right) \pi_{\nu\nu}^2 \0\\
   && \label{J2S4}
\ee
If we choose
\be
a_1=-(1+d)-a_2(d-1)\0
\ee
we get a traceless $\tilde J_{\mu\mu\nu\nu}$ 
\be
\tilde J_{\mu\mu\nu\nu}&=&{- }\frac{2^{2-2d} \,\pi ^{\frac 32-\frac d2}\, \left(k^2\right)^{\frac d2}}{(-1+e^{i \pi d})\,\Gamma (\frac{d+3}{2})}(1+a_2)(d-1) \left(\pi_{\mu\nu}^2-\frac{1}{d-1}\pi_{\mu\mu}\pi_{\nu\nu}\right)   \label{J2S5}
\ee
This is possible only for even $d$.

For spin 3 we have
\be
\tilde J_{\mu\mu\mu\nu\nu\nu}&=&-\frac{3\cdot 2^{1-2d} \,\pi ^{\frac 32-\frac d2}\, \left(k^2\right)^{\frac d2+1}}{(-1+e^{i \pi d})\,\Gamma (\frac{d+5}{2})}\pi_{\mu\nu} \left(2\pi_{\mu\nu}^2+3\pi_{\mu\mu}\pi_{\nu\nu}\right)  \label{J3S1} 
\ee
This expression is transverse but not traceless
\be
\eta^{\mu\mu}\tilde J_{\mu\mu\mu\nu\nu\nu}&=&-\frac{9\cdot 2^{2-2d} \,\pi ^{\frac 32-\frac d2}\, \left(k^2\right)^{\frac d2+1}}{(-1+e^{i \pi d})\,\Gamma (\frac{d+5}{2})}\left(3+d\right)\pi_{\mu\nu}\pi_{\nu\nu} \label{J3S2}
\ee
In even $d$ we can add local counterterms and obtain
\be
\tilde J_{\mu\mu\mu\nu\nu\nu }&=&-\frac{3\cdot 2^{1-2d} \,\pi ^{\frac 32-\frac d2}\, \left(k^2\right)^{\frac d2+1}}{(-1+e^{i \pi d})\,\Gamma (\frac{d+5}{2})}\pi_{\mu\nu} \left((2+a_1)\pi_{\mu\nu}^2+(3+a_2)\pi_{\mu\mu}\pi_{\nu\nu}\right)   \label{J3S3}
\ee
To make $\tilde J_{\mu\mu\mu\nu\nu\nu }$ traceless we must have
\be
a_1=-\frac 13 (3(3+d)+a_2(1+d))\0
\ee
A traceless $\tilde J_{\mu\mu\mu \nu\nu\nu }$ can now be written as
\be
\tilde J_{\mu\mu\mu\nu\nu\nu }&=&\frac{2^{1-2d} \,\pi ^{\frac 32-\frac d2}\, \left(k^2\right)^{\frac d2+1}}{(-1+e^{i \pi d})\,\Gamma (\frac{d+5}{2})} (3+a_2)(d+1)\pi_{\mu\nu} \left(\pi_{\mu\nu}^2-\frac{3}{d+1}\pi_{\mu\mu}\pi_{\nu\nu}\right) \label{J3S4} 
\ee

For spin 4 the two-point correlator is
\be
\tilde J_{\mu\mu\mu\mu\nu\nu\nu\nu }&=&-\frac{3\cdot 2^{-2d} \,\pi ^{\frac 32-\frac d2}\, \left(k^2\right)^{\frac d2+2}}{(-1+e^{i \pi d})\,\Gamma (\frac{d+7}{2})} \left(8\pi_{\mu\nu}^4+24\pi_{\mu\nu}^2\pi_{\mu\mu}\pi_{\nu\nu}+3\pi^2_{\mu\mu}\pi^2_{\nu\nu}\right)\label{J4S1}  
\ee
The expression for $\tilde J_{\mu_1\dots\mu_4\nu_1\dots\nu_4}$ is not traceless
\be
\eta^{\mu\mu}\tilde J_{\mu\mu\mu\mu\nu\nu\nu\nu }&=&-\frac{9\cdot 2^{2-2d} \,\pi ^{\frac 32-\frac d2}\, \left(k^2\right)^{\frac d2+2}}{(-1+e^{i \pi d})\,\Gamma (\frac{d+7}{2})}(5+d) \pi_{\nu\nu}\left(4\pi_{\mu\nu}^2+\pi_{\mu\mu}\pi_{\nu\nu}\right)\label{trJ4S1}  
\ee
Again in even $d$ we can add local counterterms
\be
\tilde J_{ \mu\mu\mu\mu\nu\nu\nu\nu}&=&-\frac{3\cdot 2^{-2d} \,\pi ^{\frac 32-\frac d2}\, \left(k^2\right)^{\frac d2+2}}{(-1+e^{i \pi d})\,\Gamma (\frac{d+7}{2})}\0\\
&& \cdot\left((8+a_1)\pi_{\mu\nu}^4+(24+a_2)\pi_{\mu\nu}^2\pi_{\mu\mu}\pi_{\nu\nu}+(3+a_3)\pi^2_{\mu\mu}\pi^2_{\nu\nu}\right) \label{J4S2} 
\ee
The tracelessness condition is now
\be
a_1=-5+a_3+\frac 43 d(3+a_3)+\frac 13 d^2 (3+a_3), \quad\quad 
a_2=-2(15+3d+(1+d)a_3)\0
\ee
so that a traceless $\tilde J_{ \mu\mu\mu\mu\nu\nu\nu\nu}$ can be written  
\be
\tilde J_{\mu\mu\mu\mu\nu\nu\nu\nu }&=&-\frac{2^{-2d} \,\pi ^{\frac 32-\frac d2}\, \left(k^2\right)^{\frac d2+2}}{(-1+e^{i \pi d})\,\Gamma (\frac{d+7}{2})}(3+a_3)(d+1)(d+3)\0\\
&&\cdot \left(\pi_{\mu\nu}^4-\frac 6{(d+3)}\pi_{\mu\nu}^2\pi_{\mu\mu}\pi_{\nu\nu}+\frac 3{(d+1)(d+3)}\pi^2_{\mu\mu}\pi^2_{\nu\nu}\right)  \label{J4S3}
\ee

As in the case of fermions, the transverse and traceless 2pt function for spin $s$ in even dimensions is proportional to
\be
\tilde J_{\mu_1\dots\mu_s\nu_1\dots\nu_s} &\sim& \sum_{l=0}^{\left\lfloor \frac{s}{2}\right\rfloor} \frac {(-1)^l}{2^{2l}l!} \frac {s!}{(s-2l)!} \frac{\Gamma\left(s+\frac{d-3}2-l\right)}{\Gamma\left(s+\frac {d-3}2\right)}\pi_{\mu\nu}^{s-2l} (\pi_{\mu\mu}\pi_{\nu\nu})^l\0\\
&=&  \,{_2F_1}\left(\frac{1-s}{2},-\frac{s}{2};\frac{5-d-2s}{2};\frac{\pi_{\mu\mu}\pi_{\nu\nu}}{\pi_{\mu\nu}^2}\right)\pi_{\mu\nu}^s\label{JsS1}
\ee

\section{4d full amplitudes}
\label{sec:4dfullamplitudes}

In view of the importance of the 4d case, we give in the following
complete explicit formulas in terms of elementary functions of the two point correlator for spin 1,2,3 in the fermionic model.

\subsection{spin 1}

\be 
&&\frac{i k^2}{\pi ^2}{(n_1\!\cdot\! \pi \!\cdot\! n_2) 
\left(-\frac{m^2}{3 k^2}+\sqrt{4 m^2-k^2} \left(\frac{m^2}{3 k^3}+\frac{1}{6 k}\right) \csc ^{-1}\left(\frac{2 m}{k}\right)+\frac{1}{12} L_1 - \frac{1}{18}\right)}\0
\ee
normalized with the understanding that in UV:
\be
&\sqrt{4 m^2-k^2} \rightarrow -i k +  i \frac{2m}{k^2} + \ldots \\
&\csc ^{-1}\left(\frac{2 m}{k}\right) \rightarrow \frac{1}{2} i \log \left(-\frac{k^2}{m^2}\right)-\frac{i m^2}{k^2} + \ldots
\ee
Here,
\be
L_n = \frac{2}{\varepsilon } + \log \left(\frac{m^2}{4 \pi }\right)+\gamma-\sum_{k=1}^{n}\frac{1}{k}
\ee

\subsection{spin 2}

\be
&&\frac{i}{\pi ^2}\frac{m^4}{4 k^4}
\left(k^2 \left(k\!\cdot\! n_2\right){}^2 (n_1\!\cdot\! \pi \!\cdot\! n_1)+2 k^2 (k\!\cdot\! n_1) (k\!\cdot\! n_2)( n_1\!\cdot\! \pi \!\cdot\! n_2)+k^2 \left(k\!\cdot\! n_1\right){}^2 (n_2\!\cdot\! \pi \!\cdot\! n_2)\right.
\0\\
&&\quad\quad\quad\quad\left.+2 \left(k\!\cdot\! n_1\right){}^2 \left(k\!\cdot\! n_2\right){}^2\right) L_2 \0\\
&&+\frac{i}{\pi ^2}k^4 \left(n_1\!\cdot\! \pi \!\cdot\! n_2\right){}^2 \left(-\frac{8 m^4}{15 k^4}+\frac{7 m^2}{360 k^2}+\left(\frac{m^4}{4 k^4}+\frac{m^2}{12 k^2}-\frac{1}{40}\right) L_2\right.\0\\
&&\quad\quad+\left.\sqrt{4 m^2-k^2} \left(\frac{8 m^4}{15 k^5}+\frac{m^2}{15 k^3}-\frac{1}{20 k}\right) \csc ^{-1}\left(\frac{2 m}{k}\right)+\frac{9}{400}\right)\0\\
 &&+ 
\frac{i}{\pi ^2}k^4 (n_1\!\cdot\! \pi \!\cdot\! n_1)( n_2\!\cdot\! \pi \!\cdot\! n_2) \left(-\frac{4 m^4}{15 k^4}+\frac{41 m^2}{360 k^2}+
\left(\frac{m^4}{4 k^4}-\frac{m^2}{12 k^2}+\frac{1}{120}\right) L_2\right.\0\\
&&\quad\quad+\left.\sqrt{4 m^2-k^2} \left(\frac{4 m^4}{15 k^5}-\frac{2 m^2}{15 k^3}+\frac{1}{60 k}\right) \csc ^{-1}\left(\frac{2 m}{k}\right)-\frac{47}{3600}\right)\0
\ee
\subsection{spin 3}

\be
&&i \left(n_1\!\cdot\! \pi \!\cdot\! n_2\right){}^3 k^6
\frac{1}{\pi ^2}
 \left(-\frac{128 m^6}{315 k^6}+\frac{32 m^4}{315 k^4}+\frac{218 m^2}{14175 k^2} + \left(\frac{8 m^6}{27 k^6}-\frac{4 m^2}{135 k^2}+\frac{4}{945}\right)L_3 \right.\0\\
&&\quad\quad\left.+\sqrt{4 m^2-k^2} \left(\frac{128 m^6}{315 k^7}-\frac{64 m^4}{945 k^5}-\frac{8 m^2}{189 k^3}+\frac{8}{945 k}\right) \csc ^{-1}\left(\frac{2 m}{k}\right)-\frac{428}{99225}\right)  \0\\
&& +
i \left(k\!\cdot\! n_1\right){}^2 \left(k\!\cdot\! n_2\right){}^2 (n_1\!\cdot\! \pi \!\cdot\! n_2) k^2
\frac{1}{\pi ^2}
{\left(\frac{m^4}{18 k^4}+ \left(\frac{37 m^6}{27 k^6}+\frac{m^4}{6 k^4}\right)L_3\right) }
\0\\
&&+ 
i \left(k\!\cdot\! n_1\right){}^3 \left(k\!\cdot\! n_2\right){}^3
\frac{1}{\pi ^2}
{\left(\frac{7 m^4}{72 k^4}+ \left(\frac{7 m^6}{9 k^6}+\frac{7 m^4}{24 k^4}\right)L_3\right) }
\0\\
&& +
i m^6 (k\!\cdot\! n_1)( k\!\cdot\! n_2) \left(n_1\!\cdot\! \pi \!\cdot\! n_2\right){}^2
\frac{8}{9 \pi ^2 k^2} { L_3 }
 \0\\
&&+
i (n_1\!\cdot\! \pi \!\cdot\! n_1)( n_1\!\cdot\! \pi \!\cdot\! n_2)( n_2\!\cdot\! \pi \!\cdot\! n_2) 
\frac{k^6}{\pi ^2}
\left(-\frac{64 m^6}{105 k^6}+\frac{257 m^4}{1512 k^4}-\frac{1877 m^2}{56700 k^2}\right.\0\\
&&\quad\quad+ \left(\frac{13 m^6}{27 k^6}-\frac{m^4}{8 k^4}+\frac{4 m^2}{135 k^2}-\frac{37}{15120}\right)L_3 \0\\
&&\quad\quad+\left.\sqrt{4 m^2-k^2} \left(\frac{64 m^6}{105 k^7}-\frac{152 m^4}{945 k^5}+\frac{187 m^2}{3780 k^3}-\frac{37}{7560 k}\right) \csc ^{-1}\left(\frac{2 m}{k}\right)+\frac{12433}{3175200}\right) 
\0\\
&&
+i \left(k^2 (k\!\cdot\! n_2)( n_2\!\cdot\! \pi \!\cdot\! n_2) \left(k\!\cdot\! n_1\right){}^3+k^2 \left(k\!\cdot\! n_2\right){}^3
 (n_1\!\cdot\! \pi \!\cdot\! n_1) (k\!\cdot\! n_1)\right)\0\\
&&\quad\quad\cdot \frac{1}{\pi ^2}
{\left(\frac{7 m^4}{144 k^4}+ \left(\frac{13 m^6}{27 k^6}+\frac{7 m^4}{48 k^4}\right)L_3\right) 
}
\0\\
&&+{13 i m^6 (k\!\cdot\! n_1) (k\!\cdot\! n_2)( n_1\!\cdot\! \pi \!\cdot\! n_1)( n_2\!\cdot\! \pi \!\cdot\! n_2)}
\frac{1}{27 \pi ^2 k^2} L_3 
\0\\
&&+
 i \left(\left(k\!\cdot\! n_2\right){}^2 (n_1\!\cdot\! \pi \!\cdot\! n_1)( n_1\!\cdot\! \pi \!\cdot\! n_2) k^4+\left(k\!\cdot\! n_1\right){}^2 (n_1\!\cdot\! \pi \!\cdot\! n_2)( n_2\!\cdot\! \pi \!\cdot\! n_2) k^4\right) \0\\
&&\quad\quad\cdot\frac{1}{\pi ^2}
 {\left(\frac{m^4}{144 k^4}+ \left(\frac{13 m^6}{27 k^6}+\frac{m^4}{48 k^4}\right)L_3\right)}\0
\ee

 \end{document}